\documentclass[twocolumn,showpacs,preprintnumbers,amsmath,amssymb]{revtex4}
\usepackage{graphicx}% Include figure files
\usepackage{dcolumn}% Align table columns on decimal point
\usepackage{bm}% bold math

\begin{document}

%\preprint{APS/123-QED}
\title{ Novel Practicable  GLHUA-i,i=1,2,3, Invisible  Cloak  Absorb the Incident Wave and Create Outgoing Wave Without Exceeding Light Speed Propagation and \\
New N Dimensional Maxwell Electromagnetic Equations}% Force line breaks with \\

\author{Jianhua Li}
 \altaffiliation[Also at ]{GL Geophysical Laboratory, USA, GLHUA@glgeo.com}%Lines break automatically or can be forced with \\
\author{  Feng Xie, Lee Xie, Ganquan Xie}%
 %\email{GLHUA-1@GLGEO.COM}
\affiliation{%
GL Geophysical Laboratory, USA
}%

\hfill\break
\author{Ganquan Xie}
\affiliation{
Chinese Dayuling Supercomputational Sciences Center, China \\
Hunan Super Computational Sciences Society, China
}%

\date{June 21, 2017}% It is always \today, today,

\begin{abstract}
In this version of our paper, we make two great breakthrough progress contribution: \\
  I. we propsed New N dimensional Maxwell Equation that is different from Gauge potential Maxwell field equation.
We propose new N dimensional Curl operator. The 3 dimensional curl operator is well defined, Howover, for N $N \ne 3$ dimensinal space, the curl operator is never  be defined in other publised papers before. As breakthough, we propose  new curl operator 
in N dimensinal space, $Curl_F  = (\nabla  \times )_N $:
\begin{equation}
Curl_F  = (\nabla  \times )_N  = \left( {\nabla \nabla  \cdot  - \nabla  \cdot \nabla } \right)^{\frac{1}{2}} ,       ( 1)
\end{equation}

and New N Dimensional Maxwell Equaions:
\begin{equation}
\begin{array}{l}
 Curl_F E = (\nabla  \times )_N E =  - \frac{{\partial B}}{{\partial t}}, \\ 
 \nabla  \cdot B = 0, \\ 
 Curl_F H = (\nabla  \times )_N H = \frac{{\partial D}}{{\partial t}} + J, \\ 
 \nabla  \cdot D = \rho ,                    ( 2) \\
 \end{array}
\end{equation}
For electric current or magnetic source, the analytic electromagnetic wave field solution of the new N dimensional Maxwell equation ( 2) are obtained.

For 3 dimension, our curl operator $Curl_F $ in ( 1) become standard 3D $curl$ 
\begin{equation}
curl_F E = (\nabla  \times )_3 E = \left| {\begin{array}{*{20}c}
   {\vec x_1 } & {\vec x_2 } & {\vec x_3 }  \\
   {\frac{\partial }{{\partial x_1 }}} & {\frac{\partial }{{\partial x_2 }}} & {\frac{\partial }{{\partial x_3 }}}  \\
   {E_{x_1 } } & {E_{x_2 } } & {E_{x_3 } }  \\
\end{array}} \right|                    ( 3) \\
\end{equation}
our N dimension Maxwell equation in ( 2) becomes standard 3D Maxwell equations.  
\begin{equation}
\begin{array}{l}
 \nabla  \times E = \left| {\begin{array}{*{20}c}
   {\vec x_1 } & {\vec x_2 } & {\vec x_3 }  \\
   {\frac{\partial }{{\partial x_1 }}} & {\frac{\partial }{{\partial x_2 }}} & {\frac{\partial }{{\partial x_3 }}}  \\
   {E_{x_1 } } & {E_{x_2 } } & {E_{x_3 } }  \\
\end{array}} \right| =  - \frac{{\partial B}}{{\partial t}}, \\ 
 \nabla  \times H = \left| {\begin{array}{*{20}c}
   {\vec x_1 } & {\vec x_2 } & {\vec x_3 }  \\
   {\frac{\partial }{{\partial x_1 }}} & {\frac{\partial }{{\partial x_2 }}} & {\frac{\partial }{{\partial x_3 }}}  \\
   {H_{x_1 } } & {H_{x_2 } } & {H_{x_3 } }  \\
\end{array}} \right| = \frac{{\partial D}}{{\partial t}} + \vec J, \\ 
 \nabla  \cdot B = 0, \\ 
 \nabla  \cdot D = \rho , \\ 
 \end{array} ( 4) \\
\end{equation}

Our N dimensional Maxwell equations have great role in the theoretical and practicable applications,  in particular, in electromgnetic invisible cloaks and in super sciences. This breakthough discover has been announced in March 15, 2020 in Hunan Super Computational Society (in Figure 51 ). The second author Dr. Feng Xie in GL Geophysical Laboratory and Stanford University made main and great contribution in this breakthough discovery
and creation

II. We propose three novel practicable GLHUA-1, GLHUA-2 and GLHUA-3 invisible  cloaks, some part in our cloak annular layer absorbs the incident wave and, in the meantimely, other part of our cloak annular layer creates outgoing wave
that make it to be invisible cloak and  analytical electromagnetic wave without exceeding light speed propagation
through it.
Ten more discoveries and creations are reported in this paper.
The copyright and patent for this paper are belong to the authors of this paper.\\
1. New practicable GLHUA-1, GLHUA-2 and GLHUA-3 annular layer electromagnetic (EM) invisible cloak with relative refractive index parameters large or equal to 1 are created.
In the GLHUA-1 cloak, all relative EM parameters ( 25) are large or equal to 1, ,$R_1 \le r \le R_2 $, $R_1>0$,$R_2>R_1$, $\varepsilon_r = \mu_r = \frac{{R_2 ^2 }}{{r^2 }}$, $\varepsilon_\theta = \varepsilon_\phi = \mu_\theta = \mu_\phi = \frac{{R_2 - R_1 }}{{r - R_1 }}$.  
In the GLHUA-2 invisible cloak, relative refractive index parameters ( 11) are large or equal to 1, the angular parameter ( 8) are large or equal to 1, the radial
parameters ( 7) are large than positive constant. The EM wave in GLHUA-1 and GLHUA-2 invisible cloak are physical bounded.
In GLHUA-3 invisible cloak, relative refractive index parameters ( 18) are large or equal to 1, the angular parameter ( 14) are large or equal to 1, the radial
parameters ( 15) are large than  0.\\
2. Analytical EM wave solutions of EM full wave equation with the GLHUA-1, GLHUA-2 and GLHUA-3 cloak materials are found in annular layer of the three cloaks.\\
 3. A GLHUA-1 expansion method  are propsed to find an exact analytical EM wave propagation in the GLHUA-1,GLHUA-2 and GLHUA-3 double layer cloaks. \\
4. We proposed new negative free space concept in arxiv:1612.02857. we define the negative free space in that the sphere radial is negative or zero. Also we define the positive free space in that sphere radial is positive or zero, we live in the positive space. We proved that the electromagnetic equations in the free positive space can analytical be continue to the negative space.\\ 
5. let  $ R $ be sphere radial variable in the basic positive and negative space, $ r $ be sphere radial variable in physical plsitive space. We propose a new GLHUANP-2 transformation from the negative space $ [ - \infty , - R_2 ] $ to spherical annual layer $ [R_1 ,R_2 ] $ in positive space, the positve space  $ [ R_2, \infty ] $ is invariant. The transformation GLHUANP-2 are total different from Pendry and ULF transformation. The new GLHUANP-2 transformation in this paper is,
\begin{equation}
r = R_1  - \frac{{(R_2  - R_1 )^2 }}{{R + R_1 }}, - \infty  < R \le  - R_2 ,R_1  \le r \le R_2 ,       ( 5)   
\end{equation}
The transformation maps negative infinity, $ R =  - \infty  $ , in negative space to  $ r = R_1 $ in the physical positive space; and maps $ R=- R_2 $ in negative space to $ r=R_2 $  in the physical positive space; 
The inverse transformation of ( 5) is

\begin{equation}
R =  - R_1  - \frac{{(R_2  - R_1 )^2 }}{{r - R_1 }},R_1  \le r \le R_2 , - \infty  < R \le  - R_2 ,    ( 6)
\end{equation}

6. Using the transformation GLHUANP-2 ( 5) and ( 6), a novel GLHUA-2 invisivle cloak is created, the relative radial electric permittivity $ \varepsilon _r $ and magnetic permeanbility $ \mu_r $ are derived to be,
\begin{equation}
\varepsilon _r  = \mu _r  = \frac{1}{{r^2 }}\frac{{(R_1 (r - R_1 ) + (R_2  - R_1 )^2 )^2 }}{{(R_2  - R_1 )^2 }},R_1  \le r \le R_2 .                         ( 7)
\end{equation}

The angular realtive electric permitivitty and magnetic permeability are 

\begin{equation}
\varepsilon _\theta   = \varepsilon _\phi   = \mu _\theta   = \mu _\phi   = \frac{{(R_2  - R_1 )^2 }}{{(r - R_1 )^2 }},R_1  \le r \le R_2 .               ( 8) 
\end{equation}
The properties of GLHUA-2 invisible cloak are as following:

(1).	The EM parameters are continuous on the $r=R_2$

\begin{equation}
\begin{array}{l}
 \varepsilon _r (R_2 ) = \mu _r (R_2 ) \\ 
  = \frac{1}{{R_2 ^2 }}\frac{{(R_1 (R_2  - R_1 ) + (R_2  - R_1 )^2 )^2 }}{{(R_2  - R_1 )^2 }} \\ 
  = \frac{1}{{R_2 ^2 }}R_2 ^2  = 1, \\        ( 9)
 \end{array}
\end{equation}

\begin{equation}
\varepsilon _r (R_2 ) = \mu _r (R_2 ) = 1,  ( 10)
\end{equation}
(2).	The refractive index 
\begin{equation}
1 \le n \le  + \infty ,  ( 11)  
\end{equation}

(3).	The EM wave propagation in the GLHUA-2   cloak without exceeding light speed difficulties.

(4) The relative radial electric permittivity $ \varepsilon _r $ and magnetic permeanbility $ \mu_r $ in ( 7) are large than zero.
The property ( 7) is important that makes the maganetic tense and electric tense in GLHUA-2 cloak are physical bounded.

(5) By GLHUA-1 analytical expansion method in this paper,we fund an exact analytical EM wave propagation in the GLHUA-2 cloak without
exceeding light speed propagation. The analytical electromaganetical wave propagation in the GLHUA-2    cloak prove that the GLHUA-2 cloak is completed practicable invisible cloak.
The radial electric wave propagation through the GLHUA-2 cloak is display in the figure 29-36.
The videos "Analytical maganetic wave propagation the practicable 
GLHUA-2 invisible cloak" are shnowing in \\

https://video.weibo.com/show?fid=1034:4412551845597236
\\
https://video.weibo.com/show?fid=1034:4413016675135960
\\
https://video.weibo.com/show?fid=1034:4413022027073708
\\
7.  GLHUANP-3 radial transformation,
\begin{equation}
r = R_1  + (R_2  - R_1 )e^{\frac{{R + R_2 }}{{R_2  - R_1 }}} ,     ( 12)
\end{equation}
which also maps negative infinity,   $- \infty $ to $R_1$, and maps $ -R_2 $ to $ R_2 $.

GLHUANP-3 inverse transformation is
\begin{equation}
R =  - R_2  + (R_2  - R_1 )\log \left( {\frac{{r - R_1 }}{{R_2  - R_1 }}} \right),       ( 13)
\end{equation}
From the GLHUANP-3 transformation ( 12), third novel GLHUA-3 invisible cloak with following relative electric permittivity and magnetic permeability are created as
\begin{equation}
\varepsilon _\theta   = \mu _\theta   = \varepsilon _\phi   = \mu _\phi   = \frac{{R_2  - R_1 }}{{r - R_1 }},      ( 14)
\end{equation}
\begin{equation}
\varepsilon _r  = \mu _r  = \left( { - R_2  + (R_2  - R_1 )\log \left( {\frac{{r - R_1 }}{{R_2  - R_1 }}} \right)} \right)^2 \frac{1}{{r^2 }}\frac{{r - R_1 }}{{R_2  - R_1 }},   ( 15)
\end{equation}

The properties of GLHUA-3 invisible cloak are as following:

(1).	The EM parameters are continuous on the $r=R_2$

\begin{equation}
\begin{array}{l}
 \varepsilon _\theta  (R_2 ) = \mu _\theta  (R_2 ) = \varepsilon _\phi  (R_2 ) = \mu _\phi  (R_2 ) \\ 
  = \frac{{R_2  - R_1 }}{{r - R_1 }}(R_2 ) = 1, \\  
 \end{array},  ( 16)   
\end{equation}
\begin{equation}
\varepsilon _r (R_2 ) = \mu _r (R_2 ) = 1,  ( 17)
\end{equation}
(2).	The refractive index 
\begin{equation}
1 \le n \le  + \infty ,  ( 18)  
\end{equation}

(3).	The EM wave propagation in the GLHUA-3 cloak without exceeding light speed difficulties.

The radial electric wave propagation through the GLHUA-3 cloak is display in the figure 43-50.
(4). We find analytical EM wave without exceeding light speed propagation through ( 26) cloak.
 8. In GLHUA-1, GLHUA-2 and GLHUA-3 cloaks, some part of the cloak layer
 absorb the incident wave and, in the meantimely, other part of
cloak layer create outgoing wave
that make they to be invisible cloak and  analytical electromagnetic wave without exceeding light speed propagation
through them.\\
9. GLHUAF transformation is proposed.\\
\begin{equation}
r = R_1  + (R_2  - R_1 )e^{\frac{{R_2  - R}}{{R_2  - R_1 }}}     ( 19)
\end{equation}
Which maps positive infinity $ + \infty $ to $R_1$, and $R_2$ is invariant.\\

10. By GLHUAF transformation, a novel GLHUAF
invisible cloak with double negative materials is
created. 

Use the GLHUAF transformation to transform the radial magnetic equation in free space from positive infinity $ + \infty $ to $R_1$, and $R_2$ is invariant.
\begin{equation}
\begin{array}{l}
 \frac{\partial }{{\partial r}}\frac{1}{{\mu _\theta  }}\frac{{\partial H}}{{\partial r}} + \frac{1}{{r^2 \mu _r }}\frac{1}{{\sin \theta }}\frac{\partial }{{\partial \theta }}\sin \theta \frac{{\partial H}}{{\partial \theta }} \\ 
  + \frac{1}{{r^2 \mu _r }}\frac{1}{{\sin ^2 \theta }}\frac{{\partial ^2 H}}{{\partial \phi ^2 }} + k^2 \varepsilon _\theta  H = M_s  \\ 
 \end{array},                     ( 20)
\end{equation}
novel double negative GLHUAF relative EM parameters are created,
\begin{equation}
\varepsilon _\theta   = \mu _\theta   = \varepsilon _\phi   = \mu _\phi   =  - \frac{{R_2  - R_1 }}{{r - R_1 }},                    ( 21)
\end{equation}
\begin{equation}
\varepsilon _r  = \mu _r  =  - \frac{{R^2 }}{{r^2 }}\frac{{r - R_1 }}{{R_2  - R_1 }},\;\;\;\;\;\;\;\;( 22)
\end{equation}
By the inverse transformation of ( 19), in the ( 21), the
\begin{equation}
R = R_2  - (R_2  - R_1 )\log \left( {\frac{{r - R_1 }}{{R_2  - R_1 }}} \right),\;\;\;\;\;\;\;\;\;( 23)
\end{equation}
\begin{equation}
\varepsilon _r  = \mu _r  =  - \frac{1}{{r^2 }}\frac{{r - R_1 }}{{R_2  - R_1 }}\left( {R_2  - (R_2  - R_1 )\log \left( {\frac{{r - R_1 }}{{R_2  - R_1 }}} \right)} \right)^2 ,\;\;\;\;\;( 24)
\end{equation}
By installing the GLHUAF relative double negative electric permittivity and magnetic permeability ( 21) and ( 24) in the sphere annular layer $R_1  < r \le R_2 $ and installing the free space in the inner sphere $r \le R_1 $, the GLHUAF invisible cloak is created. 
Novel EM wave propagation through the GLHUAF invisible cloak are presented in Figure 37 to the Figure 42.

We discover a new GLHUA-1 invisible cloak and new EM parameters as follows:
 \begin{equation}
\begin{array}{l}
 \varepsilon _r  = \mu _r  = {{R_2^2 } \mathord{\left/
 {\vphantom {{R_2^2 } {r^2 }}} \right.
 \kern-\nulldelimiterspace} {r^2 }}, \\ 
 \varepsilon _\theta   = \varepsilon _\phi   = \mu _\theta   = \mu _\phi   = {{(R_2  - R_1 )} \mathord{\left/
 {\vphantom {{(R_2  - R_1 )} {(r}}} \right.
 \kern-\nulldelimiterspace} {(r}} - R_1 ),              (( 25)) \\ 
 \end{array}
\end{equation}      
We discover an analytical exact radial electric and magnetic wave propagation to satisfy the anisotropic radial electric and magnetic wave equation in the outer annular layer  $ R_1  \le r \le R_2 $ with the above GLHUA-1 relative EM parameters and in the inner sphere $ 0 \le r \le R_1 $ with free space relative EM parameters 1, where $0 < R_1  < R_2 $. Moreover, we discover the analytical exact EM wave propagation to satisfy 3D Maxwell EM equation with GLHUA-1 anisotropic invisible cloak material in the annular layer $ R_1  \le r \le R_2 $ and free space inner sphere $ 0 \le r \le R_1 $

It is totally different from the Pendry transform , by using GL no scattering modeling and inversion in [1-3], in this paper, we discover a new GLHUA-1 outer annular layer electromagnetic (EM) invisible cloak with relative parameter not less than 1. The outer layer cloak can be any positive thickness. In GLHUA-1 outer annular layer cloak, $R_1  \le r \le R_2 $,$R_1>0$,$R_2>R_1$,
$\varepsilon _r  = \mu _r  = \frac{{R_2 ^2 }}{{r^2 }}$,
$\varepsilon _\theta   = \varepsilon _\phi   = \mu _\theta   = \mu _\phi   = \frac{{R_2  - R_1 }}{{r - R_1 }}$. We discovered an exact analytical EM wave propagation solution of the full wave Maxwell EM equation in the GLHUA-1 outer cloak. In GLHUA-1 outer annular layer cloak, $R_1  \le r \le R_2 $, for the electric source
located outside of GLHUA-1 outer layer cloak, $r_s  > R_2 $,  the exact analytic radial EM wave field solution of Maxwell radial EM differential equation is as
follows VERSION 1

\begin{equation}
\begin{array}{l}
 \left[ {\begin{array}{*{20}c}
   {E_r (\vec r,\vec r_s )}  \\
   {H_r (\vec r,\vec r)}  \\
\end{array}} \right] =  \\ 
  = \sum\limits_{l = 1}^p  {\left[ {\begin{array}{*{20}c}
   {E_{r,l,1} (r,r_s )}  \\
   {H_{r,l,1} (r,r_s )}  \\
\end{array}} \right]} \cos \left( {k(R_2  - R_1 )\log \left( {\frac{{r - R_1 }}{{R_2  - R_1 }}} \right)} \right) \\ 
 \sum\limits_{m =  - l}^l {} \left[ {\begin{array}{*{20}c}
   {D_e (\theta ,\phi )}  \\
   {D_h (\theta ,\phi )}  \\
\end{array}} \right]Y_l^m (\theta ,\phi )Y_l^{m*} (\theta _s ,\phi _s ) \\ 
  + \sum\limits_{l = 1}^\infty  {\left[ {\begin{array}{*{20}c}
   {E_{r,l,2} (r,r_s )}  \\
   {H_{r,l,2} (r,r_s )}  \\
\end{array}} \right]} \sin \left( {k(R_2  - R_1 )\log \left( {\frac{{r - R_1 }}{{R_2  - R_1 }}} \right)} \right) \\ 
 \sum\limits_{m =  - l}^l {} \left[ {\begin{array}{*{20}c}
   {D_e (\theta ,\phi )}  \\
   {D_h (\theta ,\phi )}  \\
\end{array}} \right]Y_l^m (\theta ,\phi )Y_l^{m*} (\theta _s ,\phi _s ). \\ 
 \end{array} \\ (( 26))
\end{equation}
and VERSION 2

\begin{equation}
\begin{array}{l}
 \left[ {\begin{array}{*{20}c}
   {E_r (\vec r)}  \\
   {H_r (\vec r)}  \\
\end{array}} \right] =  \\ 
  = \sum\limits_{l = 1}^\infty  {\left[ {\begin{array}{*{20}c}
   {E_{r,l,1} (\vec r)}  \\
   {H_{r,l,1} (\vec r)}  \\
\end{array}} \right]} \cos \left( {k(R_2  - R_1 )\log \left( {\frac{{r - R_1 }}{{R_2  - R_1 }}} \right) + kR_2 } \right) \\ 
 \sum\limits_{m =  - l}^l {} \left[ {\begin{array}{*{20}c}
   {D_e (\theta ,\phi )}  \\
   {D_h (\theta ,\phi )}  \\
\end{array}} \right]Y_l^m (\theta ,\phi )Y_l^{m*} (\theta _s ,\phi _s ) \\ 
  + \sum\limits_{l = 1}^\infty  {\left[ {\begin{array}{*{20}c}
   {E_{r,l,2} (\vec r)}  \\
   {H_{r,l,2} (\vec r)}  \\
\end{array}} \right]} \sin \left( {k(R_2  - R_1 )\log \left( {\frac{{r - R_1 }}{{R_2  - R_1 }}} \right) + kR_2 } \right) \\ 
 \sum\limits_{m =  - l}^l {} \left[ {\begin{array}{*{20}c}
   {D_e (\theta ,\phi )}  \\
   {D_h (\theta ,\phi )}  \\
\end{array}} \right]Y_l^m (\theta ,\phi )Y_l^{m*} (\theta _s ,\phi _s ). \\ 
 \end{array} \\ (( 27))
\end{equation}
In the free space outside of the cloak, $r > R_2 $,$E_r (\vec r) = E_{b,r} (\vec r)$ $H_r (\vec r) = H_{b,r} (\vec r)$, on the boundary $r=R_2$, $E_r (\vec r)|_{r = R_2 }  = E_{b,r} (\vec r)|_{r = R_2 } $,$H_r (\vec r)|_{r = R_2 }  = H_{b,r} (\vec r)|_{r = R_2 } $, and the derivative
${\raise0.7ex\hbox{${\partial E_r (\vec r)}$} \!\mathord{\left/
 {\vphantom {{\partial E_r (\vec r)} {\partial r}}}\right.\kern-\nulldelimiterspace}
\!\lower0.7ex\hbox{${\partial r}$}}|_{r = R_2 }  = {\raise0.7ex\hbox{${\partial E_{b,r} (\vec r)}$} \!\mathord{\left/
 {\vphantom {{\partial E_{b,r} (\vec r)} {\partial r}}}\right.\kern-\nulldelimiterspace}
\!\lower0.7ex\hbox{${\partial r}$}}|_{r = R_2 } $,
${\raise0.7ex\hbox{${\partial H_r (\vec r)}$} \!\mathord{\left/
 {\vphantom {{\partial H_r (\vec r)} {\partial r}}}\right.\kern-\nulldelimiterspace}
\!\lower0.7ex\hbox{${\partial r}$}}|_{r = R_2 }  = {\raise0.7ex\hbox{${\partial H_{b,r} (\vec r)}$} \!\mathord{\left/
 {\vphantom {{\partial H_{b,r} (\vec r)} {\partial r}}}\right.\kern-\nulldelimiterspace}
\!\lower0.7ex\hbox{${\partial r}$}}|_{r = R_2 } $. The exact analytical radial EM wave solution show that there is no scattering from the GLHUA-1 outer layer cloak to disturb the incident EM wave outside of the cloak in free space. The exact analytical radial EM wave solution show that the EM wave can not propagation penetrate into the sphere $r < R_1 $.In the free space inside of the concealment, $r < R_1 $,$E_r (\vec r) = 0$ $H_r (\vec r) = 0$. Therefore, GLHUA-1 outer layer is practicable invisible cloak with cloaked concealment sphere $r < R_1 $. We discover two novel  wave in the GLHUA-1 outer layer cloak, $R_1  \le r \le R_2$, one is enter wave  from incident which is absorbed, other one is created wave by GLHUA-1 cloak materials, the absorbtion wave and created wave make that the EM wave propagation through the GLHUA-1 cloak without exceeding light speed propagation. We propose a GLHUA-1 expansion method to find an exact analytical EM wave propagation in the GLHUA-1 double cloak and other spherical annular layer cloak, and completely solved invisible cloak problem in the spherical annular devices.

We discover negative space first in the world. The space in which the radial variable in the sphere coordinate is  negative is called negative space. The space in which the radial variable in the sphere coordinate is positive or zero is called positive space, that is real three dimensional space we are living.  The positive space is visible, but  the negative space is invisible.  
The door between the positive space and negative space is $ 0 $. In normal general sciences,the positive space and negative space are not connected, the door $ 0 $ between the positive space and negative space is closed. Up to now,
all scientists are working in the positive space, all sciences principle,
for example, conservation of energy law, are hold and satisfied in the positive space. We proved that the radial EM equation and acoustic equation and their solutions in the sphere coordinate in the positive space can be analytic continuation into negative space with negative radial $R < 0$..

We create GLHUANP-2 radial transformation  and GLHUA-2 invisible cloak. We creat GLHUANP-3 radial transformation and GLHUA-3 invisible cloak.

The idea, discovers, creations  in this paper are breakthrough progress and different from all other cloak in other research publications. Our analytical EM wave in GLHUA-1, GLHUA-2 and GLHUA-3 cloak is undisputed evidence and rigorous proof to prove that the GLHUA-1, GLHUA-2 and GLHUA-3 double layer cloak are practicable invisible cloak without exceed light speed propagation.
Copyright and patent of the GLHUA-1, GLHUA-2 and GLHUA-3 EM cloaks and GL modeling
and inversion methods and all rights are reserved by authors in GL Geophysical Laboratory.

% \vskip0.5mm {\bf DOI}:~10.2529/PIERSxxxxxxxxxxxx
%%\hskip1cm {\small \it (Received July 10, 2006, Published July %31, 2006)}
\end{abstract}
%\end {document}
\pacs{13.40.-f, 41.20.-q, 41.20.jb,42.25.Bs}% PACS, the Physics and Astronomy
\maketitle

\hfill \break

$\boldsymbol {CONTENTS}$  \\

 I.
\\ SCIENCES DISCOVER REPORT
\hfill \break

II. 	
\\ INTRODUCTION
\hfill \break

III.          
\\  PRACTICABLE GLHUA-1 OUTER LAYER
\\ CLOAKWITH RELATIVE PARAMETER NOT
\\  LESS THAN 1

\hfill \break

IV. 
\\    EXACT ANALYTICAL EM WAVE
\\ PROPAGATION IN THE GLHUA-1 OUTER
\\ LAYER CLOAK
\hfill \break

V.    
 \\ SIMULATION OF THE EXACT ANALYTIC
\\ MAGNETIC WAVE PROPAGATION THROUGH
\\ THE GLHUA-1 OUTER LAYER CLOAK
\hfill \break

VI.    
 \\  DISCUSSION AND CONCLUSION ON EXACT ANALYTIC EM WAVE IN GLHUAII CLOAK
\hfill \break

VII.      
\\ GLHUANP-4
\\ ELECTROMAGNETIC INVISIBLE CLOAK AND
\\ NOVEL GLHUANP-4 SPHERICAL RADIAL
\\ TRANSFORMATION MAPS THE NEGATIVE
\\ INFINITY $ - \infty  $ TO $R_1$ AND $R_2$ TO $R_2$

\hfill \break

VIII.
\\ NOVEL GLHUAF INVISIBLE CLOAK 
\\ WITH DOUBLE NEGATIVE EM PARAMETER 
\\ AND SQUARE OF REFRACTIVE INDEX $ n^2 \ge 1 $

IX.
\\ GLHUA-3 INVISIBLE CLOAK BY 
\\ TRANSFORMATION WHICH MAPS 
\\ NEGATIVE  INFINITY TO $R_1$ 
\\ AND MAPAS $- R_2$ TO $R_2$

  %             $\boldsymbol {Special \ Breakthrough \ Discover Report }$  \\

\section {SCIENCES DISCOVER REPORT}

\subsection {New N dimensional Maxwell Equation}
 we propsed New N dimensional Maxwell Equation that is different from Gauge potential Maxwell field equation.
We propose new N dimensional Curl operator. The 3 dimensional curl operator is well defined, Howover, for N $N \ne 3$ dimensinal space, the curl operator is never  be defined in other publised papers before. As breakthough, we propose  new curl operator 
in N dimensinal space, $Curl_F  = (\nabla  \times )_N $:
\[
%\begin{equation}
Curl_F  = (\nabla  \times )_N  = \left( {\nabla \nabla  \cdot  - \nabla  \cdot \nabla } \right)^{\frac{1}{2}} ,       ( 1)
%\end{equation}
\]

and New N Dimensional Maxwell Equaions:
\[
%\begin{equation}
\begin{array}{l}
 Curl_F E = (\nabla  \times )_N E =  - \frac{{\partial B}}{{\partial t}}, \\ 
 \nabla  \cdot B = 0, \\ 
 Curl_F H = (\nabla  \times )_N H = \frac{{\partial D}}{{\partial t}} + J, \\ 
 \nabla  \cdot D = \rho ,                    ( 2) \\
 \end{array}
%\end{equation}
\]
For electric current or magnetic source, the analytic electromagnetic wave field solution of the new N dimensional Maxwell equation ( 2) are obtained.

For 3 dimension, our curl operator $Curl_F $ in ( 1) become standard 3D $curl$ 
\[
%\begin{equation}
curl_F E = (\nabla  \times )_3 E = \left| {\begin{array}{*{20}c}
   {\vec x_1 } & {\vec x_2 } & {\vec x_3 }  \\
   {\frac{\partial }{{\partial x_1 }}} & {\frac{\partial }{{\partial x_2 }}} & {\frac{\partial }{{\partial x_3 }}}  \\
   {E_{x_1 } } & {E_{x_2 } } & {E_{x_3 } }  \\
\end{array}} \right|                    ( 3) \\
%\end{equation}
\]
our N dimension Maxwell equation in ( 2) becomes standard 3D Maxwell equations.
\[  
%\begin{equation}
\begin{array}{l}
 \nabla  \times E = \left| {\begin{array}{*{20}c}
   {\vec x_1 } & {\vec x_2 } & {\vec x_3 }  \\
   {\frac{\partial }{{\partial x_1 }}} & {\frac{\partial }{{\partial x_2 }}} & {\frac{\partial }{{\partial x_3 }}}  \\
   {E_{x_1 } } & {E_{x_2 } } & {E_{x_3 } }  \\
\end{array}} \right| =  - \frac{{\partial B}}{{\partial t}}, \\ 
 \nabla  \times H = \left| {\begin{array}{*{20}c}
   {\vec x_1 } & {\vec x_2 } & {\vec x_3 }  \\
   {\frac{\partial }{{\partial x_1 }}} & {\frac{\partial }{{\partial x_2 }}} & {\frac{\partial }{{\partial x_3 }}}  \\
   {H_{x_1 } } & {H_{x_2 } } & {H_{x_3 } }  \\
\end{array}} \right| = \frac{{\partial D}}{{\partial t}} + \vec J, \\ 
 \nabla  \cdot B = 0, \\ 
 \nabla  \cdot D = \rho , \\ 
 \end{array} ( 4) \\
%\end{equation}
\]
Our N dimensional Maxwell equations have great role in the theoretical and practicable applications,  in particular, in electromgnetic invisible cloaks and in super sciences.  This breakthough discover has been announced in March 15, 2020 in Hunan Super Computational Society (in Figure 51 ). The second author Dr. Feng Xie in GL Geophysical Laboratory and Stanford University made main and great contribution in this breakthough discovery and creation

\subsection {Super Sciences and General Sciences}
To study visible natural sciences and world and invisible
social science, thinking sciences an human society are main
objects of the general sciences. Inversely, To study the invisible natural sciences and invisible natural world and
visible social and thinking sciences are main objects of the novel Super Sciences. N $(N > 3)$ dimansional Maxwell equations is usefull to study combination sciences of natural science and social science. Our GLHUA-1 practicable double cloak and Easton LaChappelles mind control robot hands show that the new novel super science is being born.

\subsection {Negative Space and Positive Space}
We discover negative space first in the world. We proposed new negative free space concept in arXiv:1612.02583[1]. we define the negative free space in which the sphere radial is negative or zero. Also we define the positive free space in which the sphere radial is positive or zero. that positive space is real three dimensional space we are living.  The positive space is visible, but  the negative space is invisible.  
The door between the positive space and negative space is $ 0 $. In normal general sciences,the positive space and negative space are not connected, the door $ 0 $ between the positive space and negative space is closed. Up to now,
all scientists are working in the positive space, all sciences principle,for example, conservation of energy law, are hold and satisfied in the positive space. We proved that the radial EM equation and acoustic equation and their solutions in the sphere coordinate in the positive space can be analytic continuation into negative space with negative radial $R < 0$..

We proved that the homogeneous Maxwell electromagnetic equations and their solutions in the free positive space $r \ge 0$ can analyticaly be continue to the negative free space $r \le 0$. The detailed proof will given in the following statement 1.

The second order anisotropic homogeneous Maxwell electromagnetic equation system in the sphere coordinate positive space

In paper [2], the anisotropic homogeneous Maxwell electromagnetic equation system in the sphere coordinate positive space, $r \ge 0 $, can be translated to separate second order radial electric and magnetic tense wave field equations, the homogeneous radial electric and magnetic tense wave field equation (5), (6) in paper [2] is 

\begin{equation}
\begin{array}{l}
 \frac{\partial }{{\partial r}}\left[ {\begin{array}{*{20}c}
   {\frac{1}{{\varepsilon _\theta  }}\frac{{\partial r^2 \varepsilon _r E_r }}{{\partial r}}}  \\
   {\frac{1}{{\mu _\theta  }}\frac{{\partial r^2 \mu _r H_r }}{{\partial r}}}  \\
\end{array}} \right] + \frac{1}{{\sin \theta }}\frac{\partial }{{\partial \theta }}\sin \theta \frac{\partial }{{\partial \theta }}\left[ {\begin{array}{*{20}c}
   {E_r }  \\
   {H_r }  \\
\end{array}} \right] \\ 
  + \frac{1}{{\sin \theta }}\frac{{\partial ^2 }}{{\partial \phi ^2 }}\left[ {\begin{array}{*{20}c}
   {E_r }  \\
   {H_r }  \\
\end{array}} \right] + k^2 r^2 \left[ {\begin{array}{*{20}c}
   {\mu _\theta  \varepsilon _r E_r }  \\
   {\varepsilon _\theta  \mu _r H_r }  \\
\end{array}} \right] = 0, \\ 
 \end{array}  ( 28)
\end{equation}

Where $ E_r$ is the radial electric tense wave field, $H_r$ is the radial magnetic tense wave field, other angular electric tense wave field $E_\theta  $, $E_\phi  $ and angular magnetic tense wave field  $H_\theta  $, $H_\phi  $ , will be found from (7)-(10) in paper [2], the electric permittivity are $(\varepsilon _r ,\varepsilon _\theta  ,\varepsilon _\phi  )\varepsilon _0 ,$ $\varepsilon _r $,$\varepsilon _\theta   = \varepsilon _\phi  $ are relative electric permittivity, the magnetic permeability $(\mu _r ,\mu _\theta  ,\mu _\phi  )\mu _0 ,$  $\mu _r $, 
$\mu _\theta   = \mu _\phi  $ are relative magnetic permeability. $\varepsilon _0  = {\rm 0}{\rm .8854d - 11}$  is the basic electric permittivity in the free space, $\mu _0  = 4\pi  \times 10^{ - 7} $
 is the basic magnetic permeability in the free space, $c = 1/\sqrt {\varepsilon _0 \mu _0 } $ is the light speed in the free space, $k^2  = \omega ^2 \varepsilon _0 \mu _0 $ is wave number, 
$\omega ^2  = 2\pi f$ is angular frequency, $f$ is frequency.
We define a new radial electric and magnetic wave field,

\begin{equation}
\left[ {\begin{array}{*{20}c}
   {E(r,\theta ,\phi )}  \\
   {H(r,\theta ,\phi )}  \\
\end{array}} \right] = \left[ {\begin{array}{*{20}c}
   {r^2 \varepsilon _r E_r (r,\theta ,\phi )}  \\
   {r^2 \mu _r H_r (r,\theta ,\phi )}  \\
\end{array}} \right], ( 29)
\end{equation}

Substitute ( 29) into equation ( 28), the equation ( 28) becomes a new electromagnetic wave field equation

\begin{equation}
\begin{array}{l}
 \frac{\partial }{{\partial r}}\left[ {\begin{array}{*{20}c}
   {\frac{1}{{\varepsilon _\theta  }}\frac{{\partial E}}{{\partial r}}}  \\
   {\frac{1}{{\mu _\theta  }}\frac{{\partial H}}{{\partial r}}}  \\
\end{array}} \right] + \frac{1}{{r^2 \mu _r \sin \theta }}\frac{\partial }{{\partial \theta }}\sin \theta \frac{\partial }{{\partial \theta }}\left[ {\begin{array}{*{20}c}
   {\frac{E}{{\varepsilon _r }}}  \\
   {\frac{H}{{\mu _r }}}  \\
\end{array}} \right] \\ 
  + \frac{1}{{r^2 \mu _r \sin \theta }}\frac{{\partial ^2 }}{{\partial \phi ^2 }}\left[ {\begin{array}{*{20}c}
   {\frac{E}{{\varepsilon _r }}}  \\
   {\frac{H}{{\mu _r }}}  \\
\end{array}} \right] + k^2 \left[ {\begin{array}{*{20}c}
   {\mu _\theta  E}  \\
   {\varepsilon _\theta  H}  \\
\end{array}} \right] = 0 \\ 
 \end{array} ( 30)
\end{equation}

In the next, we study the electric and magnetic tense wave field equation ( 28) and new electric and magnetic wave field equation ( 30). In the free space the relative electric permittivity
 $\varepsilon _r  = \varepsilon _\theta   = \varepsilon _\phi   = 1$, the relative magnetic permeability $\mu _r  = \mu _\theta   = \mu _\phi   = 1$, the homogeneous radial electric and magnetic tense wave field equation ( 28) in the sphere coordinate with radial variable, 
$r \ge 0$, is reduced to

\begin{equation}
\begin{array}{l}
 \frac{\partial }{{\partial r}}\frac{\partial }{{\partial r}}\left[ {\begin{array}{*{20}c}
   {r^2 E_r }  \\
   {r^2 H_r }  \\
\end{array}} \right] + \frac{1}{{\sin \theta }}\frac{\partial }{{\partial \theta }}\sin \theta \frac{\partial }{{\partial \theta }}\left[ {\begin{array}{*{20}c}
   {E_r }  \\
   {H_r }  \\
\end{array}} \right] \\ 
  + \frac{1}{{\sin \theta }}\frac{{\partial ^2 }}{{\partial \phi ^2 }}\left[ {\begin{array}{*{20}c}
   {E_r }  \\
   {H_r }  \\
\end{array}} \right] + k^2 r^2 \left[ {\begin{array}{*{20}c}
   {E_r }  \\
   {H_r }  \\
\end{array}} \right] = 0 \\ 
 \end{array}  ( 31)
\end{equation}

The new magnetic wave field equation ( 32) in free space becomes to, 
\begin{equation}
\left[ {\begin{array}{*{20}c}
   {E(r,\theta ,\phi )}  \\
   {H(r,\theta ,\phi )}  \\
\end{array}} \right] = \left[ {\begin{array}{*{20}c}
   {r^2 E_r (r,\theta ,\phi )}  \\
   {r^2 H_r (r,\theta ,\phi )}  \\
\end{array}} \right] ( 32)
\end{equation} 
$E_r (r,\theta ,\phi )$ is radial electric tense, $\varepsilon _0 E_r (r,\theta ,\phi )
$ is radial electric displacement, $H_r (r,\theta ,\phi )$ is radial magnetic tense, $\mu _0 H_r (r,\theta ,\phi )$ is radial magnetic flux in the free space. 
\begin{equation}
\left[ {\begin{array}{*{20}c}
   {D_r }  \\
   {B_r }  \\
\end{array}} \right] = \left[ {\begin{array}{*{20}c}
   {\varepsilon _0 E_r (r,\theta ,\phi }  \\
   {\mu _0 H_r (r,\theta ,\phi }  \\
\end{array}} \right] ( 33)
\end{equation},                    

 Substitute ( 32) into ( 31),  the equation ( 31) becomes to electric and magnetic wave field equation in the free space,
\begin{equation}
\begin{array}{l}
 \frac{\partial }{{\partial r}}\frac{\partial }{{\partial r}}\left[ {\begin{array}{*{20}c}
   E  \\
   H  \\
\end{array}} \right] + \frac{1}{{r^2 \sin \theta }}\frac{\partial }{{\partial \theta }}\sin \theta \frac{\partial }{{\partial \theta }}\left[ {\begin{array}{*{20}c}
   E  \\
   H  \\
\end{array}} \right] \\ 
  + \frac{1}{{r^2 \sin \theta }}\frac{{\partial ^2 }}{{\partial \phi ^2 }}\left[ {\begin{array}{*{20}c}
   E  \\
   H  \\
\end{array}} \right] + k^2 \left[ {\begin{array}{*{20}c}
   E  \\
   H  \\
\end{array}} \right] = 0 \\ 
 \end{array}  ( 34)
\end{equation},      

${\boldsymbol{Statement \ 1:}}$  The homogeneous electromagnetic wave field equations and their solutions in the free positive space $r \ge 0$ can analyticaly be continue to the negative free space $r \le 0$.

Proof: Suppose that the radial electric and magnetic wave field $E(r,\theta ,\phi )$ and $H(r,\theta ,\phi ) $, satisfy the radial electric magnetic wave field equation ( 34).  By direct calculation $E(r,\theta ,\phi )$ and $H(r,\theta ,\phi ) $,  also satisfy the equation ( 34).

For $R < 0 $, Substitute $r =  - R > 0$, into the ( 34), bacause
 
\begin{equation}
\begin{array}{l}
 \frac{\partial }{{\partial ( - R)}}\frac{\partial }{{\partial ( - R)}}\left[ {\begin{array}{*{20}c}
   {E( - ( - R),\theta ,\phi )}  \\
   {H( - ( - R),\theta ,\phi )}  \\
\end{array}} \right] \\ 
  + \frac{1}{{\sin \theta }}\frac{\partial }{{\partial \theta }}\sin \theta \frac{\partial }{{\partial \theta }}\left[ {\begin{array}{*{20}c}
   {E( - ( - R),\theta ,\phi )}  \\
   {H( - ( - R),\theta ,\phi )}  \\
\end{array}} \right] \\ 
  + \frac{1}{{\sin \theta }}\frac{{\partial ^2 }}{{\partial \phi ^2 }}\left[ {\begin{array}{*{20}c}
   {E( - ( - R),\theta ,\phi )}  \\
   {H( - ( - R),\theta ,\phi )}  \\
\end{array}} \right] \\ 
  + k^2 \left[ {\begin{array}{*{20}c}
   {E( - ( - R),\theta ,\phi )}  \\
   {H( - ( - R),\theta ,\phi )}  \\
\end{array}} \right] = 0, \\ 
 \end{array}  ( 35)
\end{equation}

We have 
\begin{equation}
\begin{array}{l}
 \frac{\partial }{{\partial R}}\frac{\partial }{{\partial R}}\left[ {\begin{array}{*{20}c}
   {E(R,\theta ,\phi )}  \\
   {H(R,\theta ,\phi )}  \\
\end{array}} \right]  \\
+ \frac{1}{{R^2 \sin \theta }}\frac{\partial }{{\partial \theta }}\sin \theta \frac{\partial }{{\partial \theta }}\left[ {\begin{array}{*{20}c}
   {E(R,\theta ,\phi )}  \\
   {H(R,\theta ,\phi )}  \\
\end{array}} \right] \\ 
  + \frac{1}{{R^2 \sin \theta }}\frac{{\partial ^2 }}{{\partial \phi ^2 }}\left[ {\begin{array}{*{20}c}
   {E(R,\theta ,\phi )}  \\
   {H(R,\theta ,\phi )}  \\
\end{array}} \right] \\ 
  + k^2 \left[ {\begin{array}{*{20}c}
   {E(R,\theta ,\phi )}  \\
   {H(R,\theta ,\phi )}  \\
\end{array}} \right] = 0,forR < 0, \\ 
 \end{array}       ( 36)
\end{equation}

From equation ( 36), the homogeneous radial electric and magnetic equations ( 34) and their solutions $E(r,\theta ,\phi )$ and $H(r,\theta ,\phi )$ in the free positive space $r \ge 0$ has analyticaly been continue to the negative free space $R \le 0$.  Substitue $E(R,\theta ,\phi ) = R^2 E_r (R,\theta ,\phi )$ and
$H(R,\theta ,\phi ) = R^2 H_r (R,\theta ,\phi )$  in ( 32) into ( 34), the homogeneous radial electric and magnetic tense field equations ( 31) and their solutions $E_r (r,\theta ,\phi )$ and $H_r (r,\theta ,\phi )$ in the free positive space $r \ge 0$ can analyticaly be continue to the negative free space $R \le 0$.  From (7), (8), (9) and (10) in paper [2], the homogeneous angular electromagnetic field equations and their solutions $E_\theta  (r,\theta ,\phi )$, $E_\phi  (r,\theta ,\phi )$, $H_\theta  (r,\theta ,\phi )$, $H_\phi  (r,\theta ,\phi )$, in the free positive space $r \ge 0$ can analyticaly be continue to the negative free space $R \le 0$.  The statement is proved.

\subsection {A novel GLHUA-2 electromaganetic invisible cloak
With Refractive Index $n \ge 1$ that absorbs incoming wave and creates outgoing wave }
A novel GLHUA-2 electromaganetic invisible cloak

In this paper, we create a novel GLHUA-2 electromaganetic invisible cloak with electromagnetic parameters depended on radial r in sphere anular layer $R_1  \le r \le R_2 $.  The relative rafractive index of the invisible cloak is large or equal to one, \\
\[ n(r) = {{\left( {R_1  + {{(R_2  - R_1 )^2 } \mathord{\left/
 {\vphantom {{(R_2  - R_1 )^2 } {(r - R_1 )}}} \right.
 \kern-\nulldelimiterspace} {(r - R_1 )}}} \right)} \mathord{\left/
 {\vphantom {{\left( {R_1  + {{(R_2  - R_1 )^2 } \mathord{\left/
 {\vphantom {{(R_2  - R_1 )^2 } {(r - R_1 )}}} \right.
 \kern-\nulldelimiterspace} {(r - R_1 )}}} \right)} r}} \right.
 \kern-\nulldelimiterspace} r} \ge 1   \], \\
and the relative angular electromagnetic parameters of the cloak are large or equal to one, \\
\[\varepsilon _\theta  (r) = \varepsilon _\phi  (r) = (R_2  - R_1 )^2 /(r - R_1 )^2  \ge 1 \\ \],\\
\[\mu _\theta  (r) = \mu _\phi  (r) = (R_2  - R_1 )^2 /(r - R_1 )^2  \ge 1, \\   \]
and relative radial electromagnetic parameters of the cloak are large or equal to a positive constant \[ (R_2  - R_1 )^2 /R_1 ^2  > 0 \\ \],  if $R_2  = 2R_1 $,
\[\varepsilon _r (r) = 1\],\[\mu _r (r) = 1\]£¬\\
if $R_2  \ge 2R_1 $,$\varepsilon _r (r) \ge 1$,
$\mu _r (r) \ge 1$, if $R_2  < 2R_1 $£¬$\varepsilon _r (r) \ge (R_2  - R_1 )^2 /R_1 ^2  > 0$, $\mu _r (r) \ge (R_2  - R_1 )^2 /R_1 ^2  > 0$£¬these electromagnetic parameters and their properties make that analytical exact electromagnetic tense and magnetic flux and electric displacement wave propagation in the cloak are physical bounded and without exceeding light propagation[1-3]. The invisible cloak is created by our sphere radial transformation, $r(R) = R_1  - {{(R_2  - R_1 )^2 } \mathord{\left/
 {\vphantom {{(R_2  - R_1 )^2 } {(R + R_1 )}}} \right.
 \kern-\nulldelimiterspace} {(R + R_1 )}}$,  which maps negative infinity, $ - \infty $, to $R_1$, and maps $-R_2$ to $R_2$ and derivative of the transformation at $-R_2$ equal to 1. i.e. the transformation maps the sphere annular domain $[ - \infty , - R_2 ]$ in negative free space to sphere annular domain $[R_1 ,R_2 ]$ in the positive space; New negative free space we proposed is in which the sphere radial is negative,  $r \le 0$ . The conventional free space with sphere radial $r \ge 0$ is defined to be positive free space.  Maxwell electromagnetic homogeneous equation and its solution in the positive free space can analytically be comtinue to the negative space. The cloak material is not the conventional light transmitting material. The wave rays entering the cloak does not continuous redirection curvely transmitting the cloak. The incoming electromagnetic wave which entering the cloak are all absorbed by the cloak material. The cloak material creates outgoing wave to propagate to outside of the cloak. In simplicity, the novel GLHUA-2 cloak material absorbs incident electromagnetic wave which entering cloak,  and creates outgoing wave which propagate to outside of the cloak that make the device to be invisible cloak.  The ¡°absorption and ¡°creating¡± effects can be happen in same time but different location . The ¡°absorption and ¡°creating¡± effects can be happen in same location but different time .  ¡°absorption and ¡°creating¡± time-space action of the electromagnetic wave is light propagation in free time-space or material time-space.  Because on the boundary sphere surface $r=R_2$, the refractive index and electromagnetic parameters are equal to one and continuous with that in free space $r \ge R_2$ , by Sell Law and Fresnel equations, there is no any scattering wave ray from the cloak to reflect to the free space, moreover, on the boundary spherical surface $r=R_2$ between annular layer cloak 
$R_1  \le r \le R_2 $ and outside of cloak  $r \ge R_2$ , the cloaking electric and magnetic wave field (49) and the free space incident electric and magnetic wave ( 62) and their derivative are continuous. This continuity does ensure that there is no additional scattering from the cloak to disturb the incident wave in free space outside of cloak  $r \ge R_2$ , thus the GLHUA-2 cloak is invisible, where the cloaking electric and magnetic wave field is defined to be the electric and magnetic wave field is annular layer cloak $R_1  \le r \le R_2 $ , the free space incident electric and magnetic wave is defined to be the incident electric and magnetic wave in free space  $r \ge R_2$ . Because there is no wave propagation in outside of the full negative and positive free space, i.e outside of $( - \infty  < r <  + \infty )$,  the GLHUANP-2 transformation maps the negative spherical annular manifold domain , $ - \infty  < R \le  - R_2 $
in the negative space to the spherical annular domain, $R_1  \le r \le R_2 $  in the positive space, thus there is no any electromagnetic wave propagation in the inner sphere $r \le R_2$ , moreover, in the analytical electric and magnetic wave field (49)£¬$\mathop {\lim }\limits_{r \to R_1 } R(r) = \mathop {\lim }\limits_{r \to R_1 } \left( { - R_1  - {{(R_2  - R_1 )^2 } \mathord{\left/
 {\vphantom {{(R_2  - R_1 )^2 } {(r - R_1 )}}} \right.
 \kern-\nulldelimiterspace} {(r - R_1 )}}} \right) =  - \infty ,$  and when $r \to R_1 $, the phase velocity tents to zero, thus in any finite time the incident wave propagation can not arrive the inner boundary $r = R_1 $, thus there is no any wave propagation can be penentrated into the inner sphere $r < R_1 $, the inner sphere $r < R_1 $ is really cloaked concealment. The detailed the electromagnetic parameters, analytical electromagnetic wave propagation, and negative space are proposed in the following section. 
We propose a novel GLHUANP-2 transformation map negative space to positive space

GLHUA-2 invisible cloak is created by a novel GLHUANP-2 transformation. The GLHUANP-2 transformation, $r(R) = R_1  - \frac{{(R_2  - R_1 )^2 }}{{R + R_1 }}, - \infty  < R \le  - R_2 $,
 $\mathop {\lim }\limits_{R \to  - \infty } r(R) = R_1 $, $r( - R_2 ) = R_2 $, 
maps the negative spherical annular manifold domain , $- \infty  < R \le  - R_2 $, in the negative space to the spherical annular domain, $R_1  \le r \le R_2 $  in the positive space. The additional derivetive continuos condition,$\mathop {\lim }\limits_{R \to  - R_2 } \frac{d}{{dR}}r(R) = 1$, is neceesary. The detailed transformation will mathmatical formulated in the following section. Negative and positive free space are discovered and proposed in the next paragraph,

Negative and positive free space

We proposed new negative free space concept [1]. we define the negative free space in which the sphere radial is negative or zero. Also we define the positive free space in which the sphere radial is positive or zero. We proved that the homogeneous Maxwell electromagnetic equations and their solutions in the free positive space $r \ge 0$ can analyticaly be continue to the negative free space $r \le 0$. The detailed proof will given in the above statement 1.

The analytical electromagnetic wave propagation in the GLHUA-2 invisible cloak

By the GLHUANP-2 transformation, we find the analytical electromagnetic wave propagation in the GLHUA-2 cloak, by which we analytically proved that the GLHUA-2 cloak is invsible cloak.

The following figures and their explantions show that the GLHUA-2 invisible cloak material absorb the incident wave and create outgoing wave to propagate to outside of the cloak, that make the device which including sphere anular layer,  $R_1  \le r \le R_2 $  , with the material and free space inner sphere $r \le R_1$ to be invisible cloak. The electromagnetic wave created by the cloak materials is called the created electromagnetic wave. In the following figures, there are two magnetic wave propagation in the annular layer  $R_1  \le r \le R_2 $  , one is incident incoming wave (red line (1)), other wave is created wave (blue line (2)), which is created by the cloak materials. The both magnetic wave front propagation red line (1) and blue line (2) in the annular cloak are separated and disconnected, they are only connected in a subsurface. In following figures,$R_2  = 2R_1 $,$\varepsilon _r  = 1$,$\mu _r  = 1$
$\varepsilon _\theta   = \varepsilon _\phi   = \mu _\theta   = \mu _\phi   = (R_2  - R_1 )^2 /(r - R_1 )^2  \ge 1$, with Guass source. We define first relative time step to be in which the incident wave arrive and tangent to the sphere surface $r = R_2$ in the right side of the cloak.  
 	 	   
Figure 29, At first relative time step, the incident magneic wave front (red line (1)) coming and tangent to the sphere surface  , the novel curve wave front (blue  line (2)) is created by material.	

Figure 30, At relative 5th time step, the magneic wave front (red line (1)) incoming to right side of cloak, ,the material created curve wave front (blue  line (2)) expand up and down

Figure 31, At relative 14 time step, the magnetic wave front (red line (1)) incoming to right side of cloak, the material created curve wave front ( blue line (2)) form wave front on the left side of cloak. Both wave front near each other but no connected.	

Figure 32, At relative 15 time step, the magnetic wave front (red line (1)) incoming to right side of cloak, the material created curve wave front (blue line (2)) form wave front in the left side of the cloak, the red incoming wave (1) and blue created wave front (2) are connected.

Figure 33, At relative 16 time steps, the curve wave front created by cloak material and incident wave in outside of cloak form wave front in the left side of the cloak  (blue line (2)), the incoming magneic wave front red line (1)) right side of cloak is disconncted with created wave (blue line (2)).	

Figure 34. At relative 20 time steps, the curve wave front created by the cloak material is connected to the incident wavefront outside the cloak to form a wavefront and propagate to the left side of the cloak (blue line (2)). The incoming magnetic wave (red line (1)) is separated from the created wave (blue line (2)), and the incoming wavefront (red line (1)) is shrink, and wavefront (red line (1)) is behind from the created wavefront (blue line (2) )¡£	 

Figure 35, At relative 25 time steps, the curve wave front created by cloak material and incident wave front in outside of cloak are connected to form wave front that propagate to boundary,    the left side of the cloak ( blue line (2)); The incoming magnetic wave (red line (1)) is separated from the created wave (blue line (2)), and wavefront (red line (1)) is behind from the created wavefront (blue line (2) ), and the incoming wavefront (red line (1)) is absorbed and shrink.	

Figure 36, At relative 25 time steps, the material created curve wave front propagation
( blue line (2)) has already been out of cloak and did not disturb incident wave in outside of the cloak and make the cloak is invisible; the incoming maganetic wave front (red line (1)) is shrink to a circle and absorbed and can not be penetrated to the concealment  . The inner sphere   is really cloaked concealment.

The GLHUANP-2 transformation formulation
let  $ R $  be sphere radial variable in the negative free space, $ r $  be sphere radial variable in physical positive space. We propose a new Negative space to Positive 
space GLHUANP-2 transformation formulation,
\begin{equation}
\begin{array}{l}
 r(R) = R_1  - \frac{{(R_2  - R_1 )^2 }}{{R + R_1 }}, \\ 
  - \infty  < R \le  - R_2 ,R_1  \le r(R) \le R_2 , \\ 
 \end{array}     ( 37)
\end{equation}                 
  \begin{equation}
\begin{array}{l}
 \mathop {\lim }\limits_{R \to  - \infty } r(R) = R_1 , \\ 
 r( - R_2 ) = R_2 , \\ 
 \end{array}   ( 38)
\end{equation},                              
\begin{equation}
\frac{{dr}}{{dR}}( - R_2 ) = 1, ( 39)
\end{equation},                            
The function $r(R)$ in ( 37) is monotone increase and continuous differentiable function of radial variable $R$ in the negative space , $ \infty  < R \le  - R_2 $, the angular variable $ \theta $  and $\phi$ of in the sphere coordinate are not transformed under  the GLHUANP-2 transformation.
The inverse function of the function ( 37)
\begin{equation}
\begin{array}{l}
 R(r) =  - R_1  - \frac{{(R_2  - R_1 )^2 }}{{r - R_1 }}, \\ 
 R_1  \le r \le R_2 , - \infty  < R(r) \le  - R_2 , \\ 
 \end{array}  ( 40)
\end{equation}             
Thus, the transformation ( 37)-( 39) one to one and onto maps the domain $ [ - \infty , - R_2 ] $ in negative free space to spherical annual layer  $ [R_1 ,R_2 ] $ in positive free space.  The positive free space domain $r \ge R_2$ is not transformed under the GLHUANP-2 transformation, 
\begin{equation}
r = R,r \ge R_2 ,R \ge R_2  ( 41)
\end{equation},                             
GLHUA-2 invisible cloak with refractive index $n(r) ) \ge 1$ and without exceeding light speed. 

GLHUA-2 invisible cloak relative anisotropic materials are created by GLHUANP-2 
transformation. 

To substitue the GLHUANP-2 transformation ( 37) into ( 34) in this paper, the radial isotropic homogeneous electric and magnetic field equation ( 34) in negative free space domian 
$ -\infty  \le R \le  - R_2 $ is translated to the radial anisotropic material electric and magnetic equation in annular layer in the positive space $ R_1  \le r \le R_2 $,
\begin{equation}
\begin{array}{l}
 \frac{\partial }{{\partial r}}\left[ {\begin{array}{*{20}c}
   {\frac{1}{{\frac{{dR}}{{dr}}}}\frac{\partial }{{\partial r}}E(R,\theta ,\phi )}  \\
   {\frac{1}{{\frac{{dR}}{{dr}}}}\frac{\partial }{{\partial r}}H(R,\theta ,\phi )}  \\
\end{array}} \right]   \\
+\frac{1}{{r^2 \sin \theta }}
\frac{\partial }{{\partial \theta }}\sin \theta \frac{\partial }{{\partial \theta }}\left[ {\begin{array}{*{20}c}
   {\frac{{E(R,\theta ,\phi )}}{{\frac{{dr}}{{dR}}\frac{{R^2 }}{{r^2 }}}}}  \\
   {\frac{{H(R,\theta ,\phi )}}{{\frac{{dr}}{{dR}}\frac{{R^2 }}{{r^2 }}}}}  \\
\end{array}} \right] \\ 
  + \frac{1}{{r^2 \sin \theta }}\frac{{\partial ^2 }}{{\partial \phi ^2 }}\left[ {\begin{array}{*{20}c}
   {\frac{{E(R,\theta ,\phi )}}{{\frac{{dr}}{{dR}}\frac{{R^2 }}{{r^2 }}}}}  \\
   {\frac{{H(R,\theta ,\phi )}}{{\frac{{dr}}{{dR}}\frac{{R^2 }}{{r^2 }}}}}  \\
\end{array}} \right] \\
 + k^2 \left[ {\begin{array}{*{20}c}
   {\frac{{dR}}{{dr}}E(R,\theta ,\phi )}  \\
   {\frac{{dR}}{{dr}}H(R,\theta ,\phi )}  \\
\end{array}} \right] = 0, \\ 
  - \infty  < R \le  - R_2 ,R_1  \le r(R) \le R_2 , \\ 
 \end{array} ( 42)
\end{equation}

To compare ( 42) with ( 30), we create a novel GLHUA-2 electromagnetic anisotropic invisible cloak material. The relative radial electric permittivity $ \varepsilon _r $ and magnetic permeanbility $ \mu_r $ are,
\begin{equation}
\begin{array}{l}
 \varepsilon _r (r) = \mu _r (r) = \frac{{R^2 }}{{r^2 }}\frac{{dr}}{{dR}} \\ 
  = \frac{1}{{r^2 }}\frac{{(R_1 (r - R_1 ) + (R_2  - R_1 )^2 )^2 }}{{(R_2  - R_1 )^2 }}, \\ 
 R_1  \le r \le R_2 , \\ 
 \end{array}   ( 43)
\end{equation}              
The angular realtive electric permitivitty and magnetic permeability are 
\begin{equation}
\begin{array}{l}
 \varepsilon _\theta  (r) = \varepsilon _\phi  (r) =  \\ 
 \mu _\theta  (r) = \mu _\phi  (r) = \frac{{dR}}{{dr}} \\ 
  = \frac{{(R_2  - R_1 )^2 }}{{(r - R_1 )^2 }}, \\ 
 R_1  \le r \le R_2 , \\ 
 \end{array}  ( 44)
\end{equation}                    
\begin{equation}
\begin{array}{l}
 \varepsilon _r  = \varepsilon _r  = \varepsilon _\theta   = \varepsilon _\phi   \\ 
  = \mu _\theta   = \mu _\phi   = 1,r > R_2 , \\ 
 \end{array}   ( 45)
\end{equation}                   
The electromagnetic refractive index
                    \begin{equation}
\begin{array}{l}
 n^2 (r) = \varepsilon _\theta  (r)\mu _r (r) \\ 
  = \frac{1}{{r^2 }}\frac{{(R_1 (r - R_1 ) + (R_2  - R_1 )^2 )^2 }}{{(r - R_1 )^2 }} \\ 
  = \frac{1}{{r^2 }}\left( {R_1  + \frac{{(R_2  - R_1 )^2 }}{{(r - R_1 )}}} \right)^2 , \\ 
 R_1  \le r \le R_2 , \\ 
 \end{array}  ( 46)
\end{equation}                 
                     \begin{equation}
n = 1,r \ge R_2     ( 47)
\end{equation},                                   
The properties of GLHUA-2 invisible cloak materials
(1).	The electromagnetic parameters are continuous on the sphere surface, $r=R_2$
By equation ( 43) 
\begin{equation}
\begin{array}{l}
 \varepsilon _r (R_2 ) = \mu _r (R_2 ) \\ 
  = \frac{1}{{R_2 ^2 }}\frac{{(R_1 (R_2  - R_1 ) + (R_2  - R_1 )^2 )^2 }}{{(R_2  - R_1 )^2 }} \\ 
  = 1, \\ 
 \end{array}  ( 48)
\end{equation}
,               
By ( 44) 
                      \begin{equation}
\begin{array}{l}
 \varepsilon _r (R_2 ) = \mu _r (R_2 ) \\ 
  = \frac{1}{{R_2 ^2 }}\frac{{(R_1 (R_2  - R_1 ) + (R_2  - R_1 )^2 )^2 }}{{(R_2  - R_1 )^2 }} \\ 
  = 1, \\ 
 \end{array}   ( 49)
\end{equation}                 
(2) The refractive index are continuous on the sphere surface, $r=R_2$                            by equation ( 46), ( 48) and ( 49)
\begin{equation}
\begin{array}{l}
 n^2 (R_2 ) = n^2 (R_2 ^ -  ) = \varepsilon _\theta  (R_2 )\mu _r (R_2 ) \\ 
  = \frac{1}{{R_2 ^2 }}\frac{{(R_1 (R_2  - R_1 ) + (R_2  - R_1 )^2 )^2 }}{{(R_2  - R_1 )^2 }} \\ 
  = \frac{{R_2 ^2 }}{{R_2 ^2 }} = 1 = n^2 (R_2 ^ +  ), \\ 
 \end{array}  ( 50)
\end{equation}
                      
 (3) The refractive index large than or equal to one
\begin{equation}
\begin{array}{l}
 \frac{d}{{dr}}n^2 (r) = \frac{d}{{dr}}\varepsilon _\theta  (r)\mu _r (r) \\ 
  = \frac{d}{{dr}}\frac{1}{{r^2 }}\left( {R_1  + \frac{{(R_2  - R_1 )^2 }}{{(r - R_1 )}}} \right)^2  \\ 
  =  - 2\frac{1}{{r^3 }}\left( {R_1  + \frac{{(R_2  - R_1 )^2 }}{{(r - R_1 )}}} \right)^2  \\ 
  - \frac{2}{{r^2 }}\left( {R_1  + \frac{{(R_2  - R_1 )^2 }}{{(r - R_1 )}}} \right)\frac{{(R_2  - R_1 )^2 }}{{(r - R_1 )^2 }} \\ 
  < 0, \\ 
 \end{array}  ( 51)
\end{equation}
      
By ( 51), the square of the refractive index function in ( 46) is monotone decreasing
function in the interval $ R_1  \le r \le R_2 $,  and by ( 50) $ n^2 (R_2 ) = 1$, therefore,
we proved that refractive index large than or equal to one. 
\begin{equation}
n(r) \ge 1,R_1  \le r \le R_2 ,   ( 52)
\end{equation},                    
(4) The relative radial electric permittivity $ \varepsilon _r $ and magnetic permeanbility $\mu_r$ in ( 43) are
large than positive constant which large than zero, the electric tense and magnetic tense are bounded in the GLHUA-2 invisible cloak.
\begin{equation}
\begin{array}{l}
 \frac{d}{{dr}}\varepsilon _r (r) = \frac{d}{{dr}}\mu _r (r) \\ 
  =  - \frac{2}{{r^3 }}\frac{{(R_1 (r - R_1 ) + (R_2  - R_1 )^2 )^2 }}{{(R_2  - R_1 )^2 }} \\ 
  + \frac{{2R_1 }}{{r^2 }}\frac{{R_1 (r - R_1 ) + (R_2  - R_1 )^2 }}{{(R_2  - R_1 )^2 }} \\ 
  = \frac{{R_1 (r - R_1 ) + (R_2  - R_1 )^2 }}{{(R_2  - R_1 )^2 }} \\ 
 \frac{2}{{r^3 }}\left( { - (R_1 (r - R_1 ) + (R_2  - R_1 )^2 ) + R_1 r} \right) \\ 
  = \frac{{R_1 (r - R_1 ) + (R_2  - R_1 )^2 }}{{(R_2  - R_1 )^2 }} \\ 
 \frac{2}{{r^3 }}\left( {(R_1 ^2  - (R_2  - R_1 )^2 )} \right), \\ 
 R_1  \le r \le R_2 , \\ 
 \end{array}   ( 53)
\end{equation}
  
when        

\begin{equation} 2R_1 > R_2 ,  ( 54) \end{equation}                           
then $ R_1 ^2  - (R_2  - R_2 )^2  > 0$, in ( 53), $ \frac{d}{{dr}}\varepsilon _r (r) = \frac{d}{{dr}}\mu _r (r) > 0$, $ \varepsilon _r (r) = \mu _r (r)$ in ( 43) is monotone increasing function in interval $ R_1  \le r \le R_2$,
 \begin{equation}
\begin{array}{l}
 \varepsilon _r (r) = \mu _r (r) \ge \varepsilon _r (R_1 ) = \mu _r (R_1 ) \\ 
  = \frac{{(R_2  - R_1 )^2 }}{{R_1 ^2 }} > 0, \\ 
 \end{array}  ( 55)
\end{equation}                    
when        
\begin{equation}2R_1=R_2,  ( 56) \end{equation},                        
then                                          \[ R_1 ^2  - (R_2  - R_2 )^2  = 0 \]
\begin{equation}\varepsilon _r (r) = \mu _r (r) = 1,  ( 57) \end{equation}                           
when        
\begin{equation}2R_1<R_2   ( 58) \end{equation},
then $ R_1 ^2  - (R_2  - R_2 )^2  < 0$, in ( 53), $ \frac{d}{{dr}}\varepsilon _r (r) = \frac{d}{{dr}}\mu _r (r) > 0$, $ \varepsilon _r (r) = \mu _r (r)$ in ( 43) is monotone decreasing function in interval $ R_1  \le r \le R_2$, thus
\begin{equation}
\begin{array}{l}
 \varepsilon _r (r) = \mu _r (r) \\ 
  \ge \varepsilon _r (R_2 ) = \mu _r (R_2 ) = 1, \\ 
 \end{array}   ( 59)
\end{equation}

The angular realtive electric permitivitty and magnetic permeability in ( 44) are 
large than or equal to one
\begin{equation}
\begin{array}{l}
 \varepsilon _\theta  (r) = \varepsilon _\phi  (r) = \mu _\theta  (r) = \mu _\phi  (r) \\ 
  = \frac{{(R_2  - R_1 )^2 }}{{(r - R_1 )^2 }} \ge 1,R_1  \le r \le R_2 , \\ 
 \end{array}  ( 60)
\end{equation}                      

Analytical electromagnetic wave propagation in GLHUA-2 invisible cloak with anisotropic materials in ( 43) and ( 44)

Consider inhomogeneous electric and magnetic wave field equation in free space with source term 

\begin{equation}
\begin{array}{l}
 \frac{\partial }{{\partial r}}\frac{\partial }{{\partial r}}\left[ {\begin{array}{*{20}c}
   E  \\
   H  \\
\end{array}} \right] + \frac{1}{{r^2 \sin \theta }}\frac{\partial }{{\partial \theta }}\sin \theta \frac{\partial }{{\partial \theta }}\left[ {\begin{array}{*{20}c}
   E  \\
   H  \\
\end{array}} \right] \\ 
  + \frac{1}{{r^2 \sin \theta }}\frac{{\partial ^2 }}{{\partial \phi ^2 }}\left[ {\begin{array}{*{20}c}
   E  \\
   H  \\
\end{array}} \right] + k^2 \left[ {\begin{array}{*{20}c}
   E  \\
   H  \\
\end{array}} \right] = \left[ {\begin{array}{*{20}c}
   {J_s }  \\
   {M_s }  \\
\end{array}} \right], \\ 
 \end{array}  ( 61)
\end{equation},   

Suppose that the $J_s$, $M_s$ electromagnetic source is located at $ \vec r_s  = (r_s ,\theta _s ,\phi _s )$ outside of the cloak in the free space, $r_s > R_2 $ and in $ \vec e$ direction, by the sphere harmonic expansion of the ( 41)-( 43) in [2], the incident electromagnetic wave is

\begin{equation}
\begin{array}{l}
 \left[ {\begin{array}{*{20}c}
   {E_i (\vec r,\vec r_s )}  \\
   {H_i (\vec r,\vec r_s )}  \\
\end{array}} \right] = \sum\limits_{l = 1}^P {\left[ {\begin{array}{*{20}c}
   {E_{i,l} (r,r_s )}  \\
   {H_{i,l} (r,r_s )}  \\
\end{array}} \right]}  \\ 
 \sum\limits_{m =  - l}^l {\left[ {\begin{array}{*{20}c}
   {D_E (\theta ,\phi )}  \\
   {D_H (\theta ,\phi )}  \\
\end{array}} \right]} Y_l^m (\theta ,\phi )Y_l^{m*} (\theta _s ,\phi _s ), \\ 
 \end{array},  ( 62)
\end{equation}

the incident electric and magnetic wave field $ E_i  (r,r_s )$, $H_i  (r,r_s ) $,in ( 62) satisfy equation ( 61), 
$ E_{i,l} (r,r_s )$, $H_{i,l} (r,r_s )$ is the  $l$ harmonic sub radial wave field in the sphere harmonic 
expansion ( 62), the subcript $i$ denotes incident wave, subcript $l$ denotes $l$ harmonic 
component.
        
 ${\boldsymbol{Statement \ 2:}}$ Suppose that the electric and magnetic wave field

\begin{equation}
\left[ {\begin{array}{*{20}c}
   {E_{l,1} }  \\
   {H_{l1} }  \\
\end{array}} \right] = \left[ {\begin{array}{*{20}c}
   {Rj_l (kR)}  \\
   {Rj_l (kR)}  \\
\end{array}} \right],    ( 63)
\end{equation} ,                      
\begin{equation}
\left[ {\begin{array}{*{20}c}
   {E_{l,2} }  \\
   {H_{l,2} }  \\
\end{array}} \right] = \left[ {\begin{array}{*{20}c}
   {Rn_l (kR)}  \\
   {Rn_l (kR)}  \\
\end{array}} \right],  ( 64)
\end{equation}
,                     

are the two ?linear?independent analytical solution of the following $l$ harmonic electric and magnetic wave field equations by using the sphere harmonic decomposition of the equation ( 34)

\begin{equation}
\begin{array}{l}
 \frac{\partial }{{\partial R}}\frac{\partial }{{\partial R}}\left[ {\begin{array}{*{20}c}
   {E_l (R)}  \\
   {H_l (R)}  \\
\end{array}} \right] - \frac{{l(l + 1)}}{{R^2 }}\left[ {\begin{array}{*{20}c}
   {E_l (R)}  \\
   {H_l (R)}  \\
\end{array}} \right] \\ 
  + k^2 \left[ {\begin{array}{*{20}c}
   {E_l (R)}  \\
   {H_l (R)}  \\
\end{array}} \right] = 0 \\ 
 \end{array},   ( 65)
\end{equation},    

in the free negative space domain $- \infty  < R \le  - R_2 $, then 

\begin{equation}
\begin{array}{l}
 \left[ {\begin{array}{*{20}c}
   {E_l (r)}  \\
   {H_l (r)}  \\
\end{array}} \right] = \left( { - R_1  - \frac{{(R_2  - R_1 )^2 }}{{(r - R_1 )}}} \right)\left[ {\begin{array}{*{20}c}
   {A_{E,l} } & {B_{E,l} }  \\
   {A_{H,l} } & {B_{H,l} }  \\
\end{array}} \right] \\ 
 \left[ {\begin{array}{*{20}c}
   {j_l \left( {k\left( { - R_1  - \frac{{(R_2  - R_1 )^2 }}{{(r - R_1 )}}} \right)} \right)}  \\
   {n_l \left( {k\left( { - R_1  - \frac{{(R_2  - R_1 )^2 }}{{(r - R_1 )}}} \right)} \right)}  \\
\end{array}} \right], \\ 
 \end{array}   ( 66)
\end{equation}  

is the analytical solution of the following $l$ harmanic magnetic wave field equation ( 65) in the anualar layer $R_1  \le r \le R_2 $, with anisotropic materials ( 43) and ( 44),
\begin{equation}
\begin{array}{l}
 \frac{\partial }{{\partial r}}\left[ {\begin{array}{*{20}c}
   {\frac{1}{{\varepsilon _\theta  }}\frac{{\partial E_l (r)}}{{\partial r}}}  \\
   {\frac{1}{{\mu _\theta  }}\frac{{\partial H_l (r)}}{{\partial r}}}  \\
\end{array}} \right] - \frac{{l(l + 1)}}{{r^2 }}\left[ {\begin{array}{*{20}c}
   {\frac{{E_l (r)}}{{\varepsilon _r }}}  \\
   {\frac{{H_l (r)}}{{\mu _r }}}  \\
\end{array}} \right] \\ 
  + k^2 \left[ {\begin{array}{*{20}c}
   {\mu _\theta  E_l (r)}  \\
   {\varepsilon _\theta  H_l (r)}  \\
\end{array}} \right] = 0, \\ 
 \end{array}  ( 67)
\end{equation}             

where constants $A_{E,l} $,$B_{E,l} $ and $A_{H,l} $,$B_{H,l} $, are uniquely determined by the continuous of the radial electric and magnetic wave field and their derivatives on the boundary spherical surface $r=R_2$.

 ${\boldsymbol{Statement \ 3:}}$ The analytical electric and magnetic wave field propagation through GLHUA-2
cloak in the anualar layer $ R_1  \le r \le R_2 $ with anisotropic materials in ( 43) and ( 44) is
\begin{equation}
\begin{array}{l}
 \left[ {\begin{array}{*{20}c}
   {E(\vec r,\vec r_s )}  \\
   {H(\vec r,\vec r_s )}  \\
\end{array}} \right] = ( - R_1  - \frac{{(R_2  - R_1 )^2 }}{{r - R_1 }}) \\ 
 \sum\limits_{l = 1}^\infty  {\left( {\left[ {\begin{array}{*{20}c}
   {A_{E,l} } & {B_{E,l} }  \\
   {A_{H,l} } & {B_{H,l} }  \\
\end{array}} \right]} \right.} \left[ {\begin{array}{*{20}c}
   {j_l (k( - R_1  - \frac{{(R_2  - R_1 )^2 }}{{r - R_1 }})}  \\
   {n_l (k( - R_1  - \frac{{(R_2  - R_1 )^2 }}{{r - R_1 }})}  \\
\end{array}} \right] \\ 
 \left. {\sum\limits_{m =  - l}^l {\left[ {\begin{array}{*{20}c}
   {D_E (\theta ,\phi )}  \\
   {D_H (\theta ,\phi )}  \\
\end{array}} \right]} Y_l^m (\theta ,\phi )Y_l^{m*} (\theta _s ,\phi _s )} \right), \\ 
 \end{array}  ( 68)
\end{equation} 

the analytical electric and magnetic wave field ( 68) shows that the GLHUA-2 cloak is complete invisible cloak with anisotropic materials in ( 43) and ( 44) in the annular layer $ R_1  \le r \le R_2 $.

 \subsection {An Active GLHUA-3 Invisible Cloak With Refractive Index $n \ge 1$ that absorbs incoming wave and creates outgoing wave }

We propose other GLHUANP-3 transformation which maps $ R \to  - \infty $ in basic negative space to $ r = R_1 $ in physical positive space and maps $ R = - {R_2}  $ in basic negative space to  $ r = R_2 $  in physical positive space.

\begin{equation}
r = R_1  + (R_2  - R_1 )e^{\frac{{R + R_2 }}{{R_2  - R_1 }}} , ( 69)
\end{equation}

And inverse transformation
\begin{equation}
\begin{array}{l}
R =  - R{}_2 + (R_2  - R_1 )\log ((r - R_1 )/(R_2  - R_1 )),\\
 ( 70)\\
 \end{array} 
\end{equation}

 The invisible cloak GLHUA-3 relative electric permittivity and magnetic permeability ( 71) and  ( 72)  are created by the above  radial transformation
\begin{equation}
\begin{array}{l}
 \varepsilon _r (r) = \mu _r (r) =  \\ 
 \frac{{( - R_2  + (R_2  - R_1 )\log ({{(r - R_1 )} \mathord{\left/
 {\vphantom {{(r - R_1 )} {(R_2  - R_1 )))^2 }}} \right.
 \kern-\nulldelimiterspace} {(R_2  - R_1 )))^2 }}}}{{r^2 }}\frac{{r - R_1 }}{{R_2  - R_1 }}, \\ 
 R_1  \le r \le R_2 , \\ 
 \end{array}      ( 71)
\end{equation}
\begin{equation}
\begin{array}{l}
 \varepsilon _\theta  (r) = \varepsilon _\phi  (r) = \mu _\theta  (r) = \mu _\phi  (r) \\ 
  = \frac{{R_2  - R_1 }}{{r - R_1 }},R_1  \le r \le R_{2,}  \\ 
 \end{array}      ( 72)
\end{equation}
and the relative eletromagnetic parameters in $ r \le R_1 $ or $r \ge R_2 $ are

\begin{equation}
\begin{array}{l}
 \varepsilon _r  = \varepsilon _\phi   = \varepsilon _\phi   = \mu _r  = \mu _\theta   = \mu _\phi   = 1, \\ 
 r \le R_1 ,\;\;or\;\;r \ge R_2 , \\ 
 \end{array}  ( 73)
\end{equation}

The parameters have the following properties,

\begin{equation}
\begin{array}{l}
 \varepsilon _r (R_2 ) = \mu _r (R_2 ) = 1, \\ 
 \varepsilon _\theta  (R_2 ) = \varepsilon _\phi  (R_2 ) = \mu _\theta  (R_2 ) = \mu _\phi  (R_2 ) = 1, \\ 
 \end{array}    ( 74)
\end{equation}
\begin{equation}
\begin{array}{l}
 n^2 (r) = \varepsilon _\theta  \mu _r  = \varepsilon _r \mu _\phi   \\ 
  = \frac{{\left( { - R_2  + (R_2  - R_1 )\log ({{(r - R_1 )} \mathord{\left/
 {\vphantom {{(r - R_1 )} {(R_2  - R_1 )))}}} \right.
 \kern-\nulldelimiterspace} {(R_2  - R_1 )))}}} \right)^2 }}{{r^2 }} \ge 1, \\ 
 R_1  \le r \le R_2 , \\ 
 \end{array}  ( 75)
\end{equation}

\begin{equation}
n(r) > 1,inR_1  \le r < R_2 , ( 76)
\end{equation}                               

\begin{equation}
n(R_2 ) = 1,  ( 77)
\end{equation} 
From property  ( 77) and Sell Law and Fresnel equations , on the boundary , $ r = R_2 $, between the sphere and the free space,  there is no any reflection from the  sphere $r \le R_2 $ .
thus, the cloak is invisible.  From property  ( 75) and  ( 76), the refractive index large or equal 1, thus the electromagnetic or light wave propagation in this invisible cloak is no exceeding light speed. To substitute relative parameters  ( 71) and  ( 72) into the radial EM equations in the annular cloak[1], for each sphere Harmonic number  $ l $, the radial electric and magnetic wave equation, reduced to the $ l $ sub EM equation. 
Here the radial magnetic equation becomes the radial $ l $ sub magnetic equation ,
\begin{equation}
\begin{array}{l}
 \frac{\partial }{{\partial r}}\left( {\frac{{r - R_1 }}{{R_2  - R_1 }}} \right)\frac{{\partial H_l }}{{\partial r}} \\ 
  - \frac{{l(l + 1)(R_2  - R_1 )H_l }}{{\left( { - R_2  + (R_2  - R_1 )\log \left( {\frac{{r - R_1 }}{{R_2  - R_1 }}} \right)} \right)^2 (r - R_1 )}} \\ 
  + k^2 \frac{{(R_2  - R_1 )H_l }}{{(r - R_1 )}} = 0, \\ 
 \end{array}   ( 78)
\end{equation}
where $ H_l  = r^2 B_{r,l}  $  , $ B_{r,l} $ is $ l $  component of  the radial magnetic flux, the radial   sub electric equation is same type equation as  ( 78), using our method in [1], to solve equation
  ( 78), we obtain analytical $ l $ component radial magnetic wave solution
\begin{equation}
\begin{array}{l}
 H_l (r,r_s ) =\\
= ( - R_2  + (R_2  - R_1 )\log ({{(r - R_1 )} \mathord{\left/
 {\vphantom {{(r - R_1 )} {(R_2  - R_1 ))}}} \right.
 \kern-\nulldelimiterspace} {(R_2  - R_1 ))}} \\ 
 (a_l (r_s ) \\ 
 j_l (k( - R_2  + (R_2  - R_1 )\log ({{(r - R_1 )} \mathord{\left/
 {\vphantom {{(r - R_1 )} {(R_2  - R_1 ))))}}} \right.
 \kern-\nulldelimiterspace} {(R_2  - R_1 ))))}} \\ 
  + b_l (r_s ) \\ 
 n_l (k( - R_2  + (R_2  - R_1 )\log ({{(r - R_1 )} \mathord{\left/
 {\vphantom {{(r - R_1 )} {(R_2  - R_1 )))))}}} \right.
 \kern-\nulldelimiterspace} {(R_2  - R_1 )))))}}, \\ 
 ( 79) \\
 \end{array} 
\end{equation}
Where $ l = 1,2,3...... $ $ j_l $ is the Spherical Bessel Function of the First Kind, $ n_l $  is the Spherical Bessel Function of the Second Kind, or spherical Neumann functions, The two constants
$ a_l $ and $ b_l $ are uniquely determined by continuous conditions of the incident wave and its derivative with respect to the radii $ r $ on the boundary $ r=R_2 $ . Because the variable
\begin{equation}
\begin{array}{l}
  - \infty  < R =  - R_2  \\ 
  + (R_2  - R_1 )\log ({{(r - R_1 )} \mathord{\left/
 {\vphantom {{(r - R_1 )} {(R_2  - R_1 ))}}} \right.
 \kern-\nulldelimiterspace} {(R_2  - R_1 ))}} \\ 
  \le  - R_2 , \\ 
 \end{array} 
 ( 80)
\end{equation}
        
So, $ H_l (r,r_s ) $ in ( 79)  is bounded continuous differentiable physical wave. Because the $ n(R_2 ) = 1 $ in  ( 77)  , there is no any reflection from the sphere surface, 
$ r=R_2 $ , and thus the continuous boundary conditions uniquely determine that all incident light wave to be transmitted into the sphere $ r \le R_2 $  in  ( 79)  , in any finite time, the wave  ( 79)  can not be arrived to the inner spherical surface $ r=R_1 $ , and no wave to be propagation penetrated into the inner sphere , thus the inner sphere 
$ r \le R_1 $   is cloaked concealment.

 We discover that the electromagnetic spherical coordinate equation and its solution can analytically be continuation to a negative space with negative radii $ r \le 0 $. The space with sphere radii $ r \ge 0 $ is called the positive space. The space with sphere radii $ r \le 0 $ is called the negative space.
In the following paragraph, we describe the analytical radial magnetic wave propagation through the invisible cloak with the relative materials   ( 71)
 and    ( 72)  in the annular layer $ R_1  \le r \le R_2 $, , $ R?_1 = 0.5m $
 and  $ R_2=1.0m $
The analytical  radial magnetic wave is
\begin{equation}
\begin{array}{l}
 H(r,\theta ,\phi ) =  \\ 
 \sum\limits_{l = 1}^{LM} {H_l } (r,r_s )D_h (\theta ,\phi )\sum\limits_{m =  - l}^l {Y_l^{(m)} } (\theta ,\phi )Y_l^{(m)*} (\theta _s ,\phi _s ), \\ 
 R_1  \le r \le R_2 , \\ 
( 81) \\
 \end{array}
\end{equation}

Where $H(r) = r^2 B_r (r) $, $B_r (r) $ is the radial magnetic flux component.  In the free space, $ r \ge R_2 $ $ H(r,\theta ,\phi ) = H_i (r,\theta ,\phi ) = r^2 B_{r,i} (r,\theta ,\phi )$, in inner sphere $ r \le R_1 $  $ H(r,\theta ,\phi ) = 0 $, thus the inner sphere is cloaked concealment, the $ B_{r,i} (r,\theta ,\phi ) $ is incident radial magnetic flux wave in free space , $ r \ge R_2$  the incident magnetic wave is excited by the electric current time pulse source located in the $ (r_s ,\theta _s ,\phi _s ) = (18.3m,\pi /2,0)$ with direction $\vec y $,
$ \vec y = \left( {\sin \theta \sin \phi ,\cos \theta \sin \phi ,\cos \phi } \right) $ .
The following figures show the above radial magnetic wave propagation though the invisible cloak without exceeding light speed. There are two waves propagates in the annular layer  
 $ R_1  \le r \le R_2 $,   one is incident incoming wave, other wave is created wave which is created by the cloak materials, the wave fronts are colored by bold red. When the incident radial magnetic wave excited by right impulse source is propagation in $ r > R_2 $ ,right and outside of the cloak sphere, the invisible cloak material creates the created wave colored by red in the right part of the inner boundary $ r= R_1 $ in the annular layer.  The radial electric wave propagation through the GLHUA-3 cloak is display in the figure 43-50. 
%\end {document} 

\section{\label{sec:level1}INTRODUCTION} 
%First-level heading:\protect\\ 
%The line break was forced \lowercase{via}
In 2000, in our Global Integral and Local Differential GILD EM modeling and inversion simulation [4], we observed a strange double layered cloak phenomena in Fig.2 in [5] and in Fig.11 of [5]. The double layered cloak phenomena was called ¡°GILD phenomena¡± has been published in SEG Expanded Abstracts, Vol. 21 , No. 1, 692-695, 2002, by Jianhua Li at al. [5]. The double layer cloak ¡°GILD phenomena¡± is nonzero variation solution imaging of the zero scattering inversion which cloaked the object of the scattering inversion imager. Since then, we created the Global and Local modeling and inversion [6] in 2006, and used it to create our novel GLLH [11] and GLHUA [1-3] invisible EM cloak. A ¡°New computational mirage ¡°has been published in PIERS abstract 2005 in Hang Zhou of China, 296[13].  In 2006, using transform optics, Pendry et al proposed first invisible cloak [7] with infinite speed and exceeding light speed propagation. Ulf Leonhardt et al. [27] proposed a cloak with finite speed based on a Euclid and non Euclid joint transform optics. However, ULF cloak [27] still has exceeding light speed difficulty and without theoretical full wave proof and without full wave simulation verification. In 2009, we published paper in arXiv[10], ¡°A Double Layer Electromagnetic Cloak And GL EM Modeling¡±, in this paper, the inner layer cloak with relative refractive index not less than 1 was proposed first in the world[10]. In 2010, first in the world, we published  ¡°GLLH EM Invisible Cloak with Novel Front Branching and without Exceed Light Speed Violation¡± in ArXiv1005.3999 in the May of 2010 [11] and in PIERS 2010[12]. Next year, in 2011, Ulf Leonhardt et al. published paper ¡°Invisibility cloaking without superluminal propagation¡± in Arxiv: 1105.0164v3 [14]. Ulf did write a complete review and references of cloak history in his paper[14], he did criticism for many papers about Pendry¡¯s cloak and all published cloak papers and wrote that ¡°The fundamental problem is that perfect invisibility (Pendry cloak[7] )£¬requires that light should propagate in certain cloaking regions with a superluminal phase velocity that tends to infinity.¡± In his paper [14] in 2011, Ulf cited our paper ArXiv:1005.3999 in 2010 as his reference [35] and wrote that ¡°The preprint [35], proposes a different method for cloaking without superluminal propagation.¡± However, it is needed more communication for understand between our GLHUA cloak and ULF cloak. Up to now, ULF¡¯s cloak is still needed to verify by full wave theoretical analysis and is still needed to verify by  full wave computational simulation. Pendry cloak [7] is created by a linear transformation which maps zero to $R_1$ and $R_2$ to $R_2$ in positive space.
Therefore, Pendry cloak can not be realized because Pendry cloak has infinite speed and exceeding light speed propagation that is fundamental  and can not be solved difficults. 
 
 We created GILD modeling and inversion method [4][5][6][8] and used it to discover double cloak phenomena in 2000, We did create GL double layer cloak and inner layer cloak without exceeding light speed in 2009[10], We used our Global and Local  GL no scattering and inversion to  create GLLH double layer cloak with relative refractive index not less than 1 and without exceeding light speed propagation that published in 2010 [11] first in the world.  practicable GLHUA double layer invisible cloak with relative EM parameter not less than 1 in 2016
arxiv:1612.02857[1] and its theoretical proof in arxiv:1701.00534{2], we create an GLHUA analytical modeling method and a GLHUA-1 invisible cloak and an exact analytical EM wave propagation in the GLHUA-1 double layer cloak in this paper version 1 in 2017. We propose a negative space concept in version 13 of our paper arxiv:1612.02857. We proved statement 1 that "The homogeneous electromagnetic wave field equations and their solutions in the free positive space $r \ge 0$ can analyticaly be continue to the negative free space $r \le 0$" in this paper. We propose Novel
thransformation GLHUUANP-1 which maps negative infinity, $ - \infty $ in negative space, to $R_1$ in positive space, and maps $R_2$ to $R_2$ . We propose Novel
thransformation GLHUUANP-2 and GLHUANP-3 which map negative infinity, $ - \infty $ in negative space, to $R_1$ in positive space, and maps $-R_2$ in negative space to $R_2$ in positive
space. We create practicable GLHUA-2 and GLHUA-3 invisible cloak with the relative refractive index large or equal to one and without exceeding light speed propagation difficult. When $2R_1=R_2$, the GLHUA-2 cloak materials are
GLHUA cloak materials with $\alpha =2 $, GLHUA-2 cloak materials is created by GLHUANP-2 transformation in this paper, GLHUA cloak materials is created by GL no scattering modeling inversion in [1]and [2]. The relative angular EM parameters 
\[
\begin{array}{l}
 \varepsilon _\theta   = \varepsilon _\phi   = \mu _\theta   = \mu _\phi   =  \\ 
  = {{(R_2  - R_1 )} \mathord{\left/
 {\vphantom {{(R_2  - R_1 )} {(r - R_1 )}}} \right.
 \kern-\nulldelimiterspace} {(r - R_1 )}} \\ 
 \end{array}
\] of GLHUA cloak with $\alpha=1$ and GLHUA-1 cloaks can be created by
GLHUANP-3  transformation. 
The relative angular EM parameters of GLLH invisible cloak in (5) of arxiv:1005.3999[11]
\[
\begin{array}{l}
 \varepsilon _\theta   = \varepsilon _\phi   = \mu _\theta   = \mu _\phi   \\ 
  = \frac{{R_o }}{{2\log \left( {e^{{1 \mathord{\left/
 {\vphantom {1 {R_o }}} \right.
 \kern-\nulldelimiterspace} {R_o }}} {{(R_0  - R_1 )} \mathord{\left/
 {\vphantom {{(R_0  - R_1 )} {(r - R_1 )}}} \right.
 \kern-\nulldelimiterspace} {(r - R_1 )}}} \right)(r - R_1 )}} \\ 
  + \frac{1}{{2R_0 \log ^3 \left( {e^{{1 \mathord{\left/
 {\vphantom {1 {R_o }}} \right.
 \kern-\nulldelimiterspace} {R_o }}} {{(R_0  - R_1 )} \mathord{\left/
 {\vphantom {{(R_0  - R_1 )} {(r - R_1 )}}} \right.
 \kern-\nulldelimiterspace} {(r - R_1 )}}} \right)(r - R_1 )}} \\ 
 \end{array} (GLLH (5))
\]
can be created from other nove GLLHNP transformation which map negative infinity, $ - \infty $ in negative space, to $R_1$ in positive space, and maps $-R_2$ in negative space to $R_2$ in positive space. 
There is typo in (5) of the our paper arxiv:1005.3999 [11]. The corrected 
(5) of [11] is the above formula GLLH(5).
   
\hfill \break

\section {Practicable GLHUA-1 outer layer cloakwith relative parameter not less than 1}
In this section, we created a novel practicable GLHUA-1 outer annular layer EM invisible cloak with relative parameter not less than 1.

In the outer annular layer $R_1  \le r \le R_2 $ the radial relative electric permittivity and magnetic permeability are
\begin{equation}
\varepsilon _r (r) = \mu _r (r) = \frac{{R_2 ^2 }}{{r^2 }},   
\end{equation}
They are monotone decease function and
\begin{equation} 
 \frac{{R_2 ^2 }}{{R_1 ^2 }} \ge \varepsilon _r (r) = \mu _r (r) = \frac{{R_2 ^2 }}{{r^2 }} \ge 1,
\end{equation}
the angular relative electric permittivity and magnetic permeability are
\begin{equation} 
\varepsilon _\phi  (r) = \mu _\theta  (r) = \mu _\phi  (r) = \frac{{R_2  - R_1 }}{{r - R_1 }},
\end{equation} 
They are monotone decease function and 
\begin{equation} 
  \begin{array}{l}
 \infty  \ge \varepsilon _\theta  (r) = \varepsilon _\phi  (r) = \mu _\theta  (r) = \mu _\phi  (r) \\ 
  = \frac{{R_2  - R_1 }}{{r - R_1 }} \ge \frac{{R_2  - R_1 }}{{R_2  - R_1 }} = 1. \\ 
 \end{array}
\end{equation} 
From ( 82) and ( 84), on the boundary $r = R_2 $,The relative parameters in GLHUA-1 outer layer cloak are equal to 1 and are continuous across the outer boundary $r = R_2 $. Summary, we have the following theorem 1.
\hfill\break \\

${\boldsymbol{Theorem \ 1:}}$ \  In GLHUA-1 outer layer cloak, $R_1  \le r \le R_2 $ the radial relative electric permittivity and magnetic permeability in ( 82) and the angular relative electric permittivity and magnetic permeability in ( 84) are not less than 1 and are continuous across the outer boundary $r = R_2 $.
\\
\section {  Exact analytical EM wave propagation in the GLHUA-1 outer layer cloak}

In this section, we create an analytical EM wave field propagation in the GLHUA-1 outer cloak with relative parameters in ( 82) and ( 84). For simplicity, we create a new analytical expansion of the solution of the magnetic wave equation in the GLHUA-1 outer layer cloak domain $R_1  \le r \le R_2 $. 

\subsection { The spherical harmonic expansion of incident radial GL magnetic wave field}
 
Suppose that the incident radial magnetic wave field ${H^b}_r$,

\begin{equation}
H^b {}_r = \cos \phi \frac{1}{r}\frac{{\partial g}}{{\partial \theta }} - \frac{{\cos \theta \sin \phi }}{{r\sin \theta }}\frac{{\partial g}}{{\partial \phi }} 
\end{equation}
is incident magnetic wave for GLHUA-1 outer layer cloak, where    
\begin{equation}                  
g(\vec r,\vec r_s ) =  - \frac{1}{{4\pi }}\frac{{e^{  ik\left| {\vec r - \vec r_s } \right|} }}{{\left| {\vec r - \vec r_s } \right|}},
\end{equation}
$H^b_r$ is excited by the electric point source
\begin{equation}
\vec J_s  = \frac{1}{{r^2 \sin \theta }}\delta (r - r_s )\delta (\theta  - \theta _s )\delta (\phi  - \phi _s )\vec e_y ,
\end{equation}
located in the point $\vec r_s  = (r_s ,\theta _s ,\phi _s )$,$r_s  > R_{out}  > R_2 $. $H^b_r$ is the solution of the
magnetic wave equation
\begin{equation}
\begin{array}{l}
 \frac{\partial }{{\partial r}}\frac{1}{{\mu _\theta  }}\frac{{\partial \left( {r^2 \mu _r H_r } \right)}}{{\partial r}} + \frac{1}{{\sin \theta }}\frac{\partial }{{\partial \theta }}\sin \theta \frac{{\partial H_r }}{{\partial \theta }} \\ 
  + \frac{1}{{\sin ^2 \theta }}\frac{{\partial H_r }}{{\partial \phi ^2 }} + k^2 r^2 \varepsilon _\theta \mu _r  H_r  = 0, \\ 
 \end{array}
\end{equation}

in the annular layer domain $R_2  < r < R_{out} $. in free space, with relative parameters are equal to 1, $\mu _\theta   = \varepsilon _\theta   = 1$, $\mu _\phi   = \varepsilon _\phi   = 1$ $\mu _r  = \varepsilon _r  = 1$.

Define Global and Local - GL magnetic wave
       \begin{equation}      
      H(\vec r) = r^2 \mu _r H_r (\vec r),                               
       \end{equation}
The magnetic wave equation ( 89) is translated to the following GL magnetic wave equation

       \begin{equation}     
\begin{array}{l}
 \frac{\partial }{{\partial r}}\frac{1}{{\mu _\theta  }}\frac{{\partial H}}{{\partial r}} + \frac{1}{{r^2 \mu _r }}\frac{1}{{\sin \theta }}\frac{\partial }{{\partial \theta }}\sin \theta \frac{{\partial H}}{{\partial \theta }} \\ 
  + \frac{1}{{r^2 \mu _r }}\frac{1}{{\sin ^2 \theta }}\frac{{\partial H}}{{\partial \phi ^2 }} + k^2 \varepsilon _\theta  H = 0, \\ 
 \end{array}
       \end{equation}     
Using the GL electric wave field $E(\vec r) = r^2 \varepsilon _r E_r (\vec r),$ to replace GL magnetic wave field $H(\vec r) = r^2 \mu _r H_r (\vec r)$ in ( 91), the equation ( 91) become the GL electric wave equation.
The GL EM wave propagation in free space satisfies the equation ( 91) with relative parameter equal to 1; GL EM wave propagation in GLHUA-1 outer layer cloak satisfies the equation ( 91) with GLHUA-1 relative parameters not less than 1 in ( 82) and ( 84).

Define the operator 
\begin{equation}
\Re (\theta ,\phi ) = \frac{1}{{\sin \theta }}\frac{\partial }{{\partial \theta }}\sin \theta \frac{\partial }{{\partial \theta }} + \frac{1}{{\sin ^2 \theta }}\frac{\partial }{{\partial \phi ^2 }}
\end{equation}
\begin{equation}
D_h (\theta ,\phi ) = \cos \phi \frac{\partial }{{\partial \theta }} - \frac{{\cos \theta \sin \phi }}{{\sin \theta }}\frac{\partial }{{\partial \phi }}
\end{equation}
By direct calculation, we can prove
\begin{equation}
\Re (\theta ,\phi )D_h (\theta ,\phi ) = D_h (\theta ,\phi )\Re (\theta ,\phi ),
\end{equation}

The spherical harmonic expansion of the incident GL magnetic wave
$H^{(b)} (\vec r) = r^2 H^b {}_r(\vec r)$ is
\begin{equation}
\begin{array}{l}
 H^{(b)} (r,\theta ,\phi ) = r^2 H^{(b)} _r (r,\theta ,\phi ) \\ 
  =  - ik\sum\limits_{l = 1}^\infty  {rj_l (kr)} h^{(1)} (kr_s )\sum\limits_{m =  - l}^l {} D_h (\theta ,\phi )Y_l^m (\theta ,\phi )Y_l^{m*} (\theta _s ,\phi _s ) \\ 
  = \sum\limits_{l = 1}^\infty  {} H^{(b)} _l (r)\sum\limits_{m =  - l}^l {} D_h (\theta ,\phi )Y_l^m (\theta ,\phi )Y_l^{m*} (\theta _s ,\phi _s ), \\ 
 \end{array}
\end{equation}
Because the electric point source $(r_s ,\theta _s ,\phi _s )$ in outside of the GLHUA-1 cloak, $r_s  > R_{out}  > R_2 $ and operator $D_h (\theta ,\phi )$ and $\Re (\theta ,\phi )$ can be commutative ( 94),
substitute ( 95) into the GL magnetic equation ( 91) in the sphere annular domain  ,without source, the equation ( 91) is split into the second ordinary differential equation for every $l$, $l = 1,2, \cdots$, as follows,
\begin{equation}
\frac{\partial }{{\partial r}}\frac{{\partial H^{(b)} _l (r)}}{{\partial r}} - \frac{{l(l + 1)}}{{r^2 }}H^{(b)} _l (r) + k^2 H^{(b)} _l (r) = 0,
\end{equation}

\subsection{Analytical EM Wave propagation in the GLHUA-1 outer layer cloak}

We propose GLHUA-1 expansion method to create analytic expansion of
the magnetic wave field in GLHUA-1 outer layer cloak which satisfies the magnetic equation ( 91) in the outer layer domain with relative parameters ( 82) and ( 84). The GLHUA-1 analytic magnetic wave field and incident magnetic wave and their derivative satisfy the boundary continuous condition across the boundary $r = R_2 $.

Substitute the spherical harmonic expansion of the magnetic wave field
   \begin{equation}    
 \begin{array}{l}
 H(r,\theta ,\phi ) =  \\ 
  = \sum\limits_{l = 1}^\infty  {H_l (r)} \sum\limits_{m =  - l}^l {} D_h (\theta ,\phi )Y_l^m (\theta ,\phi )Y_l^{m*} (\theta _s ,\phi _s ), \\ 
 \end{array}
\end{equation}
into the GL magnetic equation ( 91) in the GLHUA-1 outer annular layer domain $R_1  \le r \le R_2 $, the equation ( 91) is split into the second ordinary differential equation for every $l$, $l = 1,2, \cdots$,  
\begin {equation}
\frac{\partial }{{\partial r}}\frac{1}{{\varepsilon _\theta  }}\frac{{\partial H_l (r)}}{{\partial r}} - \frac{{l(l + 1)}}{{\mu _r r^2 }}H_l (r) + k^2 \varepsilon _\theta  H_l (r) = 0,
\end{equation}
By comparison between magnetic equation ( 98) in GLHUA-1 outer layer cloak and 
incident magnetic wave equation ( 96), the $l$ component magnetic field $H_l (r)$ and $l$ component incident magnetic field $H^{(b)} _l (r)$ are continuous across boundary $r = R_2 $,
\begin{equation}
H_l (R_2 ) = H^{(b)} _l (R_2 ),
\end{equation}
and their radial derivative are continuous across boundary $r = R_2 $,

\begin{equation}
\frac{\partial }{{\partial r}}H_l (R_2 ) = \frac{\partial }{{\partial r}}H^{(b)} _l (R_2 ),
\end{equation}
Substitute GLHUA-1 cloak relative materials $\varepsilon _r (r) = \mu _r (r) = \frac{{R_2 ^2 }}{{r^2 }}$  in ( 82) and relative angular parameter in ( 84) into the equation ( 91), we have
\begin{equation}
\begin{array}{l}
\frac{\partial }{{\partial r}}\frac{{r - R_1 }}{{R_2  - R_1 }}\frac{{\partial H_l (r)}}{{\partial r}} \\
 - \frac{{l(l + 1)}}{{R_1 ^2 }}H_l (r) + k^2 \frac{{R_2  - R_1 }}{{r - R_1 }}H_l (r) = 0,
\end{array}
\end{equation}
We create GLHUA-1 expansion of the $l$ component GL magnetic field
$H_l (r)$

\begin{equation}
\begin{array}{l}
 H_l (r) = g_{l,1} (r)\cos (k(R_2  - R_1 )\log (\frac{{r - R_1 }}{{R_2  - R_1 }})) \\ 
  + g_{l,2} (r)\sin (k(R_2  - R_1 )\log (\frac{{r - R_1 }}{{R_2  - R_1 }})), \\ 
 \end{array}
\end{equation}

By calculation,

\begin{equation}
\begin{array}{l}
 \frac{\partial }{{\partial r}}\frac{{r - R_1 }}{{R_2  - R_1 }}\frac{\partial }{{\partial r}}H_l (r) \\ 
  = \left( {\frac{\partial }{{\partial r}}\frac{{r - R_1 }}{{R_2  - R_1 }}\frac{\partial }{{\partial r}}g_{l,1} (r) + 2k\frac{\partial }{{\partial r}}g_{l,2} (r)} \right) \\ 
 \cos (k(R_2  - R_1 )\log (\frac{{r - R_1 }}{{R_2  - R_1 }})) \\ 
  + \left( {\frac{\partial }{{\partial r}}\frac{{r - R_1 }}{{R_2  - R_1 }}\frac{\partial }{{\partial r}}g_{l,2} (r) - 2k\frac{\partial }{{\partial r}}g_{l,1} (r)} \right) \\ 
 \sin (k(R_2  - R_1 )\log (\frac{{r - R_1 }}{{R_2  - R_1 }})) \\ 
  - k^2 g_{l,1} (r)\cos (k(R_2  - R_1 )\log (\frac{{r - R_1 }}{{R_2  - R_1 }}))\frac{{R_2  - R_1 }}{{r - R_1 }} \\ 
  - k^2 g_{l,2} (r)\sin (k(R_2  - R_1 )\log (\frac{{r - R_1 }}{{R_2  - R_1 }}))\frac{{R_2  - R_1 }}{{r - R_1 }}, \\ 
 \end{array}
\end{equation}

Substitute ( 102) and ( 103) into the equation ( 101), we obtain a new GLHUA-1 two equation system

\begin{equation}
\begin{array}{l}
\frac{\partial }{{\partial r}}\frac{{r - R_1 }}{{R_2  - R_1 }}\frac{\partial }{{\partial r}}g_{l,1} (r) \\
+ 2k\frac{\partial }{{\partial r}}g_{l,2} (r) + \frac{{l(l + 1)}}{{R_2 ^2 }}g_{l,1} (r) = 0,
\end{array}
\end{equation}

\begin{equation}
\begin{array}{l}
\frac{\partial }{{\partial r}}\frac{{r - R_1 }}{{R_2  - R_1 }}\frac{\partial }{{\partial r}}g_{l,2} (r) \\
- 2k\frac{\partial }{{\partial r}}g_{l,1} (r) + \frac{{l(l + 1)}}{{R_2 ^2 }}g_{l,2} (r) = 0,
\end{array}
\end{equation}

We expand solution $(g_{l,1} (r),g_{l,2} (r))$ of the GLHUA-1 two equation system ( 104) and ( 105) as analytical power series,
\begin{equation}
 g_{l,1} (r) = \sum\limits_{j = 0}^\infty  {a_{l,j} } (r - R_1 )^j ,
\end{equation}
\begin{equation}
g_{l,2} (r) = \sum\limits_{j = 0}^\infty  {b_{l,j} } (r - R_1 )^j ,
\end{equation}

By calculation,
\begin{equation}
\begin{array}{l}
 \frac{\partial }{{\partial r}}g_{l,1} (r) = \sum\limits_{j = 1}^\infty  {a_{l,j} } j(r - R_1 )^{j - 1}  \\ 
  = \sum\limits_{j = 0}^\infty  {a_{l,j + 1} } (j + 1)(r - R_1 )^j , \\ 
 \end{array}
\end{equation}

\begin{equation}
\begin{array}{l}
\frac{\partial }{{\partial r}}g_{l,2} (r) = \sum\limits_{j = 1}^\infty  {b_{l,j} } j(r - R_1 )^{j - 1}  \\ 
= \sum\limits_{j = 0}^\infty  {b_{l,j + 1} } (j + 1)(r - R_1 )^j ,
 \end{array}
\end{equation} 

\begin{equation} 
 \begin{array}{l}
 (r - R_1 )\frac{\partial }{{\partial r}}\frac{\partial }{{\partial r}}g_{l,1} (r) \\
= \sum\limits_{j = 2}^\infty  {a_{l,j} } j(j - 1)(r - R_1 )^{j - 1}  \\ 
  = \sum\limits_{j = 1}^\infty  {a_{l,j + 1} } (j + 1)j(r - R_1 )^j , \\ 
 \end{array}
\end{equation} 

\begin{equation} 
\begin{array}{l}
 (r - R_1 )\frac{\partial }{{\partial r}}\frac{\partial }{{\partial r}}g_{l,2} (r) \\
= \sum\limits_{j = 2}^\infty  {b_{l,j} } j(j - 1)(r - R_1 )^{j - 1}  \\ 
  = \sum\limits_{j = 1}^\infty  {b_{l,j + 1} } (j + 1)j(r - R_1 )^j ,
 \end{array}
\end{equation}

Substitute (106-111) into GLHUA-1 equation ( 104), we have
\begin{equation} 
\begin{array}{l}
 \frac{1}{{R_2  - R_1 }}\sum\limits_{j = 1}^\infty  {a_{l,j + 1} (j + 1)} j(r - R_1 )^j  \\ 
  + \frac{1}{{R_2  - R_1 }}\sum\limits_{j = 0}^\infty  {a_{l,j + 1} (j + 1)} (r - R_1 )^j  \\ 
  + 2k\sum\limits_{j = 0}^\infty  {b_{l,j + 1} (j + 1)} (r - R_1 )^j  \\ 
  - \frac{{l(l + 1)}}{{R_2 ^2 }}\sum\limits_{j = 0}^\infty  {a_{l,j} } (r - R_1 )^j = 0 \\ 
 \end{array}
\end{equation} 
Substitute (106-111) into GLHUA-1 equation ( 105), we have

\begin{equation} 
\begin{array}{l}
 \frac{1}{{R_2  - R_1 }}\sum\limits_{j = 1}^\infty  {b_{l,j + 1} (j + 1)} j(r - R_1 )^j  \\ 
  + \frac{1}{{R_2  - R_1 }}\sum\limits_{j = 0}^\infty  {b_{l,j + 1} (j + 1)} (r - R_1 )^j  \\ 
  - 2k\sum\limits_{j = 0}^\infty  {a_{l,j + 1} (j + 1)} (r - R_1 )^j  \\ 
  - \frac{{l(l + 1)}}{{R_2 ^2 }}\sum\limits_{j = 0}^\infty  {b_{l,j} } (r - R_1 )^j  = 0, \\ 
 \end{array}
\end{equation} 

When  $j=0$ in ( 112), we have 
\begin{equation}
  \frac{1}{{R_2  - R_1 }}a_{l,1}  + 2kb_{l,1}  = \frac{{l(l + 1)}}{{R_2 ^2 }}a_{l,0} ,      
\end{equation}

When  $j=0$ in ( 113), we have

\begin{equation}
\frac{1}{{R_2  - R_1 }}b_{l,1}  - 2ka_{l,1}  = \frac{{l(l + 1)}}{{R_2 ^2 }}b_{l,0} ,
\end{equation} 

Solve the linear equation ( 114) and ( 115), we obtain
\begin{equation}
\begin{array}{l}
 \left[ {\begin{array}{*{20}c}
   {a_{l,1} }  \\
   {b_{l,1} }  \\
\end{array}} \right] = \frac{{l(l + 1)}}{{(1 + 4k(R_2  - R_1 )^2 )R_2^2 }} \\
\left[ {\begin{array}{*{20}c}
   {(R_2  - R_1 )} & { - 2k(R_2  - R_1 )^2 }  \\
   {2k(R_2  - R_1 )^2 } & {(R_2  - R_1 )}  \\
\end{array}} \right]\left[ {\begin{array}{*{20}c}
   {a_{l,0} }  \\
   {b_{l,0} }  \\
\end{array}} \right] \\ 
 \end{array}
\end{equation}
For every $j \ge 1$, in ( 112) and ( 113), we have
\begin{equation}
\begin{array}{l}
 \frac{1}{{R_2  - R_1 }}(j + 1)^2 a_{l,j + 1}  + 2k(j + 1)b_{l,j + 1}  \\ 
  = \frac{{l(l + 1)}}{{R_2 ^2 }}a_{l,j} , 
 \end{array}
\end{equation}
\begin{equation}
\begin{array}{l}
 \frac{1}{{R_2  - R_1 }}(j + 1)^2 b_{l,j + 1}  - 2k(j + 1)a_{l,j + 1}  \\ 
  = \frac{{l(l + 1)}}{{R_2 ^2 }}b_{l,j} , 
 \end{array}
\end{equation}

Solve the linear equation ( 117) and ( 118), we obtain

\begin{equation}
\begin{array}{l}
 \left[ {\begin{array}{*{20}c}
   {a_{l,j + 1} }  \\
   {b_{l,j + 1} }  \\
\end{array}} \right] = \frac{{l(l + 1)}}{{((j + 1)^3  + 4k(R_2  - R_1 )^2 (j + 1))R_2^2 }} \\ 
 \left[ {\begin{array}{*{20}c}
   {(j + 1)(R_2  - R_1 )} & { - 2k(R_2  - R_1 )^2 }  \\
   {2k(R_2  - R_1 )^2 } & {(j + 1)(R_2  - R_1 )}  \\
\end{array}} \right]\left[ {\begin{array}{*{20}c}
   {a_{l,j} }  \\
   {b_{l,j} }  \\
\end{array}} \right], \\ 
 \end{array}
\end{equation} 
There are two linearly independent vector function solution of the GLHUA-1 two equations system ( 104) and ( 105), $(g_{l,1,p} (r),g_{l,2,p} (r)),p = 1,2,$ When  $a_{l,0}^{(1)}  = 1,b_{l,0}^{(1)}  = 0$ from ( 116) and ( 119), we obtain  
$(a_{l,j}^{(1)} ,b_{l,j}^{(1)} ),j = 1,2, \cdots$ and from analytical power series ( 106) and ( 107), we obtain the first solution of GLHUA-1 quations ( 110) and ( 111),$(g_{l,1,1} (r),g_{l,2,1} (r))$  . When 
$a_{l,0}^{(2)}  = 0,b_{l,0}^{(2)}  = 1$ , from ( 116) and ( 119), we obtain 
$(a_{l,j}^{(2)} ,b_{l,j}^{(2)} ),j = 1,2, \cdots$, and from analytical power series ( 106) and ( 107), we obtain the second solution of GLHUA-1 equations ( 104) and ( 105).$(g_{l,1,2} (r),g_{l,2,2} (r))$  ,for $l = 1,2, \cdots $ .

 For arbitrary two constants $A_1$   and  $A_2$, 
 the general solution of the GLHUA-1 equation ( 104) and ( 105) is

\begin{equation} 
H_{l,1} (r) = A_1 g_{l,1,1} (r) + A_2 g_{l,1,2} (r),
\end{equation} 

\begin{equation} 
H_{l,2} (r) = A_1 g_{l,2,1} (r) + A_2 g_{l,2,2} (r),
\end{equation} 

Substitute ( 120) and ( 121) into ( 102), we obtain

\begin{equation} 
\begin{array}{l}
 H_l (r) \\ 
  = H_{l,1} (r)\cos \left( {k(R_2  - R_1 )\log \left( {\frac{{r - R_1 }}{{R_2  - R_1 }}} \right)} \right) \\ 
  + H_{l,2} (r)\sin \left( {k(R_2  - R_1 )\log \left( {\frac{{r - R_1 }}{{R_2  - R_1 }}} \right)} \right) \\ 
  = (A_1 g_{l,1,1} (r) + A_2 g_{l,1,2} (r)) \\ 
 \cos \left( {k(R_2  - R_1 )\log \left( {\frac{{r - R_1 }}{{R_2  - R_1 }}} \right)} \right) \\ 
  + (A_1 g_{l,2,1} (r) + A_2 g_{l,2,2} (r)) \\ 
 \sin \left( {k(R_2  - R_1 )\log \left( {\frac{{r - R_1 }}{{R_2  - R_1 }}} \right)} \right). \\ 
\end{array}    
\end{equation}

From the continuous condition of  $H_l (r)$  and $H^{(b)}_l(r)$  on  $r = R_2$  ( 99) and their derivative continuous condition ( 100), we determinate the constants $A_1$   and  $A_2$ in ( 122). Substitute ( 122) into ( 97), we obtained the GL analytic magnetic wave $H(\vec r)$, in the GLHUA-1 outer layer with relative parameter ( 82) and ( 84) not less than 1. 
Summary, we have proved the following theorem.
\hfill\break \\
\begin{figure}[b]
%\begin{m$$inipage}[t]{0.3\linewidth}
\centering
\includegraphics[width=0.86\linewidth,draft=false]{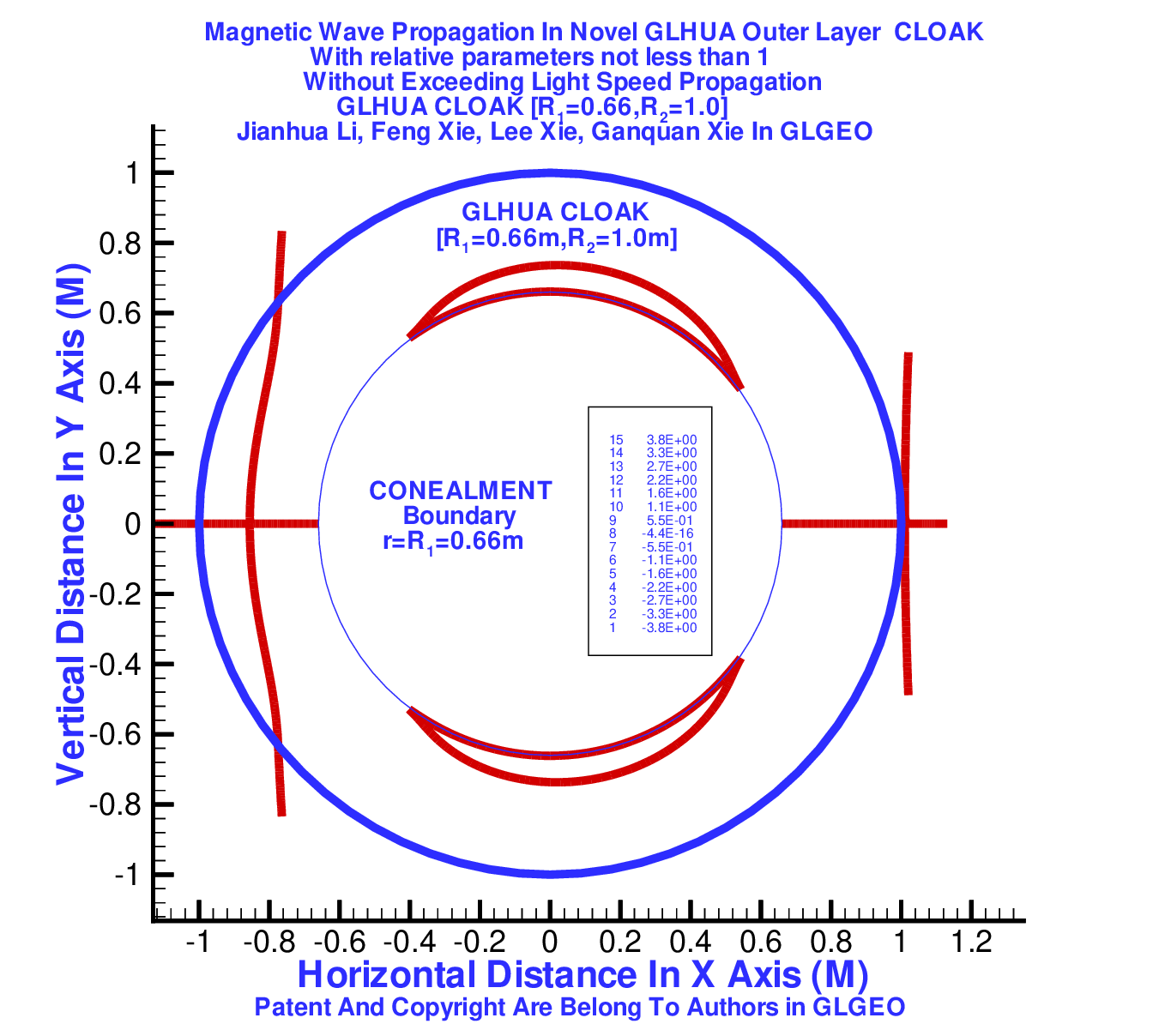}
\caption{ (color online) 
the magnetic wave is being tangent to the outer boundary $r=R_2$ without any scattering. It is very strangle, there is a down crescent type GL wave in the upper part of the inner spherical surface, $r=R_1$ , and upward crescent type GL wave in the lower part of the same inner spherical surface, $r=R_1$. 
.}\label{fig1}
\end{figure}
\begin{figure}[b]
%\begin{minipage}[t]{0.3\linewidth}
\centering
\includegraphics[width=0.86\linewidth,draft=false]{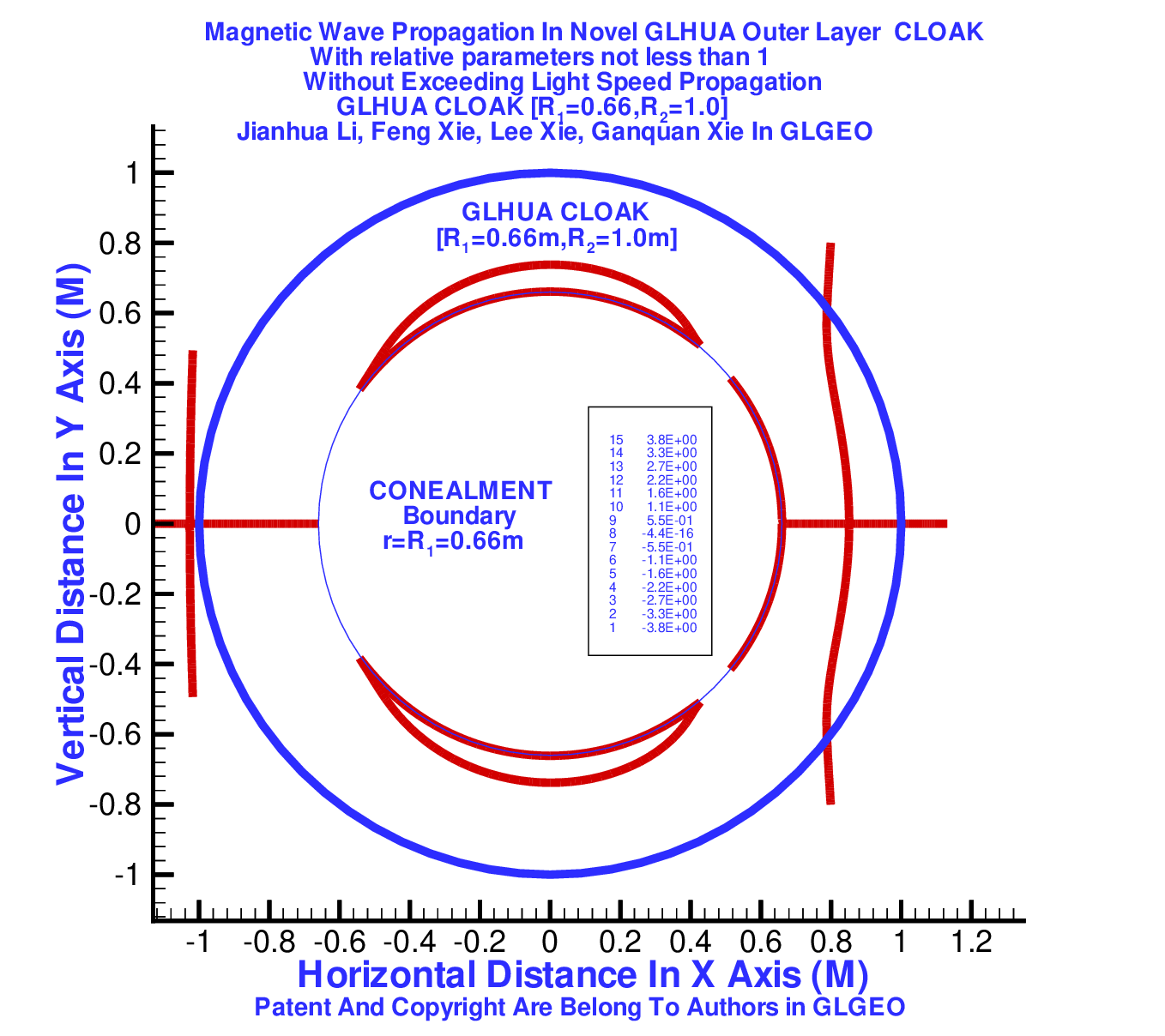}
\caption{ (color online) 
Magnetic wave propagation through GLHUA-1 
outer layer cloak, wave front at 31 step.}\label{fig2}
\end{figure}
\begin{figure}[b]
%\hskip5mm
%\begin{minipage}[t]{0.3\linewidth}
\centering
\includegraphics[width=0.86\linewidth,draft=false]{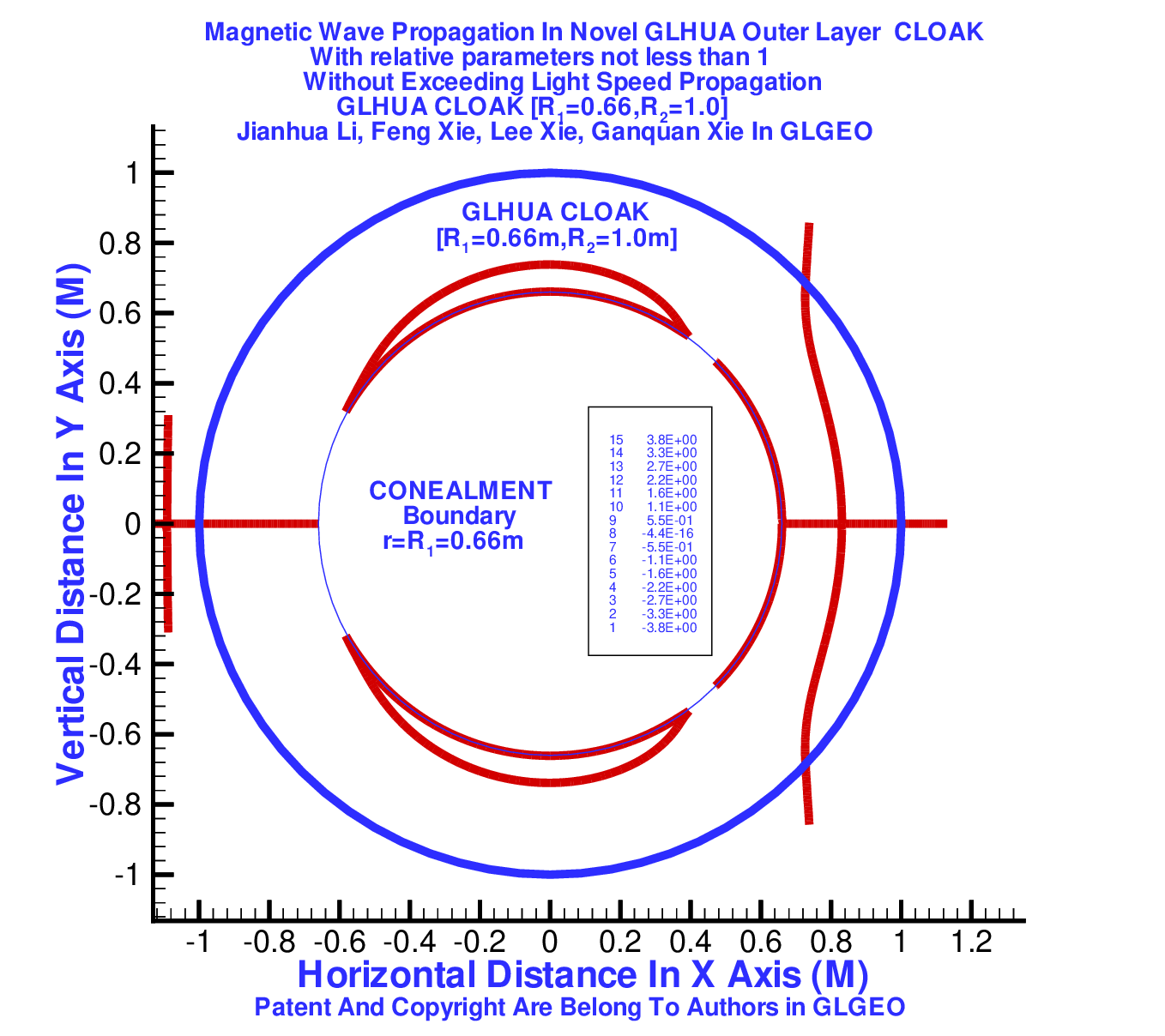}
\caption{ (color online)
Magnetic wave propagation through GLHUA-1 
outer layer cloak, wave front at 37 step. }\label{fig3}
%\end{minipage}
\end{figure}
\begin{figure}[b]
%\hskip5mm
%\begin{minipage}[t]{0.3\linewidth}
\centering
\includegraphics[width=0.86\linewidth,draft=false]{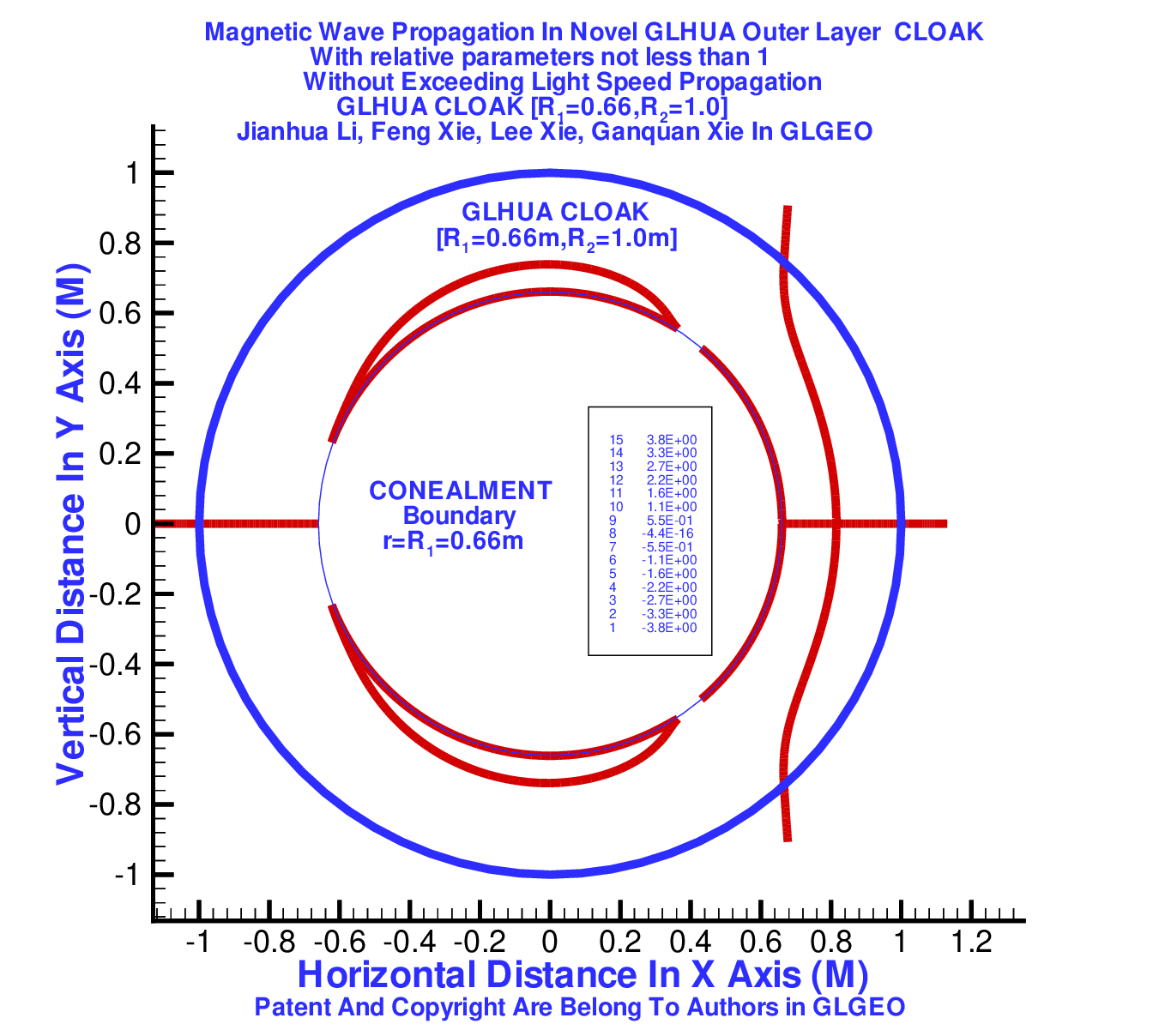}
\caption{ (color online)
Magnetic wave propagation through GLHUA-1 outer layer cloak, wave front at 43 step. 
.}\label{fig4}
\end{figure}
\begin{figure}[h]
%\begin{minipage}[t]{0.3\linewidth}
\centering
\includegraphics[width=0.86\linewidth,draft=false]{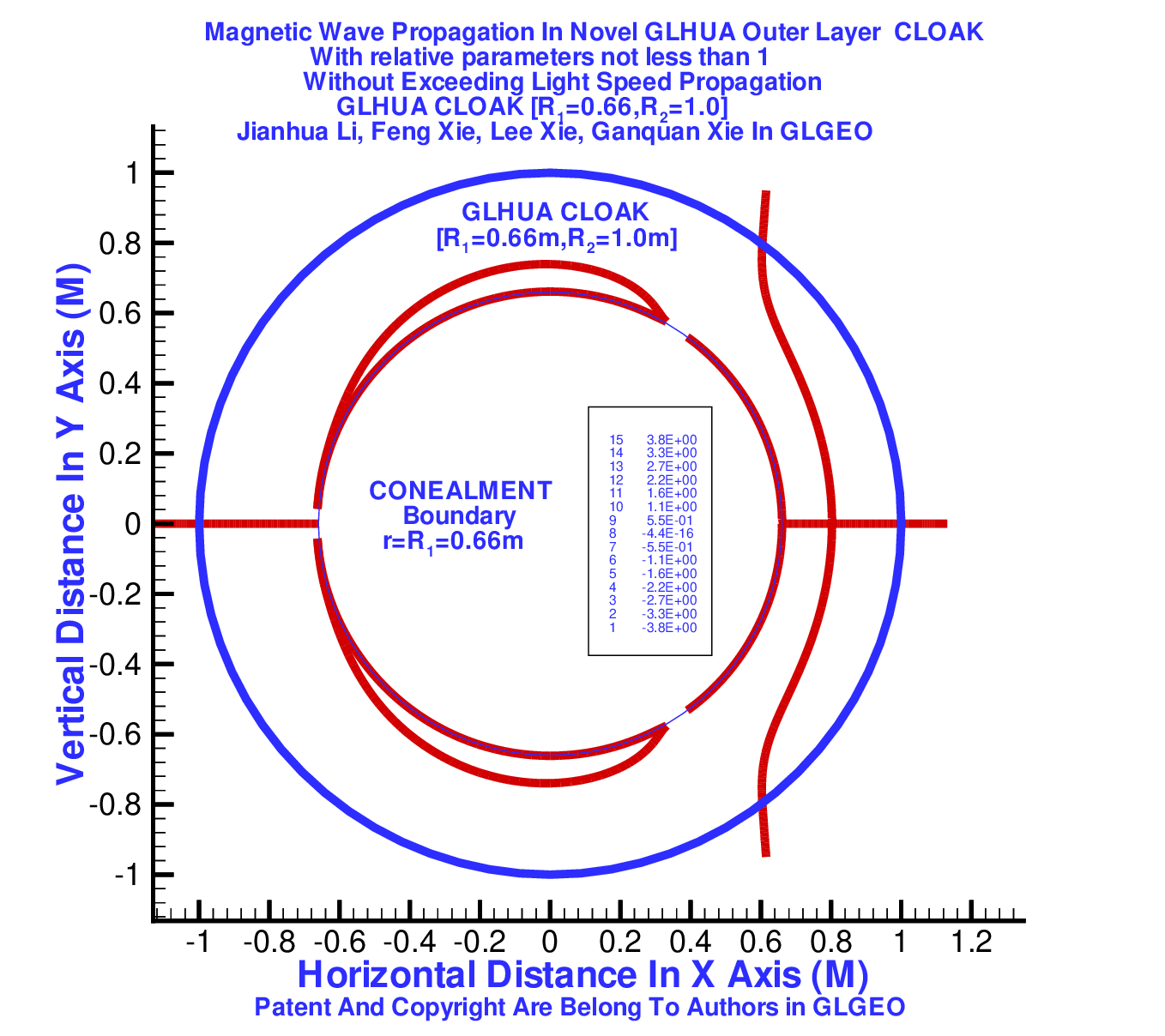}
\caption{ (color online) 
Magnetic wave propagation through GLHUA-1 outer layer cloak, wave front at 49 step.}\label{fig5}
%\end{minipage}
\end{figure}
\begin{figure}[h]
%\hskip5mm
%\begin{minipage}[t]{0.3\linewidth}
\centering
\includegraphics[width=0.85\linewidth,draft=false]{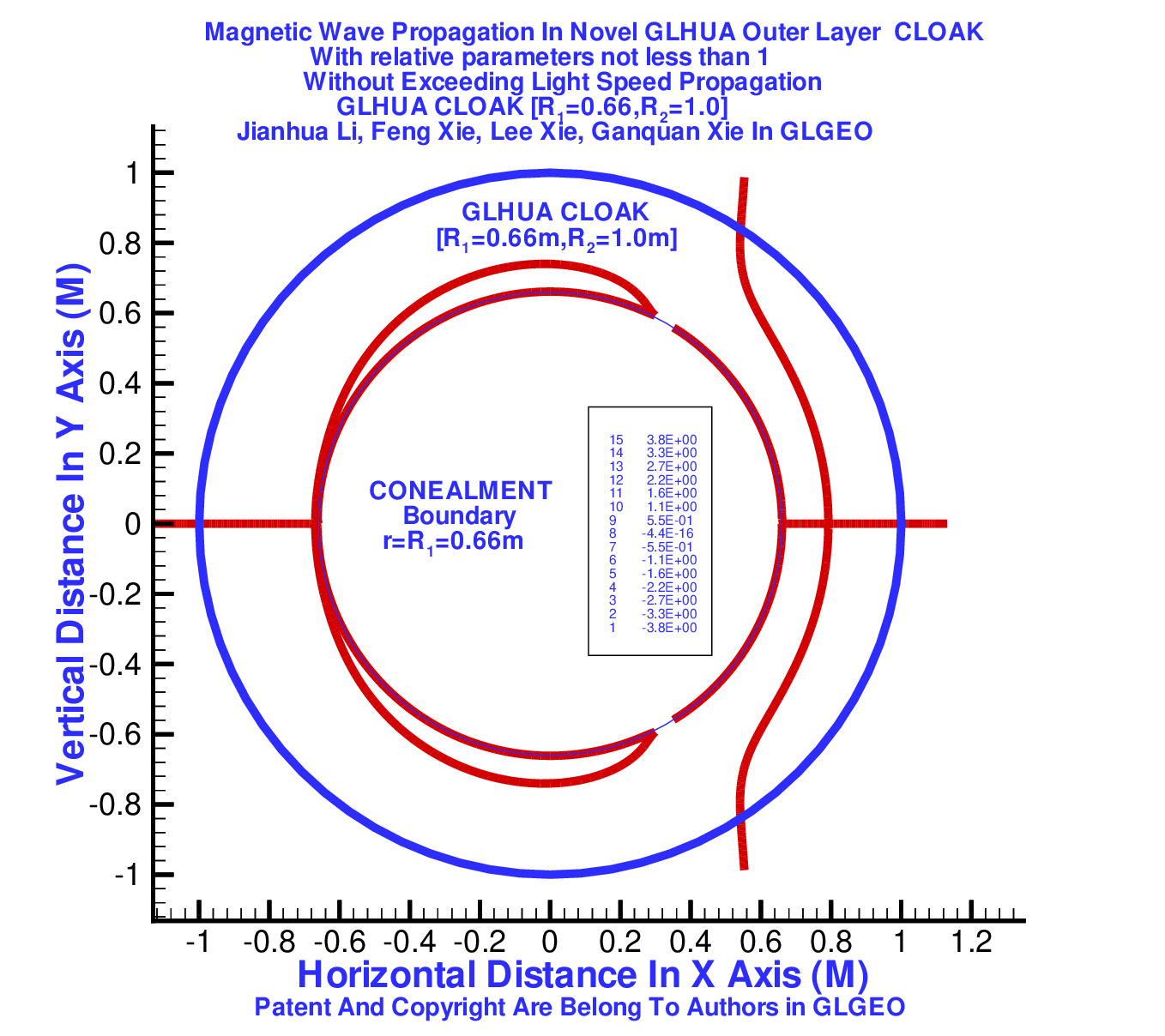}
\caption{ (color online) 
Magnetic wave propagation through GLHUA-1 outer layer cloak, wave front at 55 step.}\label{fig6}
\end{figure}
${\boldsymbol{Theorem \ 2:}}$ \  In the outer annular layer cloak, $R_1  \le r \le R_2 $, the radial relative electric permittivity and magnetic permeability, 
$\varepsilon _r  = \mu _r  = \frac{{R_2 ^2 }}{{r^2 }}$ in ( 82), and the angular relative electric permittivity and magnetic permeability, 
$\varepsilon _\theta   = \varepsilon _\phi   = \mu _\theta   = \mu _\phi   = \frac{{R_2  - R_1 }}{{r - R_1 }},$ in ( 84).
The incident magnetic wave is excited by the electric point source in the point  $\vec r_s  = (r_s ,\theta _s ,\phi _s )$, $r_s  > R_{out}  > R_2 $ . The analytic GL radial magnetic wave field solution of 3D GL radial magnetic equation ( 91) and on the boundary $r=R_2$ with field and derivative continuous to the incident magnetic wave in ( 95) is

\begin{equation} 
\begin{array}{l}
 H(\vec r) =  \\ 
  = \sum\limits_{l = 1}^\infty  {H_{l,1} (r)} \cos \left( {k(R_2  - R_1 )\log \left( {\frac{{r - R_1 }}{{R_2  - R_1 }}} \right)} \right) \\ 
 \sum\limits_{m =  - l}^l {} D_h (\theta ,\phi )Y_l^m (\theta ,\phi )Y_l^{m*} (\theta _s ,\phi _s ) \\ 
  + \sum\limits_{l = 1}^\infty  {H_{l,2} (r)} \sin \left( {k(R_2  - R_1 )\log \left( {\frac{{r - R_1 }}{{R_2  - R_1 }}} \right)} \right) \\ 
 \sum\limits_{m =  - l}^l {} D_h (\theta ,\phi )Y_l^m (\theta ,\phi )Y_l^{m*} (\theta _s ,\phi _s ), \\ 
 \end{array}
\end{equation} 

Where operator
\begin{equation} 
D_h (\theta ,\phi ) = \cos \phi \frac{\partial }{{\partial \theta }} - \frac{{\cos \theta \sin \phi }}{{\sin \theta }}\frac{\partial }{{\partial \phi }},
\end{equation} 
\begin{equation}
\left[ {\begin{array}{*{20}c}
   {H_{l,1} (r)}  \\
   {H_{l,2} (r)}  \\
\end{array}} \right] = \left[ {\begin{array}{*{20}c}
   {g_{l,1,1} (r)} & {g_{l,1,2} (r)}  \\
   {g_{l,2,1} (r)} & {g_{l,2,2} (r)}  \\
\end{array}} \right]\left[ {\begin{array}{*{20}c}
   {A_{l,1} }  \\
   {A_{l,2} }  \\
\end{array}} \right],            
\end{equation}

\begin{equation} 
\begin{array}{l}
 \left[ {\begin{array}{*{20}c}
   {g_{l,1,1} (r)} & {g_{l,1,2} (r)}  \\
   {g_{l,2,1} (r)} & {g_{l,2,2} (r)}  \\
\end{array}} \right] =  \\ 
  = \sum\limits_{j = 0}^\infty  {} \left[ {\begin{array}{*{20}c}
   {a_{l,j}^{(1)} } & {b_{l,j}^{(1)} }  \\
   {a_{l,j}^{(2)} } & {b_{l,j}^{(2)} }  \\
\end{array}} \right](r - R_1 )^j , 
 \end{array}
\end{equation}
\begin{equation}
\begin{array}{l}
 \left[ {\begin{array}{*{20}c}
   {a_{l,j + 1}^{(1)} } & {b_{l,j + 1}^{(1)} }  \\
   {a_{l,j + 1}^{(2)} } & {b_{l,j + 1}^{(2)} }  \\
\end{array}} \right] = \frac{{l(l + 1)}}{{((j + 1)^3  + 4k(R_2  - R_1 )^2 (j + 1))R_2^2 }} \\ 
 \left[ {\begin{array}{*{20}c}
   {(j + 1)(R_2  - R_1 )} & { - 2k(R_2  - R_1 )^2 }  \\
   {2k(R_2  - R_1 )^2 } & {(j + 1)(R_2  - R_1 )}  \\
\end{array}} \right]\left[ {\begin{array}{*{20}c}
   {a_{l,j}^{(1)} } & {b_{l,j}^{(1)} }  \\
   {a_{l,j}^{(2)} } & {b_{l,j}^{(2)} }  \\
\end{array}} \right], \\ 
 \end{array}
\end{equation}
\begin{equation}
\left[ {\begin{array}{*{20}c}
   {a_{l,0}^{(1)} } & {b_{l,0}^{(1)} }  \\
   {a_{l,0}^{(2)} } & {b_{l,0}^{(2)} }  \\
\end{array}} \right] = \left[ {\begin{array}{*{20}c}
   1 & 0  \\
   0 & 1  \\
\end{array}} \right],
\end{equation}
\begin{equation}
\begin{array}{l}
 \left[ {\begin{array}{*{20}c}
   {g_{l,1,1} (R_2 )} & {g_{l,1,2} (R_2 )}  \\
   {\left( {\frac{{\partial g_{l,1,1} }}{{\partial r}} + kg_{l,2,1} } \right)|(R_2 )} & {\left( {\frac{{\partial g_{l,1,2} }}{{\partial r}} + kg_{l,2,2} } \right)|(R_2 )}  \\
\end{array}} \right]\left[ {\begin{array}{*{20}c}
   {A_{l,1} }  \\
   {A_{l,2} }  \\
\end{array}} \right] \\ 
  = \left[ {\begin{array}{*{20}c}
   {H^{(b)} _l (R_2 )}  \\
   {\frac{\partial }{{\partial r}}H^{(b)} _l (R_2 )}  \\
\end{array}} \right], \\ 
 \end{array}
 \end{equation}
 \begin{equation}      
\begin{array}{l}
 \left[ {\begin{array}{*{20}c}
   {H^{(b)} _l (R_2 )}  \\
   {\frac{\partial }{{\partial r}}H^{(b)} _l (R_2 )}  \\
\end{array}} \right] =  \\ 
 \left[ {\begin{array}{*{20}c}
   { - ikR_2 j_l (kR_2 )h^{(1)} (kr_s )}  \\
   { - ik((l + 1)j_l (kR_2 ) - kR_2 j_{l + 1} (kR_2 ))h_l ^{(1)} (kr_s )}  \\
\end{array}} \right], \\ 
 \end{array}
 \end{equation}
The $j_l (kr)$ is spherical Bessel function, $h^{(1)} _l (kr_s ) = j_l (kr_s ) + iy_l (kr_s )$  is first type spherical Hankel function,$y_l (kr)$  is spherical Neumann function.
Now, we prove that GLHUA-1 exact analytical radial magnetic wave ( 123) does satisfy the GL radial magnetic equation ( 91).

Substitute the new GLHUA-1 outer layer invisible cloak ( 82) and ( 84) into the GL
radial magnetic differential equation ( 91), we obtain GL radial magnetic eqution in GLHUA-1 
cloak domain 
\begin{equation}
\begin{array}{l}
 \frac{\partial }{{\partial r}}\frac{{r - R_1 }}{{R_2  - R_1 }}\frac{\partial }{{\partial r}}H(\vec r) \\ 
  + \frac{1}{{R_2 ^2 }}\left( {\frac{1}{{\sin \theta }}\frac{\partial }{{\partial \theta }}\sin \theta \frac{\partial }{{\partial \theta }}} \right)H\left( {\vec r} \right) \\ 
  + \frac{1}{{R_2 ^2 }}\frac{1}{{\sin ^2 \theta }}\frac{{\partial ^2 }}{{\partial ^2 \phi }}H\left( {\vec r} \right) \\ 
  + k^2 \frac{{R_2  - R_1 }}{{r - R_1 }}H(\vec r) = 0, \\ 
 \end{array}
\end{equation}      
Substitute the GLHUA-1 analytical EM wave ( 123) into the equation ( 131),and
By definition of operator $\Re (\theta ,\phi )$ in ( 92) and definition of operator $D_h (\theta ,\phi )$
in ( 93), and $\Re (\theta ,\phi )D_h (\theta ,\phi ) = D_h (\theta ,\phi )\Re (\theta ,\phi )$ in ( 94), we find that
\begin{equation}
 \begin{array}{l}
 \frac{\partial }{{\partial r}}\frac{{r - R_1 }}{{R_2  - R_1 }}\frac{\partial }{{\partial r}}H(\vec r) \\ 
  + \frac{1}{{R_2 ^2 }}\left( {\frac{1}{{\sin \theta }}\frac{\partial }{{\partial \theta }}\sin \theta \frac{\partial }{{\partial \theta }}} \right)H\left( {\vec r} \right) \\ 
  + \frac{1}{{R_2 ^2 }}\frac{1}{{\sin ^2 \theta }}\frac{{\partial ^2 }}{{\partial ^2 \phi }}H\left( {\vec r} \right) \\ 
  + k^2 \frac{{R_2  - R_1 }}{{r - R_1 }}H(\vec r) \\ 
  = \sum\limits_{l = 1}^\infty  {\left( {\frac{\partial }{{\partial r}}\frac{{r - R_1 }}{{R_2  - R_1 }}\frac{\partial }{{\partial r}}H_{l,1} (r)} \right.}  \\ 
 \left. { + 2k\frac{\partial }{{\partial r}}H_{l,2} (r) - \frac{{l(l + 1)}}{{R_2 ^2 }}H_{l,1} (r)} \right) \\ 
 \cos \left( {k(R_2  - R_1 )\log \left( {\frac{{r - R_1 }}{{R_2  - R_1 }}} \right)} \right) \\ 
 \sum\limits_{m =  - l}^l {} D_h (\theta ,\phi )Y_l^m (\theta ,\phi )Y_l^{m*} (\theta _s ,\phi _s ) \\ 
  + \sum\limits_{l = 1}^\infty  {\left( {\frac{\partial }{{\partial r}}\frac{{r - R_1 }}{{R_2  - R_1 }}\frac{\partial }{{\partial r}}H_{l,2} (r)} \right.}  \\ 
 \left. { - 2k\frac{\partial }{{\partial r}}H_{l,1} (r) - \frac{{l(l + 1)}}{{R_2 ^2 }}H_{l,2} (r)} \right) \\ 
 \sin \left( {k(R_2  - R_1 )\log \left( {\frac{{r - R_1 }}{{R_2  - R_1 }}} \right)} \right) \\ 
 \sum\limits_{m =  - l}^l {} D_h (\theta ,\phi )Y_l^m (\theta ,\phi )Y_l^{m*} (\theta _s ,\phi _s ), \\ 
 \end{array}
\end{equation}
By ( 126)-( 128), $g_{l,1,1} $ and $g_{l,2,1} $ satisfy the GLHUA-1 two equation system ( 104) and ( 105)
With initial $(1,0)$,
$g_{l,1,2} $ and $g_{l,2,2}$ satisfy the GLHUA-1 two equation system ( 104) and ( 105)
With initial $(0,1)$.,
By ( 125), we know that $H_{l,1} $ and $H_{l,2} $ satisfy the GLHUA-1 two equation system ( 104) and ( 105). Substitute $H_{l,1} $ and $H_{l,2} $  in ( 125) into ( 132). We have proved that exact analytical GL radial magnetic $H(\vec r)$ in ( 125) is solution of the GL radial magnetic equation ( 82). Based on the
Condition ( 129) and ( 130), on the outer spherical boundary $r=R_2$, The magnetic $H(\vec r)$ and its derivative satisfy the continuous boundary conditions ( 99) and ( 100). Therefore, the theorem 9 is
proved. 

Suppose that the incident electric wave is excited by the electric point source ( 92) located in the point  $\vec r_s  = (r_s ,\theta _s ,\phi _s )$ ,$r_s  > R_{out}  > R_2 $. by same analytic process, from the continuous of the incident GL electric wave and its derivative on the boundary, $r = R_2 $ , we can create the exact GL analytic electric wave in the GLHUA-1 outer layer.$E(\vec r)$  .
Suppose that the incident EM wave is plane wave generated by the plane
wave  $\vec p(\vec r) = e^{i\left| k \right|r\cos (\vec k \cdot \vec r)} \vec e$ , by same analytic process in section 4, from the continuous of incident GL EM plane wave and its derivative on the boundary, $r = R_2 $ , we can create the exact GL analytic EM plane wave in the GLHUA-1 outer layer. $E(\vec r)$  
$H(\vec r)$in ( 129).
Finally, we obtain the analytic radial electric wave 
 \begin{equation}
E_r (\vec r) = \frac{1}{{R_2^2 }}E(\vec r),
 \end{equation}
and the analytic radial magnetic wave 
   \begin{equation}
H_r (\vec r) = \frac{1}{{R_2^2 }}H(\vec r),
 \end{equation}                            
which satisfy the radial Maxwell EM equation ( 91) in the GLHUA-1 outer layer cloak with relative parameters in ( 82) and ( 84) not less than 1.
Summary, we have the following 
\begin{figure}[h]
\centering
\includegraphics[width=0.85\linewidth,draft=false]{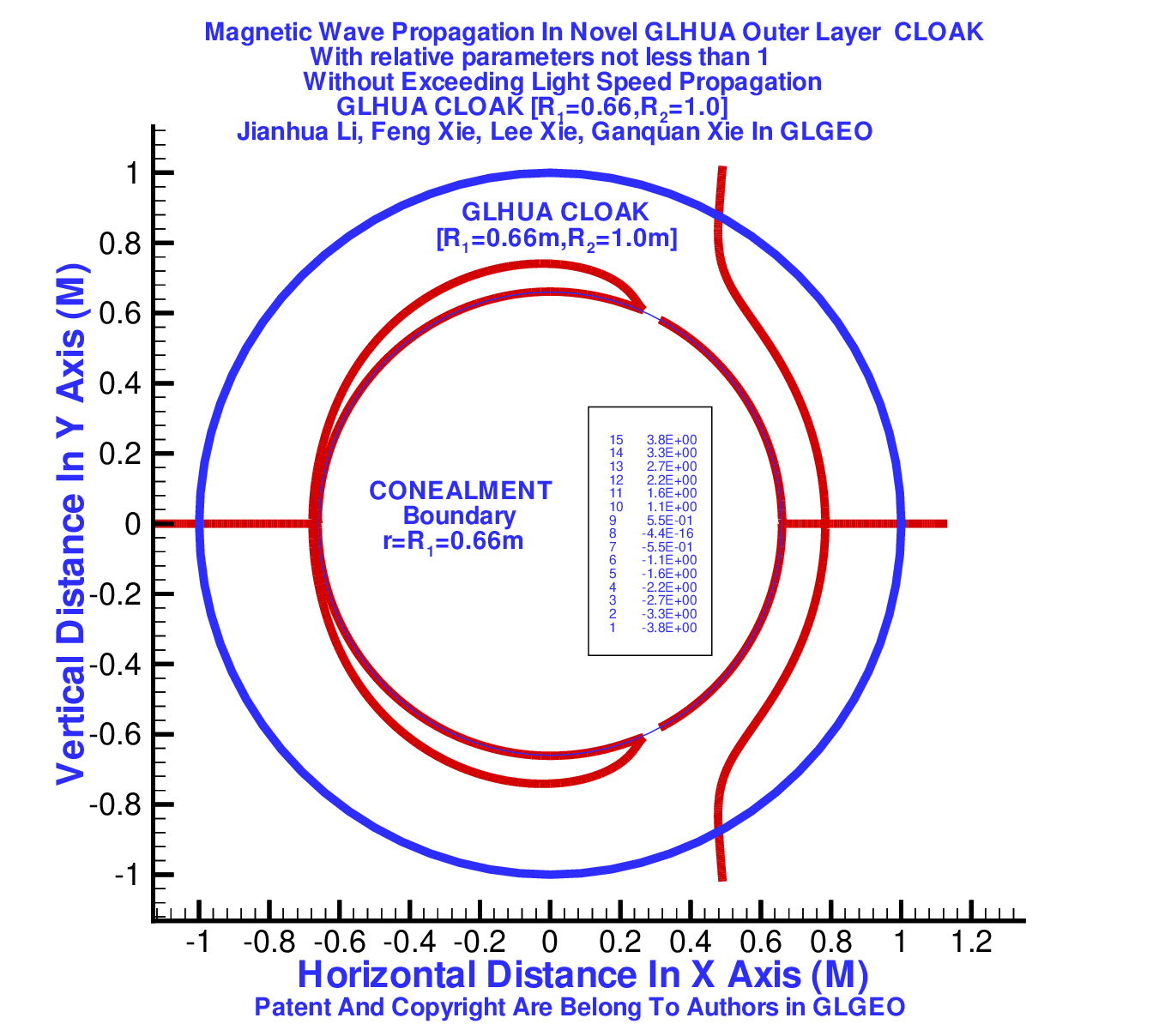}
\caption{ (color online)
Magnetic wave propagation through GLHUA-1 outer layer cloak, wave front at 61 step.
}\label{fig7}
%\end{minipage}
\end{figure}
\begin{figure}[h]
%\begin{minipage}[t]{0.3\linewidth}
\centering
\includegraphics[width=0.85\linewidth,draft=false]{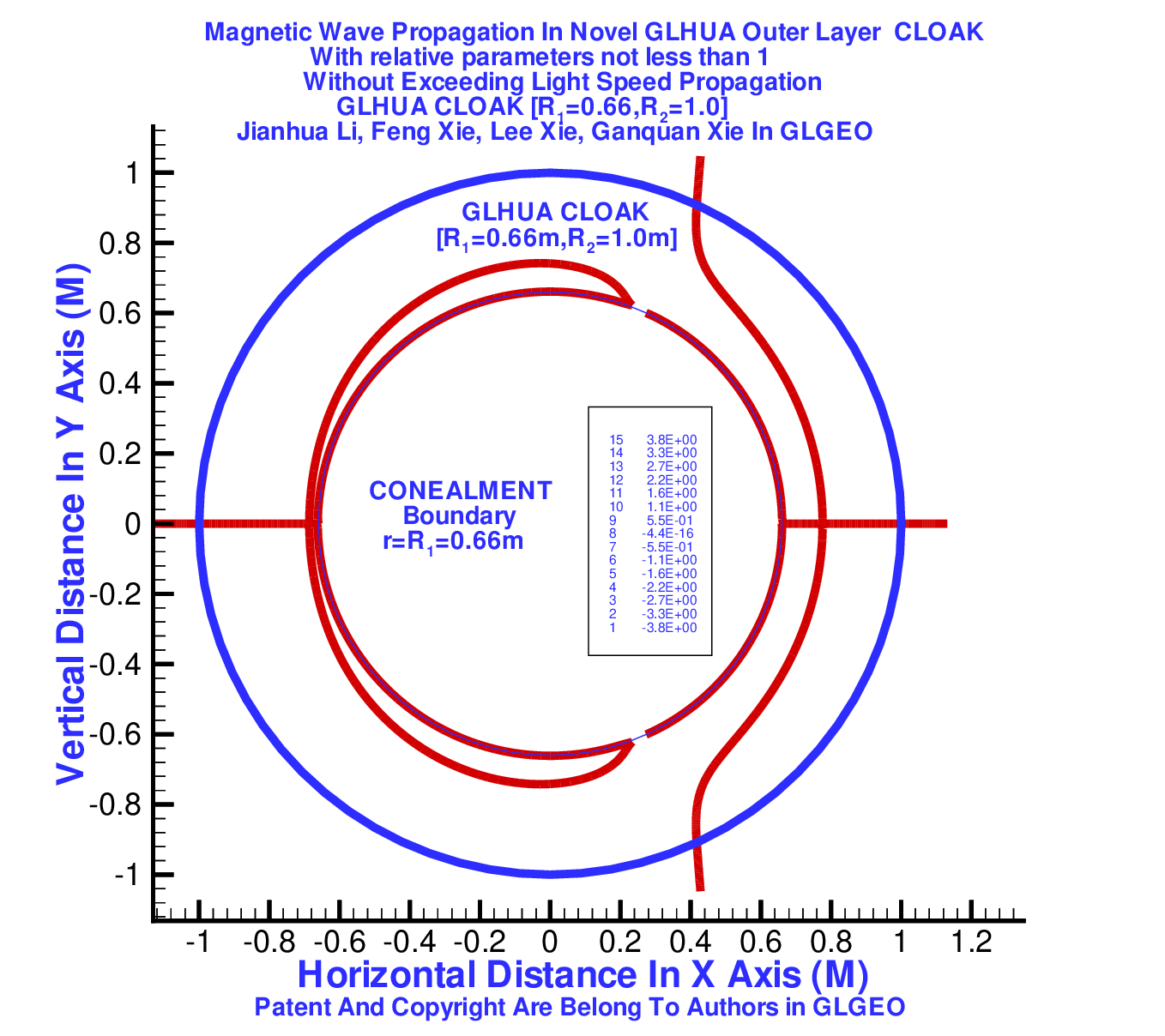}
\caption{ (color online) 
Magnetic wave propagation through GLHUA-1 
outer layer cloak, wave front at 67 step.
}\label{fig8}
%\end{minipage}
%\hskip5mm
\end{figure}
\begin{figure}[h]
%\begin{minipage}[t]{0.3\linewidth}
\centering
\includegraphics[width=0.85\linewidth,draft=false]{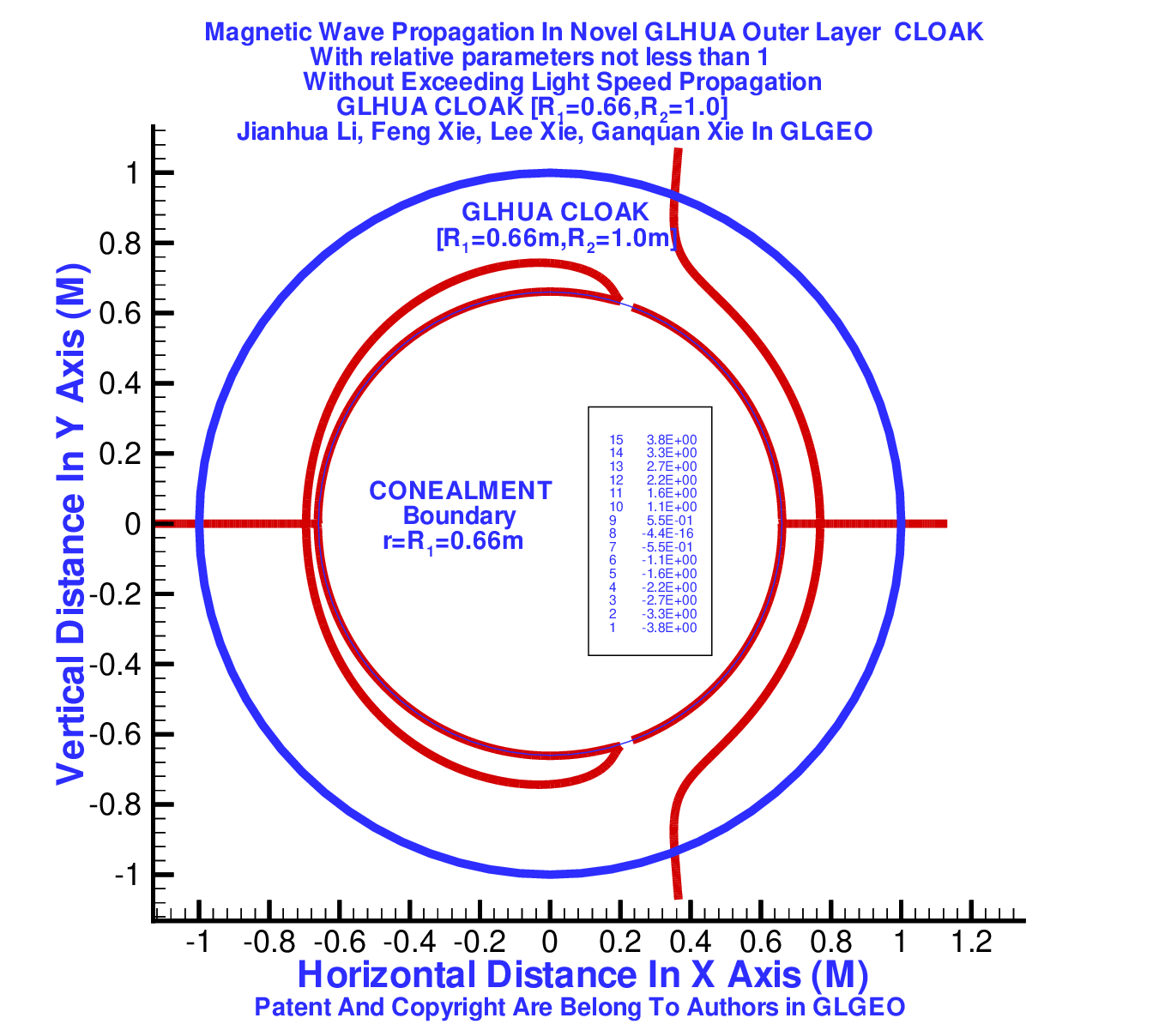}
\caption{ (color online) 
Magnetic wave propagation through GLHUA-1 
outer layer cloak, wave front at 73 step.
 }\label{fig9}
%\end{minipage}
%\hskip5mm
%begin{minipage}[t]{0.3\linewidth}
\end{figure}
\begin{figure}[h]
\centering
\includegraphics[width=0.85\linewidth,draft=false]{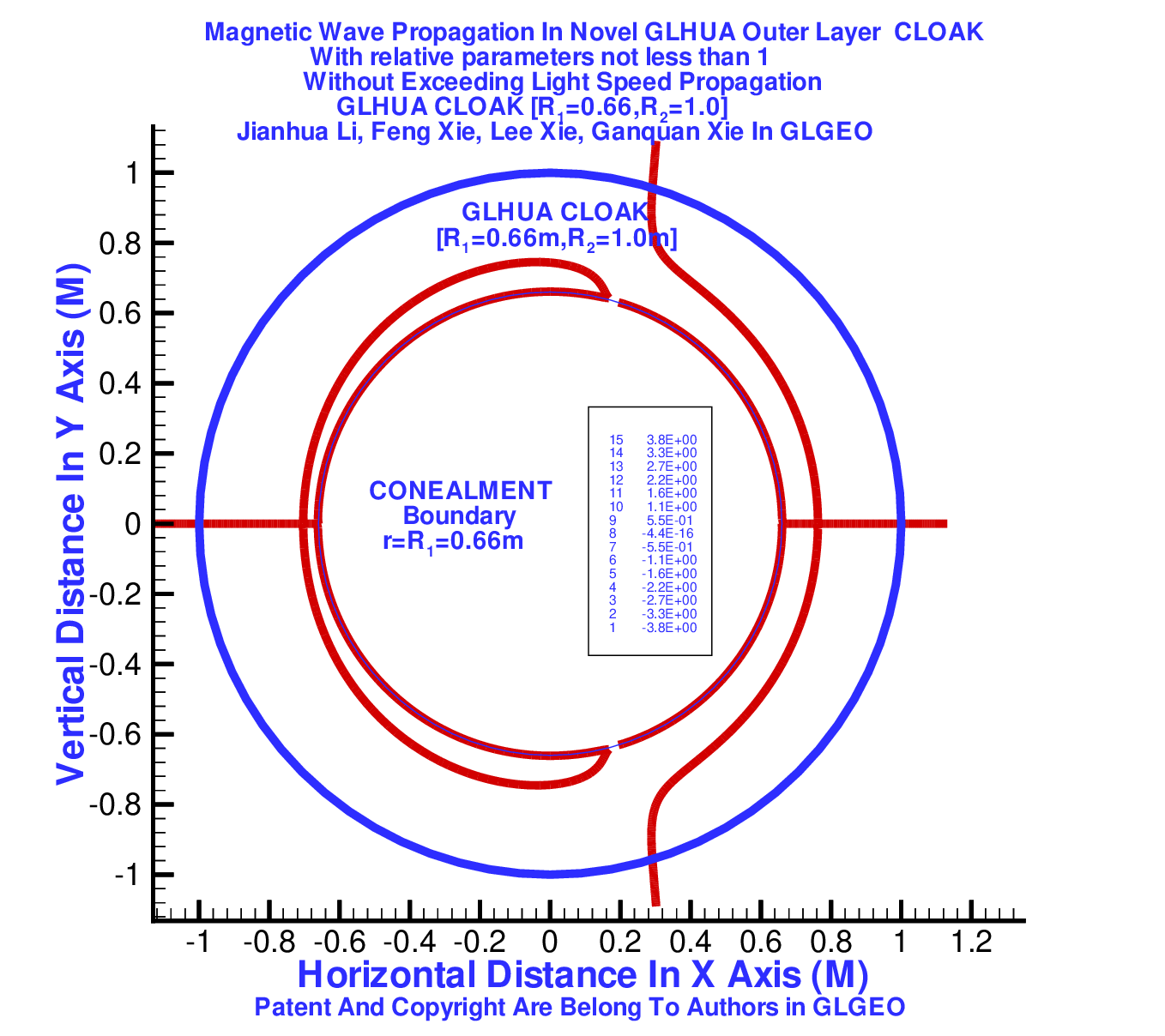}
\caption{ (color online) 
Magnetic wave propagation through GLHUA-1 
outer layer cloak, wave front at 79 step.
}\label{fig10}
%\end{minipage}
\end{figure}

${\boldsymbol{Theorem \ 3:}}$ \  In the outer annular layer cloak, $R_1  \le r \le R_2 $, the radial relative electric permittivity and magnetic permeability, 
$\varepsilon _r  = \mu _r  = \frac{{R_2 ^2 }}{{r^2 }}$ in ( 82), and the angular relative electric permittivity and magnetic permeability, 
$\varepsilon _\theta   = \varepsilon _\phi   = \mu _\theta   = \mu _\phi   = \frac{{R_2  - R_1 }}{{r - R_1 }},$ in ( 84).
The incident magnetic wave is excited by the electric point source in the point  $\vec r_s  = (r_s ,\theta _s ,\phi _s )$, $r_s  > R_{out}  > R_2 $ . The analytic GL radial electromagnetic wave field solution of 3D Maxwell radial electromagnetic equation in GLHUA-1 outer layer cloak domain and generated by the incident EM wave on the boundary $r=R_2$ is

 \begin{equation}
\begin{array}{l}
 \left[ {\begin{array}{*{20}c}
   {E_r (\vec r)}  \\
   {H_r (\vec r)}  \\
\end{array}} \right] =  \\ 
  = \sum\limits_{l = 1}^\infty  {\left[ {\begin{array}{*{20}c}
   {E_{r,l,1} (\vec r)}  \\
   {H_{r,l,1} (\vec r)}  \\
\end{array}} \right]} \cos \left( {k(R_2  - R_1 )\log \left( {\frac{{r - R_1 }}{{R_2  - R_1 }}} \right)} \right) \\ 
 \sum\limits_{m =  - l}^l {} \left[ {\begin{array}{*{20}c}
   {D_e (\theta ,\phi )}  \\
   {D_h (\theta ,\phi )}  \\
\end{array}} \right]Y_l^m (\theta ,\phi )Y_l^{m*} (\theta _s ,\phi _s ) \\ 
  + \sum\limits_{l = 1}^\infty  {\left[ {\begin{array}{*{20}c}
   {E_{r,l,2} (\vec r)}  \\
   {H_{r,l,2} (\vec r)}  \\
\end{array}} \right]} \sin \left( {k(R_2  - R_1 )\log \left( {\frac{{r - R_1 }}{{R_2  - R_1 }}} \right)} \right) \\ 
 \sum\limits_{m =  - l}^l {} \left[ {\begin{array}{*{20}c}
   {D_e (\theta ,\phi )}  \\
   {D_h (\theta ,\phi )}  \\
\end{array}} \right]Y_l^m (\theta ,\phi )Y_l^{m*} (\theta _s ,\phi _s ), \\ 
 \end{array}
  \end{equation}
\begin{figure}[h]
\centering
\includegraphics[width=0.85\linewidth,draft=false]{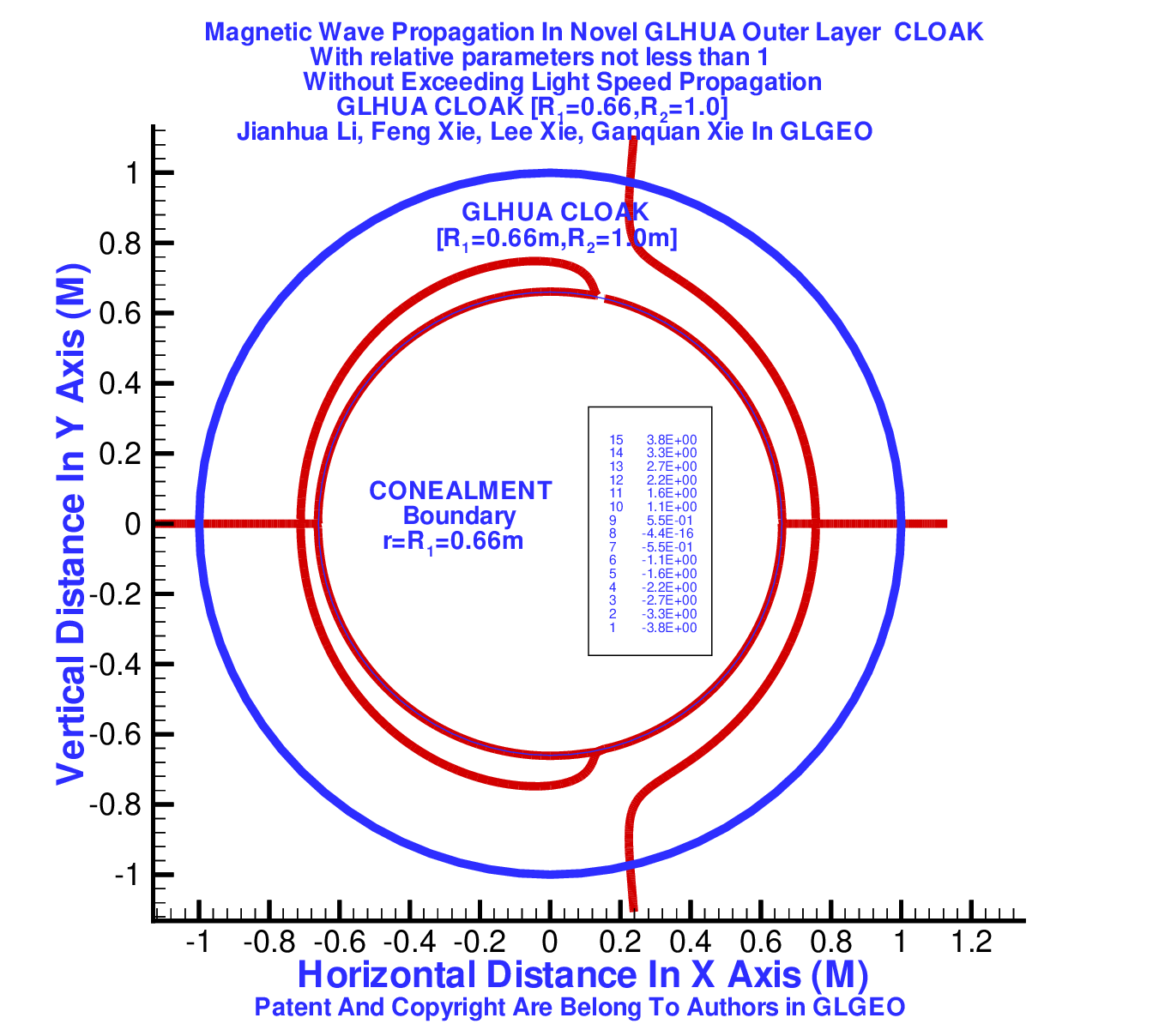}
\caption{ (color online) 
Magnetic wave propagation through GLHUA-1 
outer layer cloak, wave front at 85 step.
   }\label{fig14}
%\end{minipage}
\end{figure}
\begin{figure}[h]
\centering
\includegraphics[width=0.85\linewidth,draft=false]{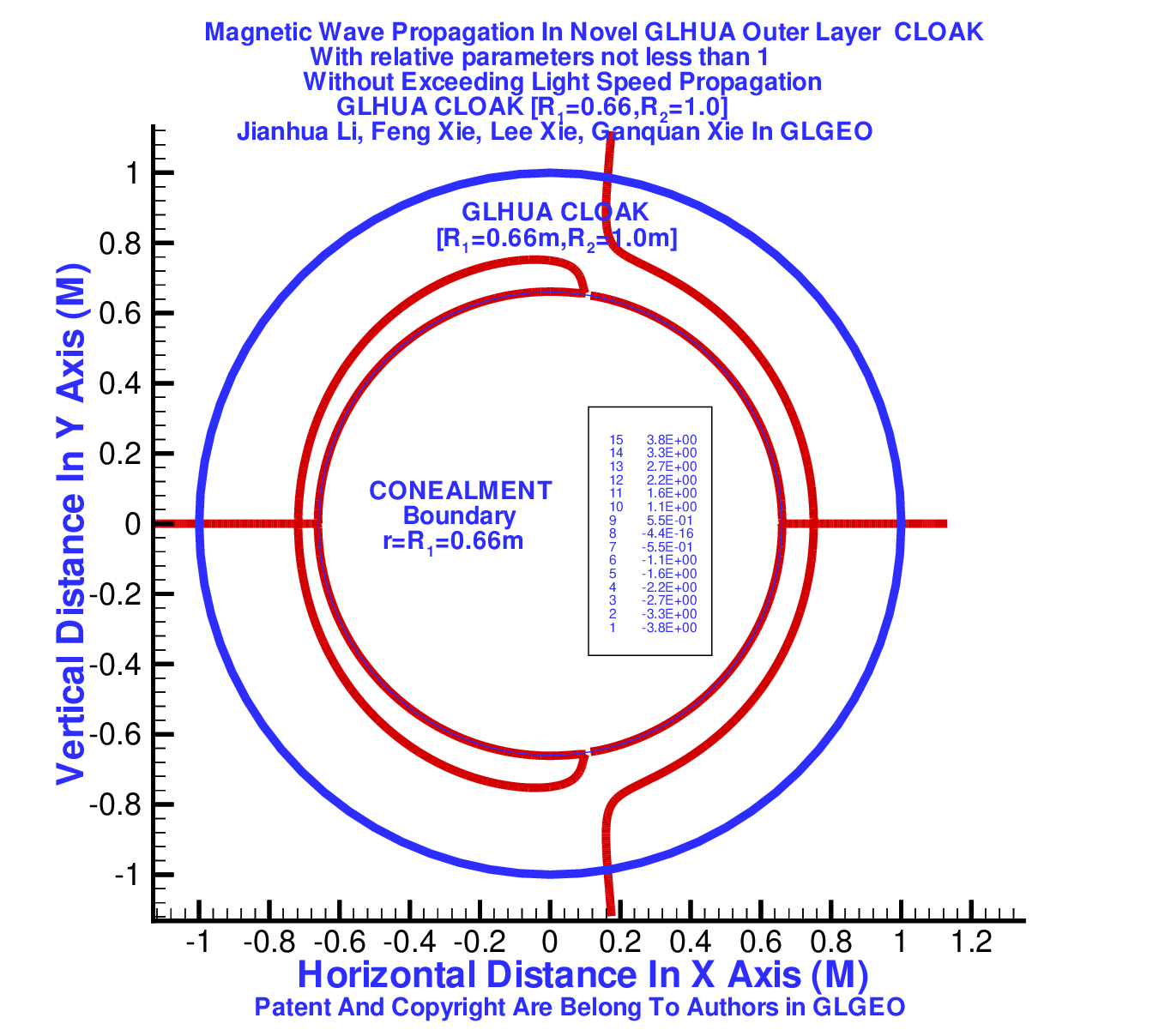}
\caption{ (color online) 
Magnetic wave propagation through GLHUA-1 
outer layer cloak, wave front at 91 step.
 }\label{fig15}
%\end{minipage}
\end{figure}
\begin{figure}[h]
\centering
\includegraphics[width=0.86\linewidth,draft=false]{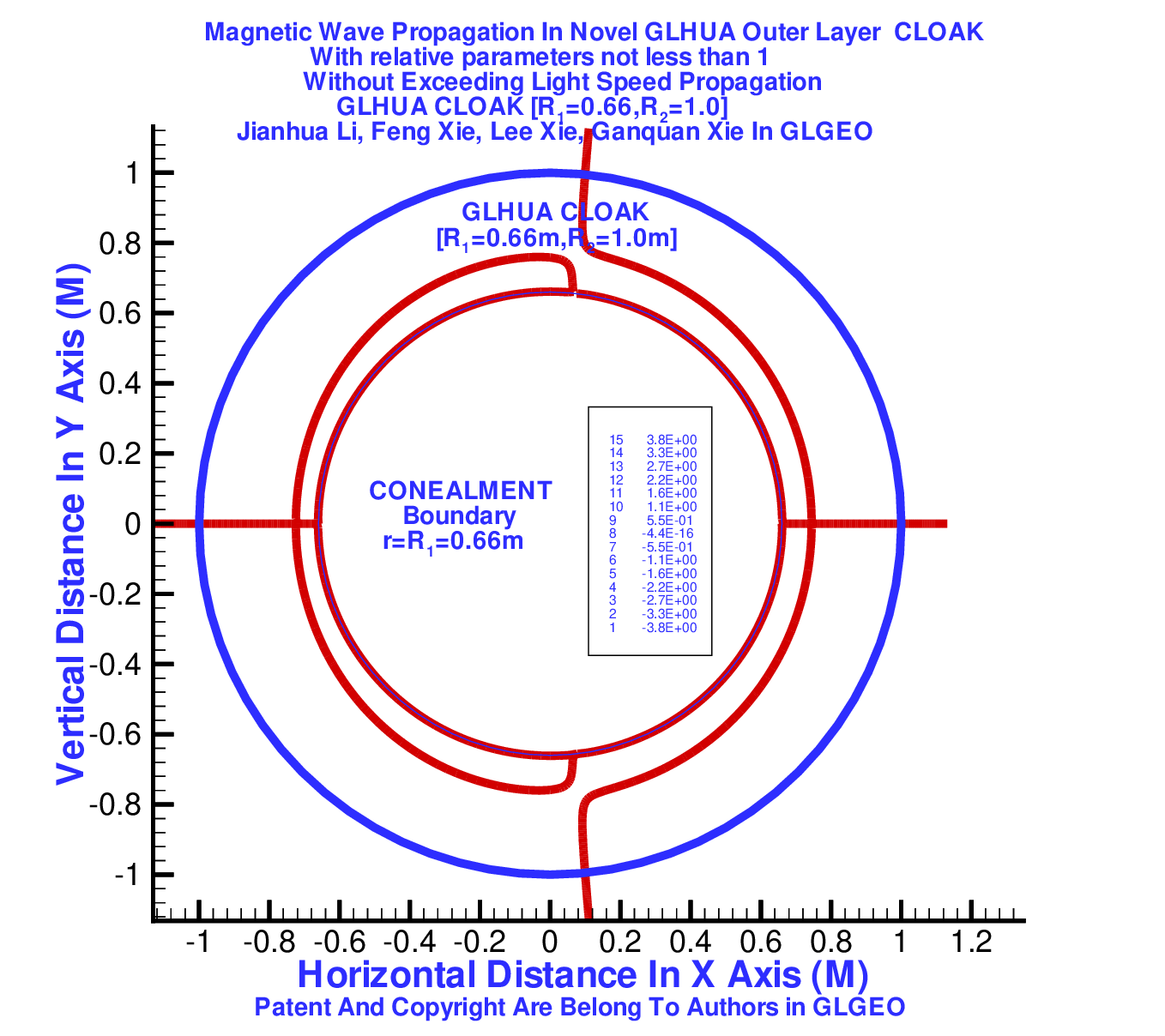}
\caption{ (color online) 
Magnetic wave propagation through GLHUA-1 
outer layer cloak, wave front at 97 step.
 }\label{fig16}
%\end{minipage}
\end{figure}
\begin{figure}[h]
\centering
\includegraphics[width=0.85\linewidth,draft=false]{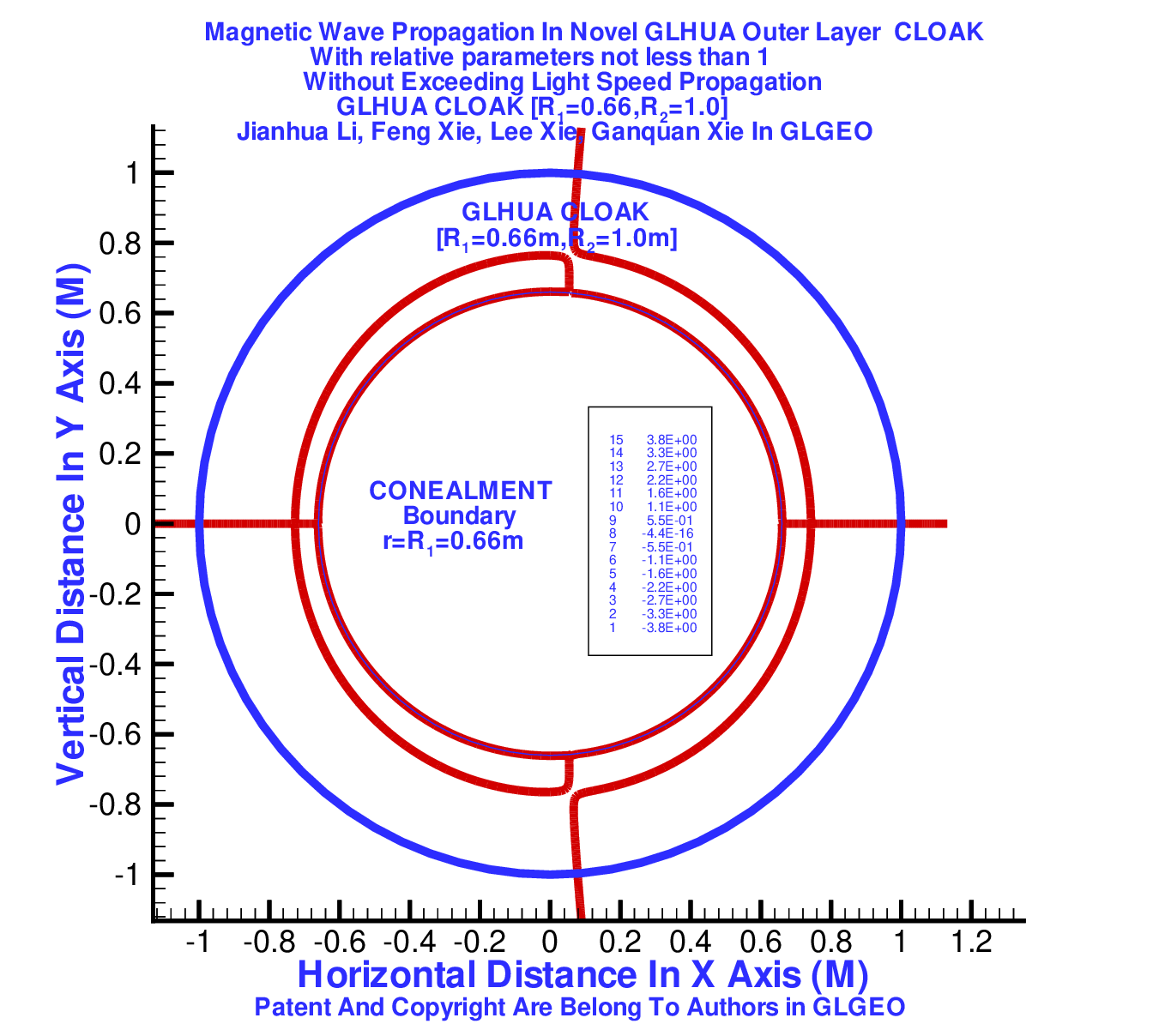}
\caption{ (color online) 
Magnetic wave propagation through GLHUA-1 outer layer cloak, wave front at 99 step. 
}\label{fig17}
%\end{minipage}
\end{figure}
\begin{figure}[h]
\centering
\includegraphics[width=0.86\linewidth,draft=false]{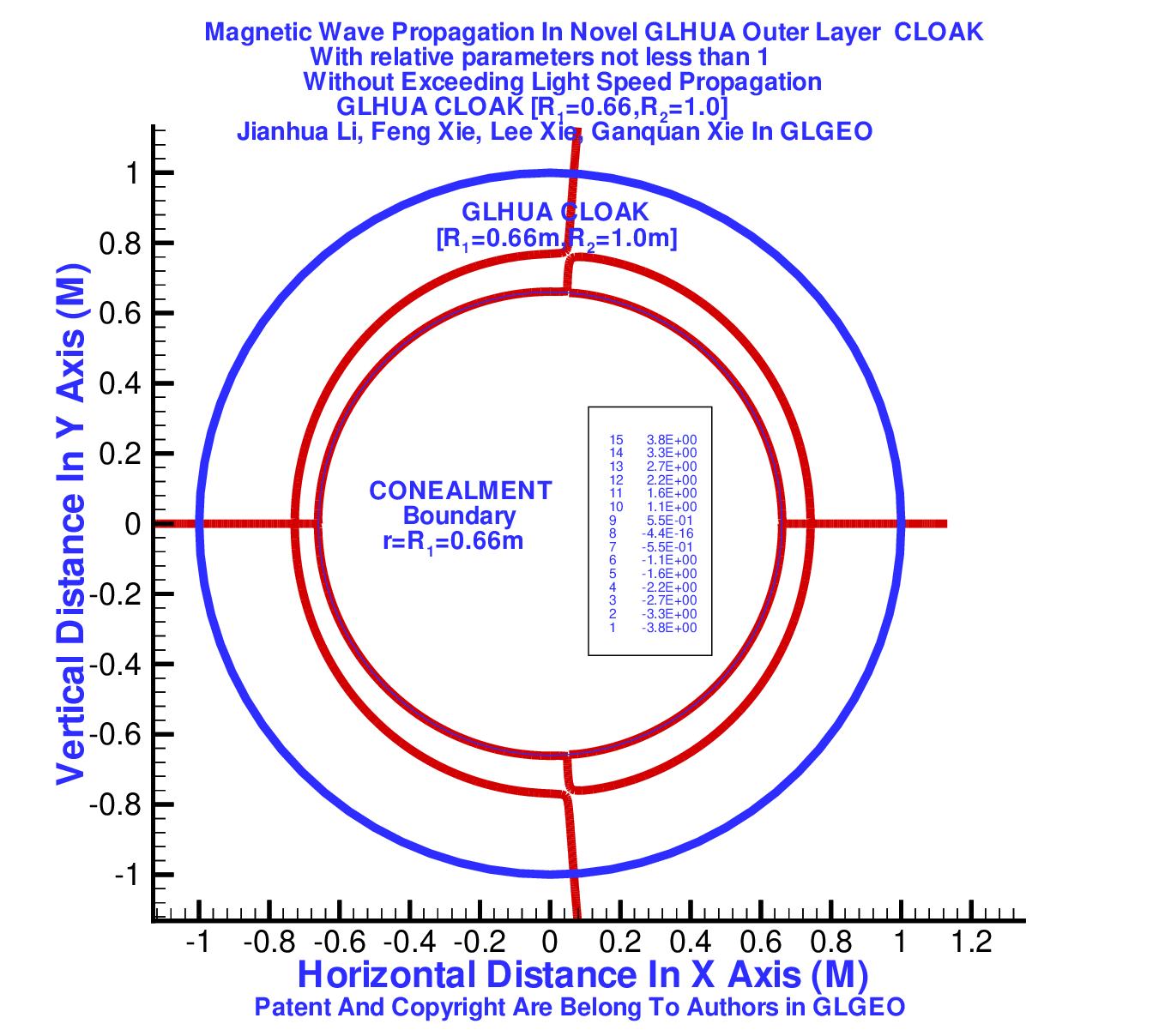}
\caption{ (color online) 
Magnetic wave propagation through GLHUA-1 outer layer cloak, wave front at 100 step.
 }\label{fig18}
%\end{minipage}
\end{figure}
\begin{figure}[h]
\centering
\includegraphics[width=0.86\linewidth,draft=false]{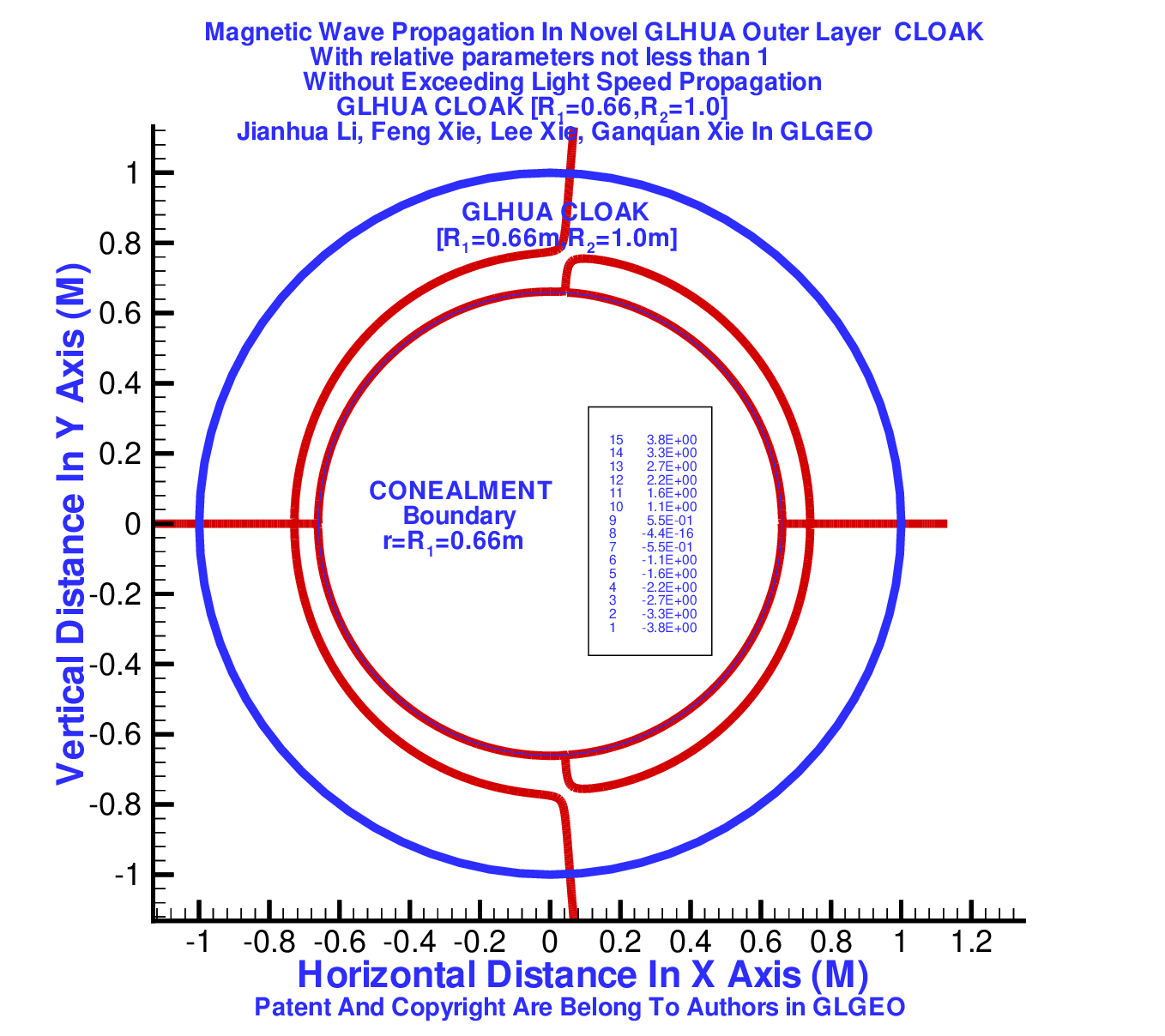}
\caption{ (color online) 
Magnetic wave propagation through GLHUA-1 outer layer cloak, wave front at 101 step.
 }\label{fig19}
%\end{minipage}
\end{figure}
\begin{figure}[h]
\centering
\includegraphics[width=0.86\linewidth,draft=false]{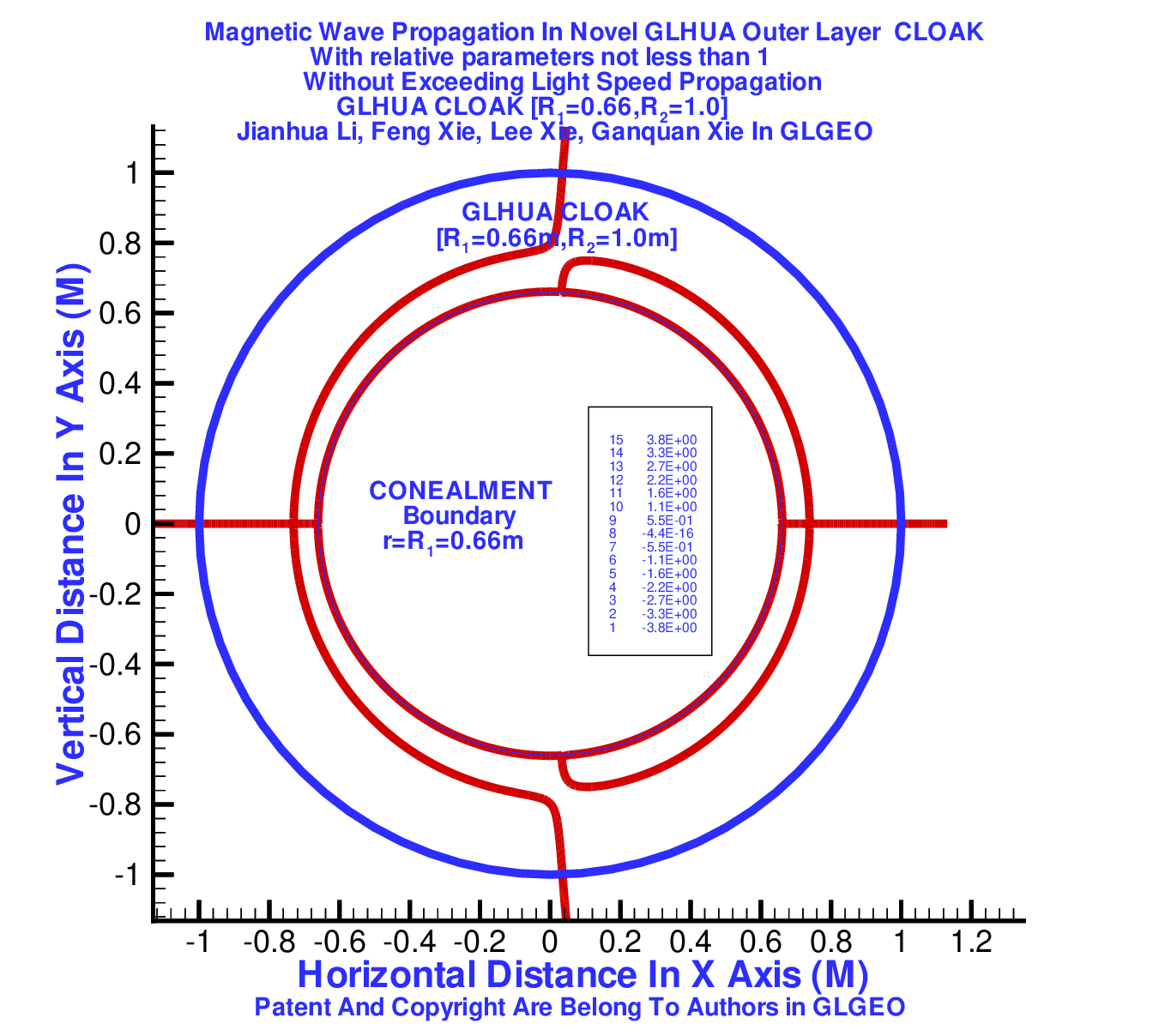}
\caption{ (color online) 
Magnetic wave propagation through GLHUA-1 outer layer cloak, wave front at 103 step..
}\label{fig31}
%\end{minipage}
\end{figure}
\begin{figure}[h]
\centering
\includegraphics[width=0.86\linewidth,draft=false]{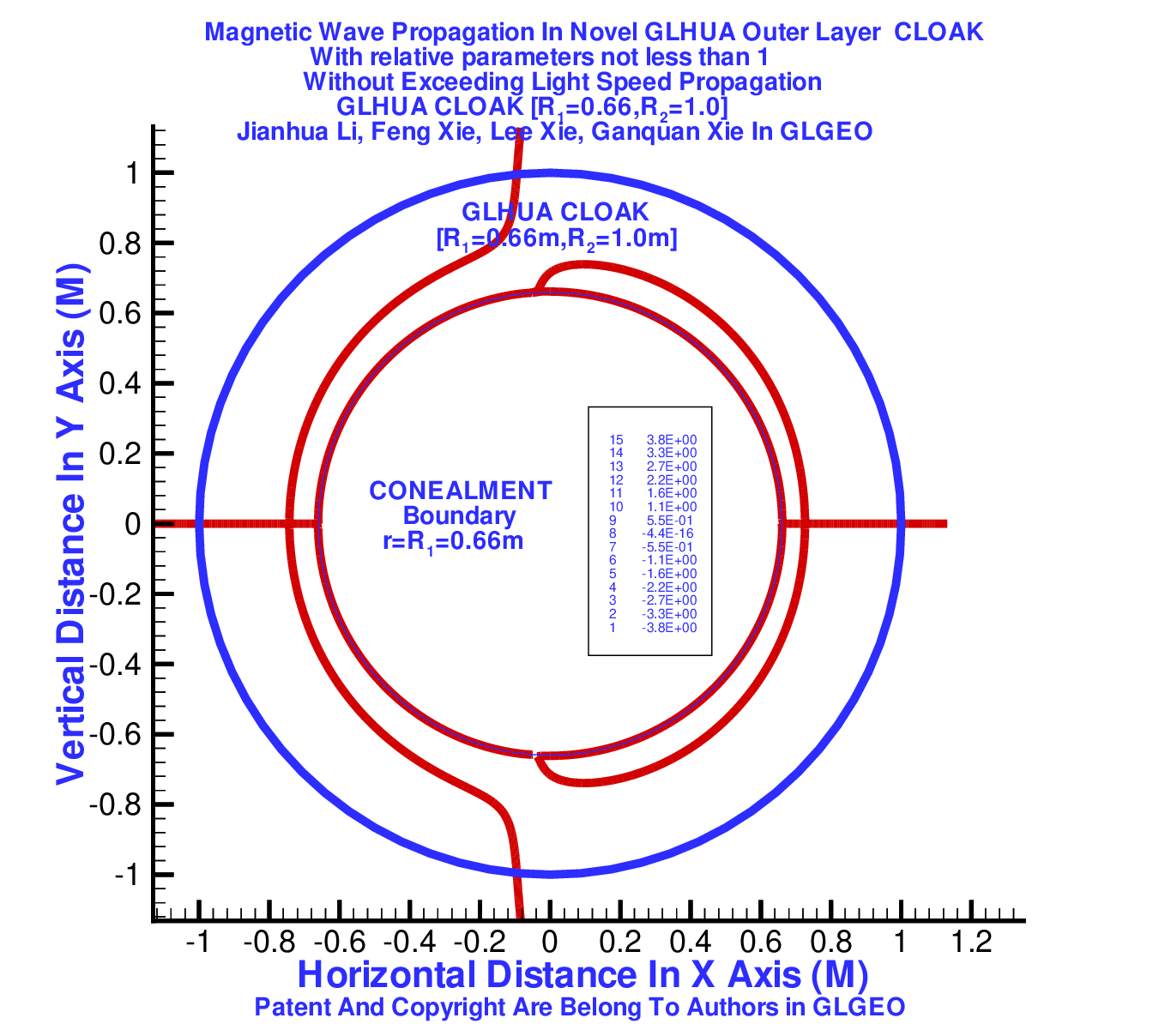}
\caption{ (color online) 
Magnetic wave propagation through GLHUA-1 outer layer cloak, wave front at 115 step.
}\label{fig31}
%\end{minipage}
\end{figure}

In the outside of the cloak,$r > R_2 $,$E_r (\vec r) = E_{b,r} (\vec r)$, $H_r (\vec r) = H_{b,r} (\vec r)$, on the boundary $r = R_2 $, 
$E_r (\vec r)|_{r = R_2 }  = E_{b,r} (\vec r)|_{r = R_2 } $,
$H_r (\vec r)|_{r = R_2 }  = H_{b,r} (\vec r)|_{r = R_2 } $
, 
  \begin{equation}
{\raise0.7ex\hbox{${\partial E_r (\vec r)}$} \!\mathord{\left/
 {\vphantom {{\partial E_r (\vec r)} {\partial r}}}\right.\kern-\nulldelimiterspace}
\!\lower0.7ex\hbox{${\partial r}$}}|_{r = R_2 }  = {\raise0.7ex\hbox{${\partial E_{b,r} (\vec r)}$} \!\mathord{\left/
 {\vphantom {{\partial E_{b,r} (\vec r)} {\partial r}}}\right.\kern-\nulldelimiterspace}
\!\lower0.7ex\hbox{${\partial r}$}}|_{r = R_2 } 
  \end{equation}
  \begin{equation}
{\raise0.7ex\hbox{${\partial H_r (\vec r)}$} \!\mathord{\left/
 {\vphantom {{\partial H_r (\vec r)} {\partial r}}}\right.\kern-\nulldelimiterspace}
\!\lower0.7ex\hbox{${\partial r}$}}|_{r = R_2 }  = {\raise0.7ex\hbox{${\partial H_{b,r} (\vec r)}$} \!\mathord{\left/
 {\vphantom {{\partial H_{b,r} (\vec r)} {\partial r}}}\right.\kern-\nulldelimiterspace}
\!\lower0.7ex\hbox{${\partial r}$}}|_{r = R_2 } ,
  \end{equation}
 The exact analytical radial EM wave solution show that there is no scattering from the GLHUA-1 outer layer cloak to disturb the incident EM wave in free space outside of the cloak.
%\end{document}
Proof: Our theorem 3 is proved by using similar proof in theorem 2.

\section {Simulation of the exact analytic magnetic wave propagation through the GLHUA-1 outer layer cloak}
\begin{figure}[h]
\centering
\includegraphics[width=0.86\linewidth,draft=false]{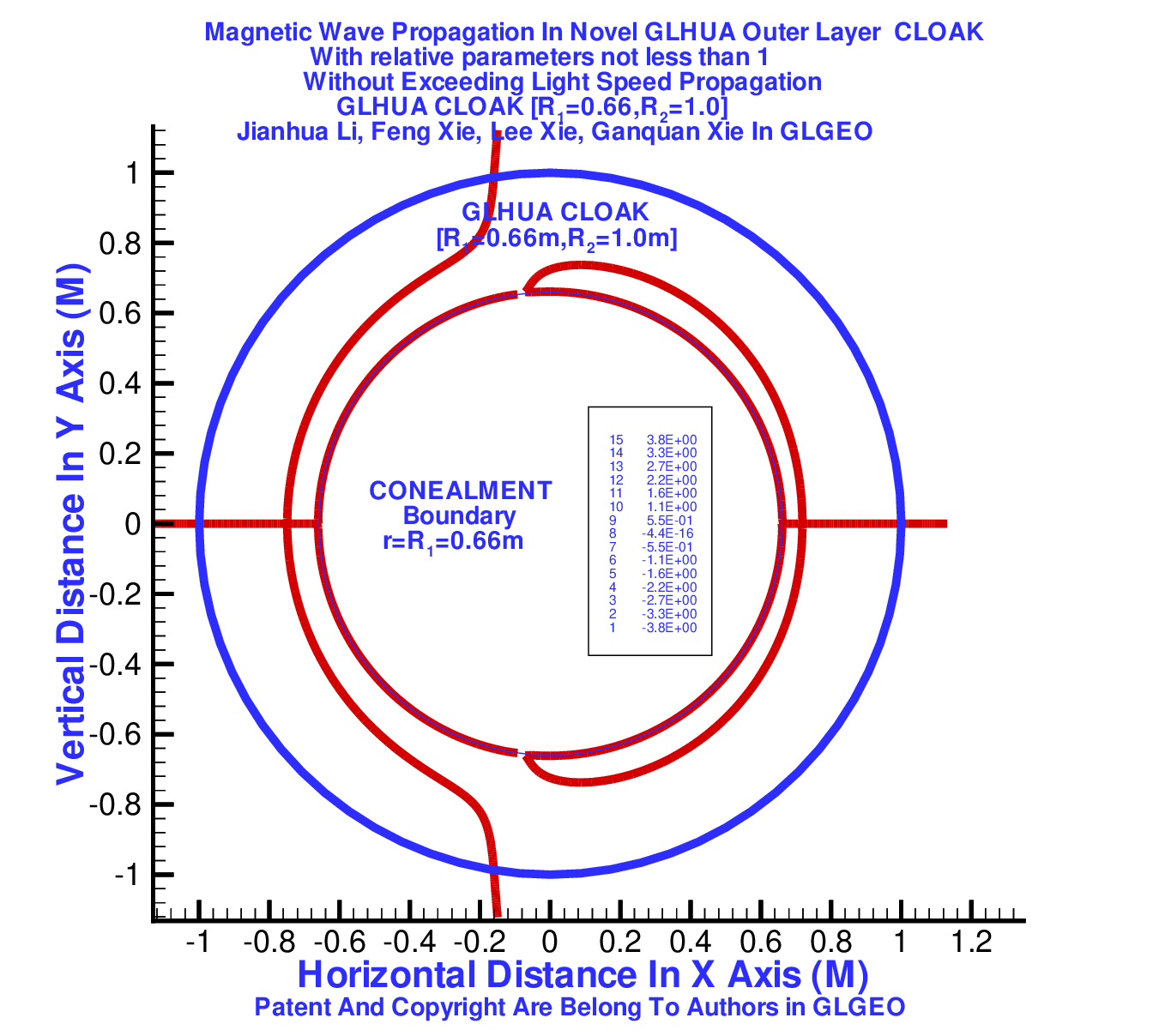}
\caption{ (color online) 
Magnetic wave propagation through GLHUA-1 outer layer cloak, wave front at 121 step.
}\label{fig31}
%\end{minipage}
\end{figure}
\begin{figure}[h]
\centering
\includegraphics[width=0.86\linewidth,draft=false]{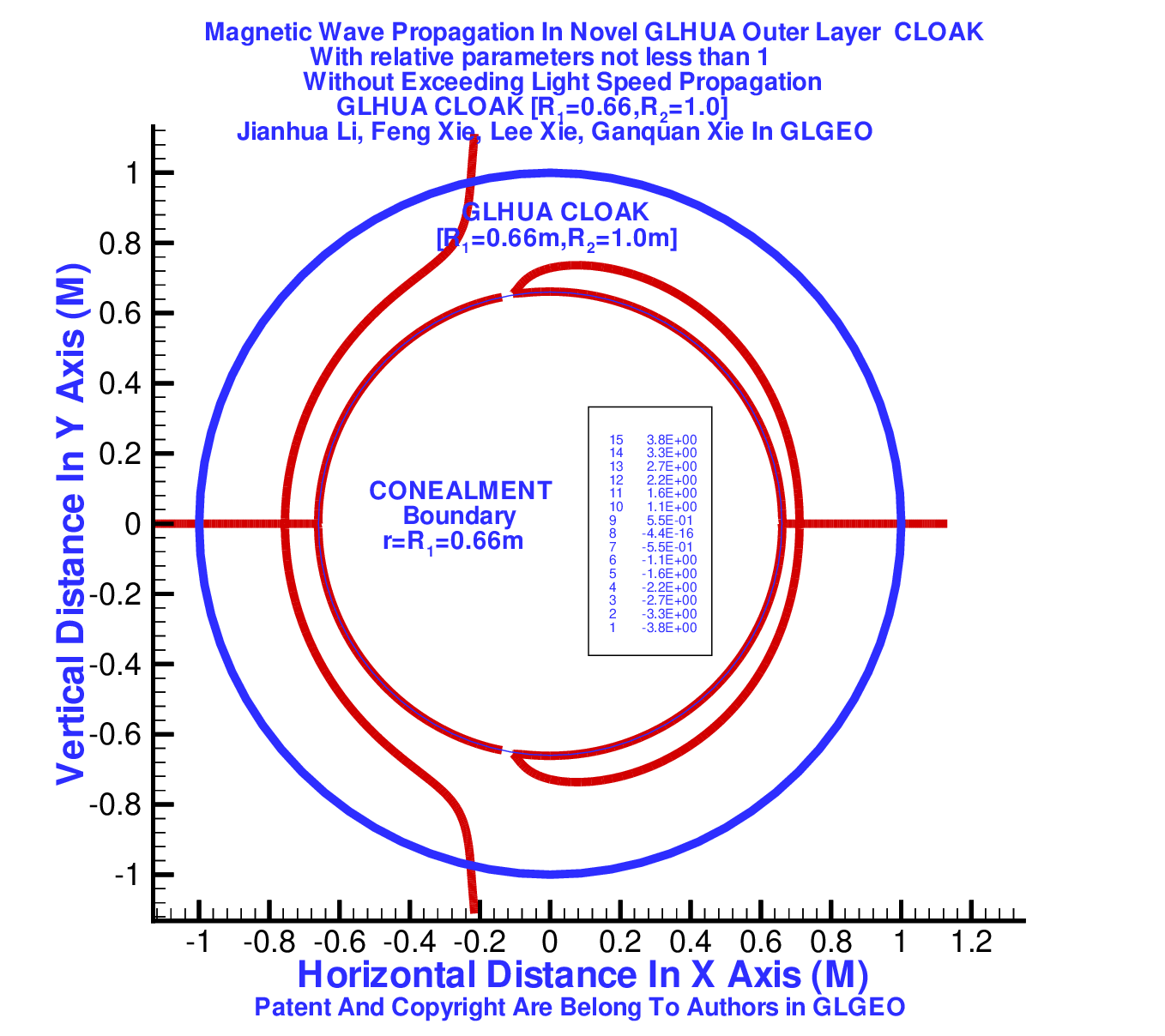}
\caption{ (color online) 
Magnetic wave propagation through GLHUA-1 outer layer cloak, wave front at 127 step.
}\label{fig31}
%\end{minipage}
\end{figure}
\begin{figure}[h]
\centering
\includegraphics[width=0.86\linewidth,draft=false]{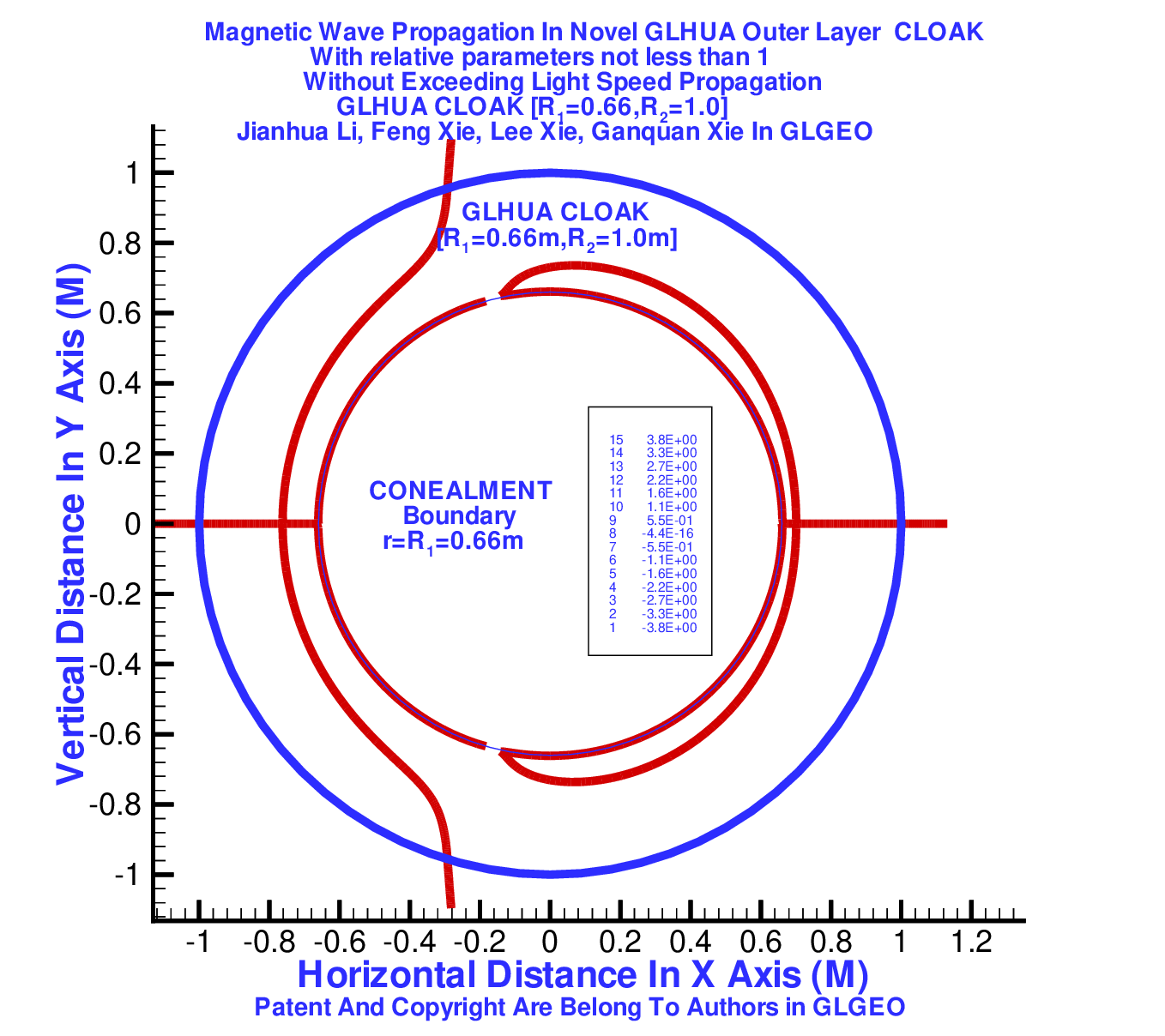}
\caption{ (color online) 
Magnetic wave propagation through GLHUA-1 outer layer cloak, wave front at 133 step.
}\label{fig31}
%\end{minipage}
\end{figure}
\begin{figure}[h]
\centering
\includegraphics[width=0.86\linewidth,draft=false]{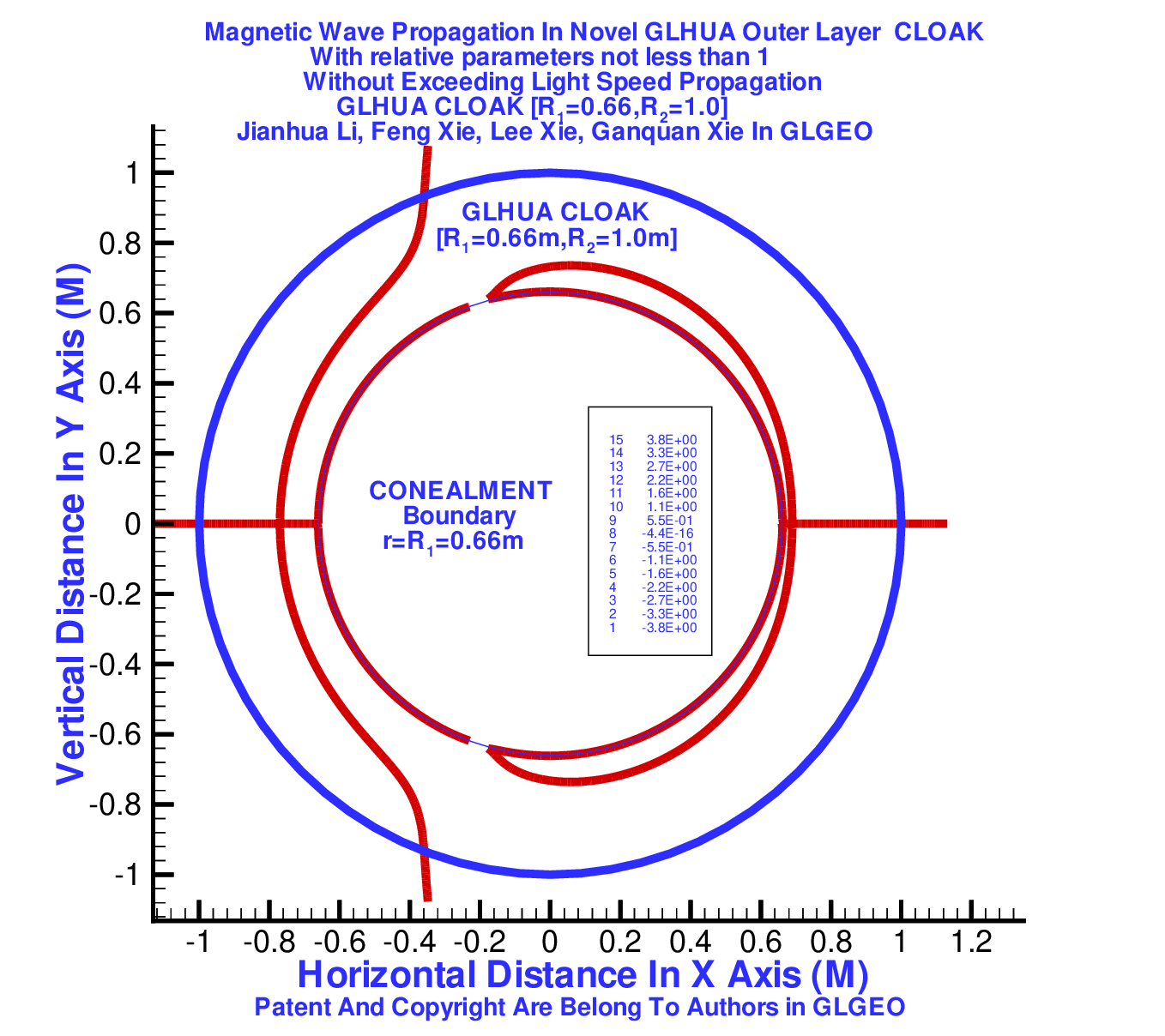}
\caption{ (color online) 
Magnetic wave propagation through GLHUA-1 outer layer cloak, wave front at 139 step.
}\label{fig31}
%\end{minipage}
\end{figure}
\begin{figure}[h]
\centering
\includegraphics[width=0.86\linewidth,draft=false]{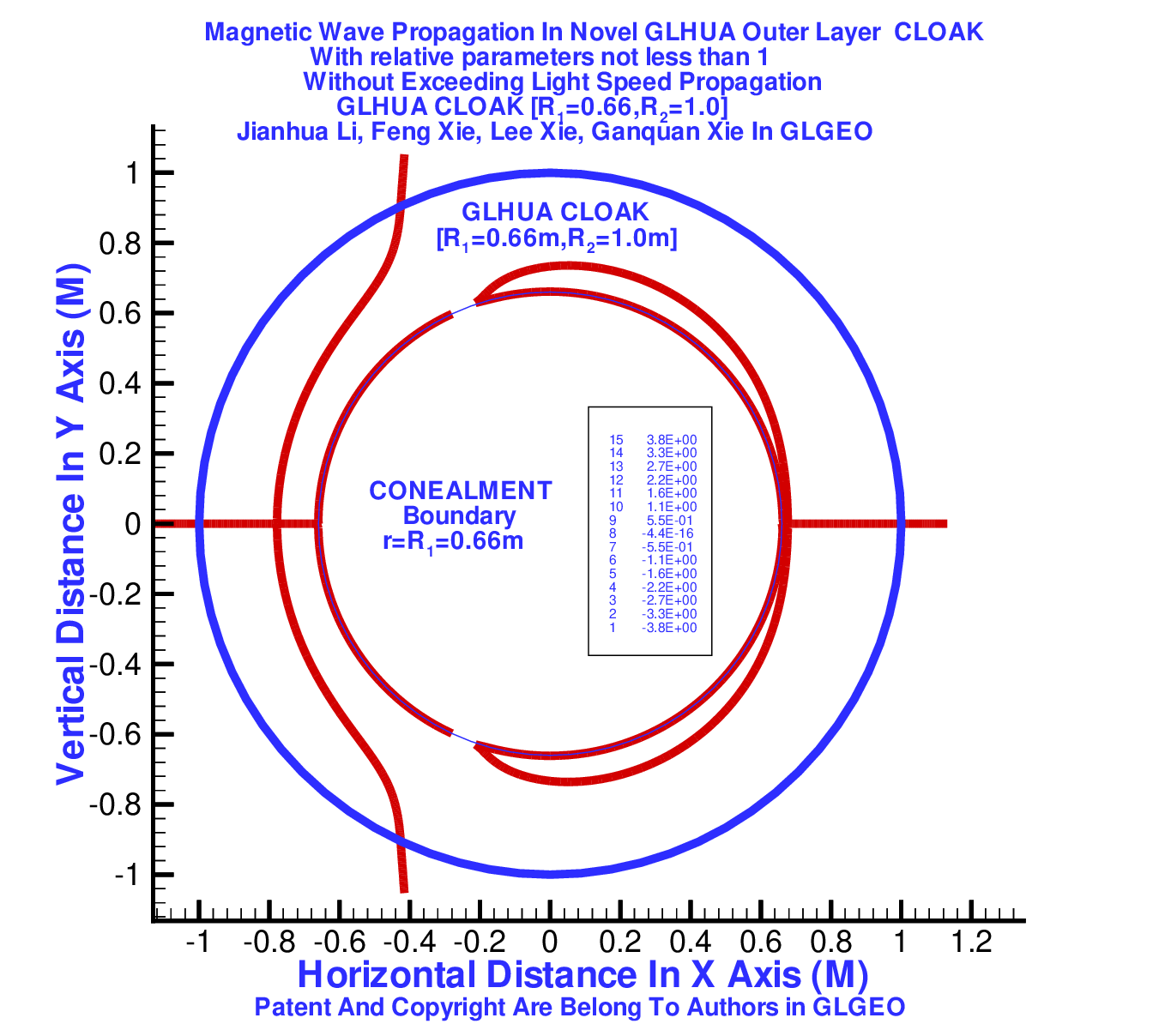}
\caption{ (color online) 
Magnetic wave propagation through GLHUA-1 outer layer cloak, wave front at 145 step.
}\label{fig31}
%\end{minipage}
\end{figure}
\begin{figure}[h]
\centering
\includegraphics[width=0.86\linewidth,draft=false]{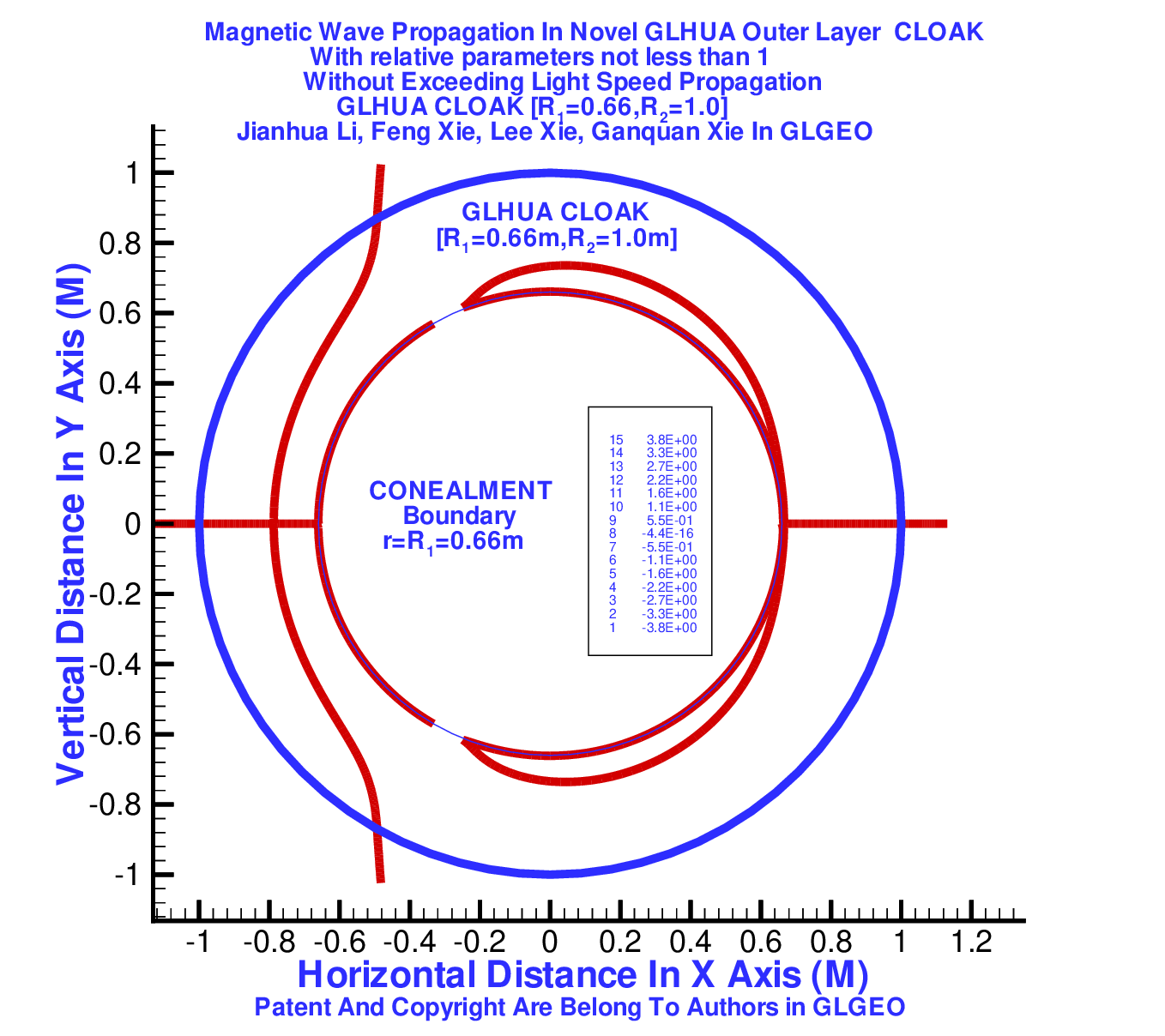}
\caption{ (color online) 
Magnetic wave propagation through GLHUA-1 outer layer cloak, wave front at 151 step.
}\label{fig31}
%\end{minipage}
\end{figure}
\begin{figure}[h]
\centering
\includegraphics[width=0.86\linewidth,draft=false]{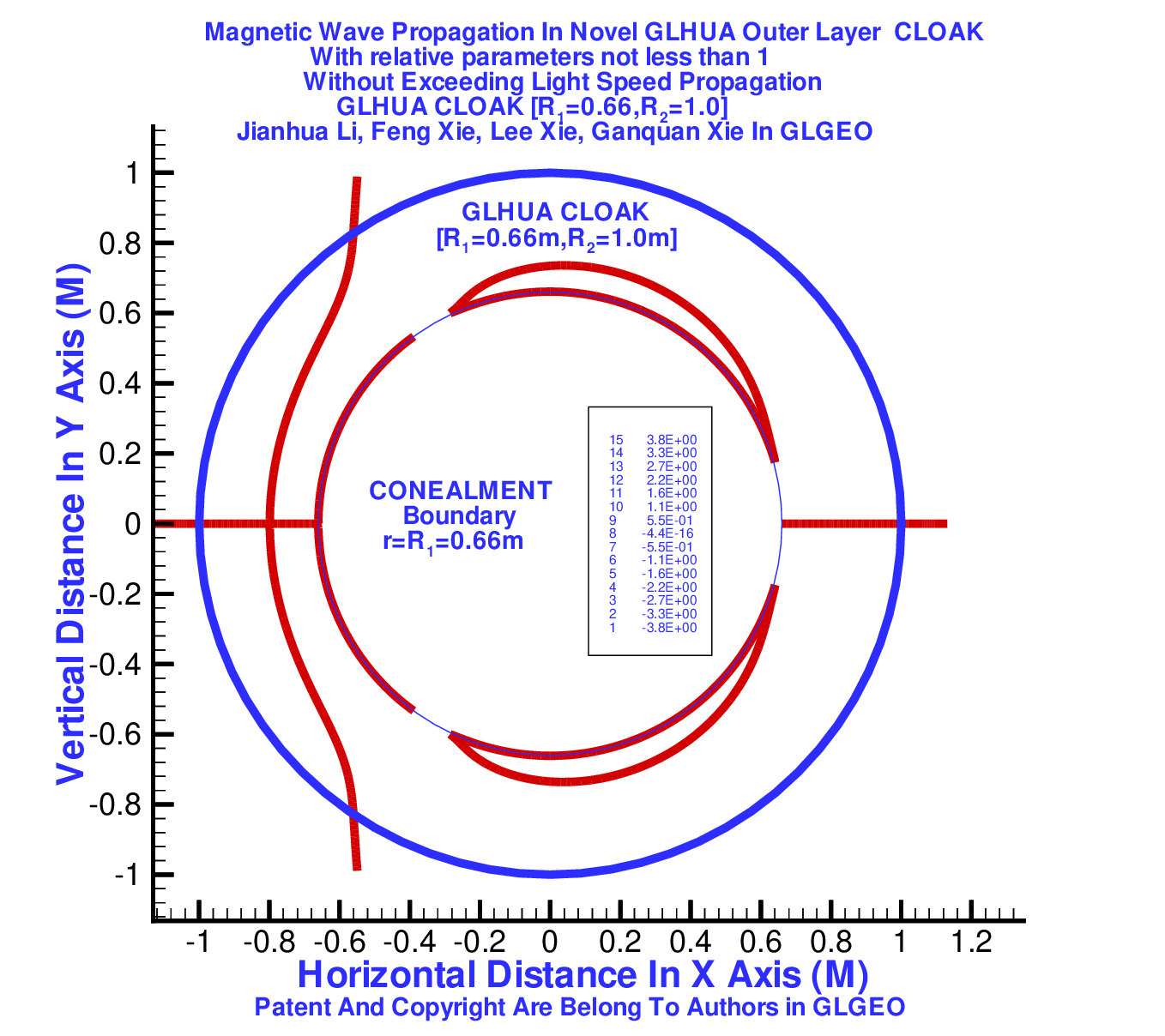}
\caption{ (color online) 
Magnetic wave propagation through GLHUA-1 outer layer cloak, wave front at 157 step.
}\label{fig25}
%\end{minipage}
\end{figure}

In GLHUA-1 outer cloak, $R_1  \le r \le R_2 $, the incident EM wave field is excited by the electric current point source located in $\vec r_s  = (13.0m,0.5\pi ,0)$, the electric current point source is
\begin{equation}
S(\vec r,\vec r_s ) = e^{ - i\omega t} \delta (\vec r - \vec r_s )\vec e_y ,
\end{equation}
where
\begin{equation}
\begin{array}{l}
 \vec e = \vec e_y  = (0,1,0) \\ 
  = \left( {\begin{array}{*{20}c}
   {\sin \theta \sin \phi } & {\cos \theta \sin \phi } & {\cos \phi }  \\
\end{array}} \right), \\ 
 \end{array}
\end{equation}
In the free space $r>R_2$, the angular frequency $\omega  = 2\pi f$, $f = 0.{\rm 1669555} \times {\rm 10}^{\rm 9} Hz$, 

  In GLHUA-1 outer cloak, $R_1  \le r \le R_2 $, $R_1  = 0.66m$,$R_2  = 1.0m$, the incident radial magnetic wave excited by the above electric current source is
                  \begin{equation}
H^b {}_r = \cos \phi \frac{1}{r}\frac{{\partial g}}{{\partial \theta }} - \frac{{\cos \theta \sin \phi }}{{r\sin \theta }}\frac{{\partial g}}{{\partial \phi }},
\end{equation}

The GLHUA-1 exact analytical EM wave propagation show that on the inner spherical surface  boundary $r=R_1$, the GLHUA-1 cloak material does generate a strange wave. We call the strange wave as ¡°GLHUA-1 created wave by cloak materials wave¡±, for simply, we call it "GL wave". In Figure 1 ( 90), the magnetic wave is being tangent to the outer boundary $r=R_2$ without any scattering. It is very strangle, there is a down crescent type GL wave in the upper part of the inner spherical surface, $r=R_1$ , and upward crescent type GL wave in the lower part of the same inner spherical surface, $r=R_1$. In the next step, the magnetic wave smoothly enter the outer annular layer $R_1  < r < R_2 ,$ is presented in Figure 2,( 112) The upper and lower two crescent type GL wave attached on the inner spherical surface, $r=R_1$ , maintain the shape and propagation from left to right; along the same inner
spherical surface, a red small part of the arc GL arc wave generated; The wave propagation phase image in figure 3( 124) is similar with the image in figure 2., the magnetic wave in the GLHUA-1 outer layer, the red arc and crescent GL wave on the inner spherical surface
are propagation to left, the arc and crescent GL wave phase shape is growing along on
the same inner spherical surface, $r=R_1$, and the arc GL wave is catching to crescent
GL wave. In figure 4 ( 136), the wave propagation phase image is similar with the image in figure 3, the phase image in figure 4 is propagated to left of image in figure 3, the upper and lower crescent GL wave along on the inner spherical surface, $r=R_1$ is connected
into double crescent GL wave. 
\begin{figure}[h]
\centering
\includegraphics[width=0.86\linewidth,draft=false]{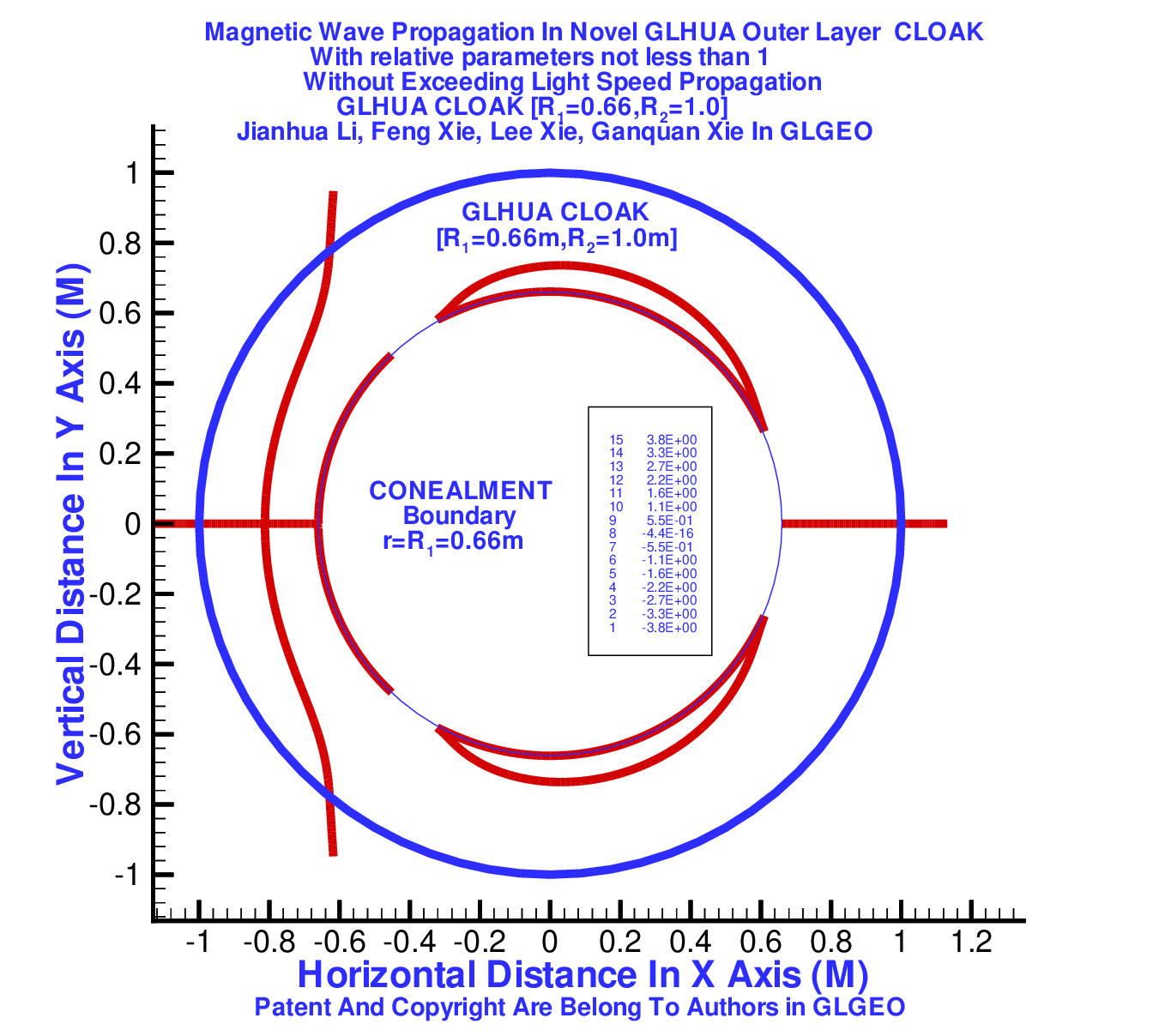}
\caption{ (color online) 
Magnetic wave propagation through GLHUA-1 outer layer cloak, wave front at 163 step.
}\label{fig26}
%\end{minipage}
\end{figure}
\begin{figure}[h]
\centering
\includegraphics[width=0.86\linewidth,draft=false]{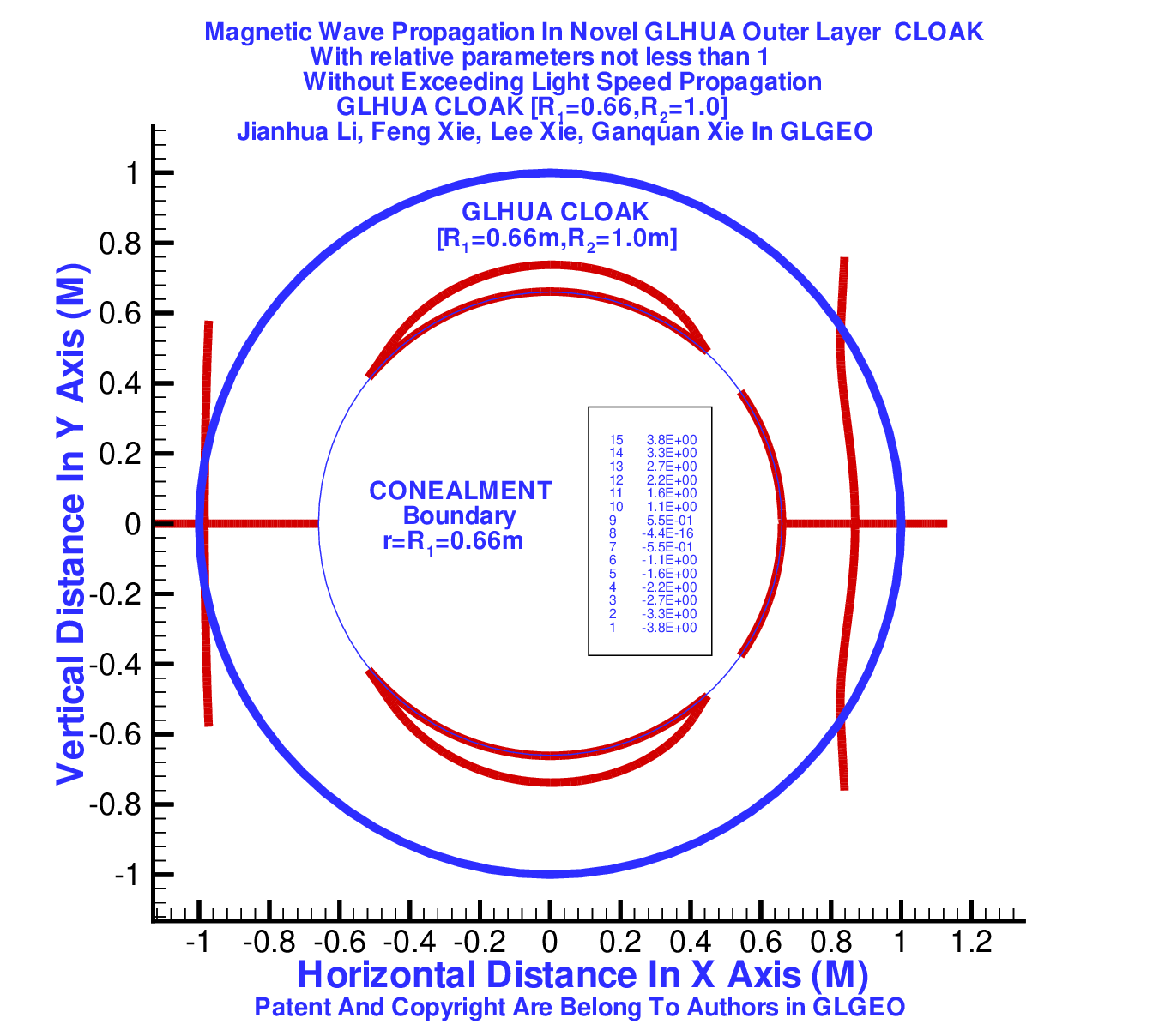}
\caption{ (color online) 
Magnetic wave propagation through GLHUA-1 outer layer cloak, wave front at 194 step.
}\label{fig27}
%\end{minipage}
\end{figure}
\begin{figure}[h]
\centering
\includegraphics[width=0.86\linewidth,draft=false]{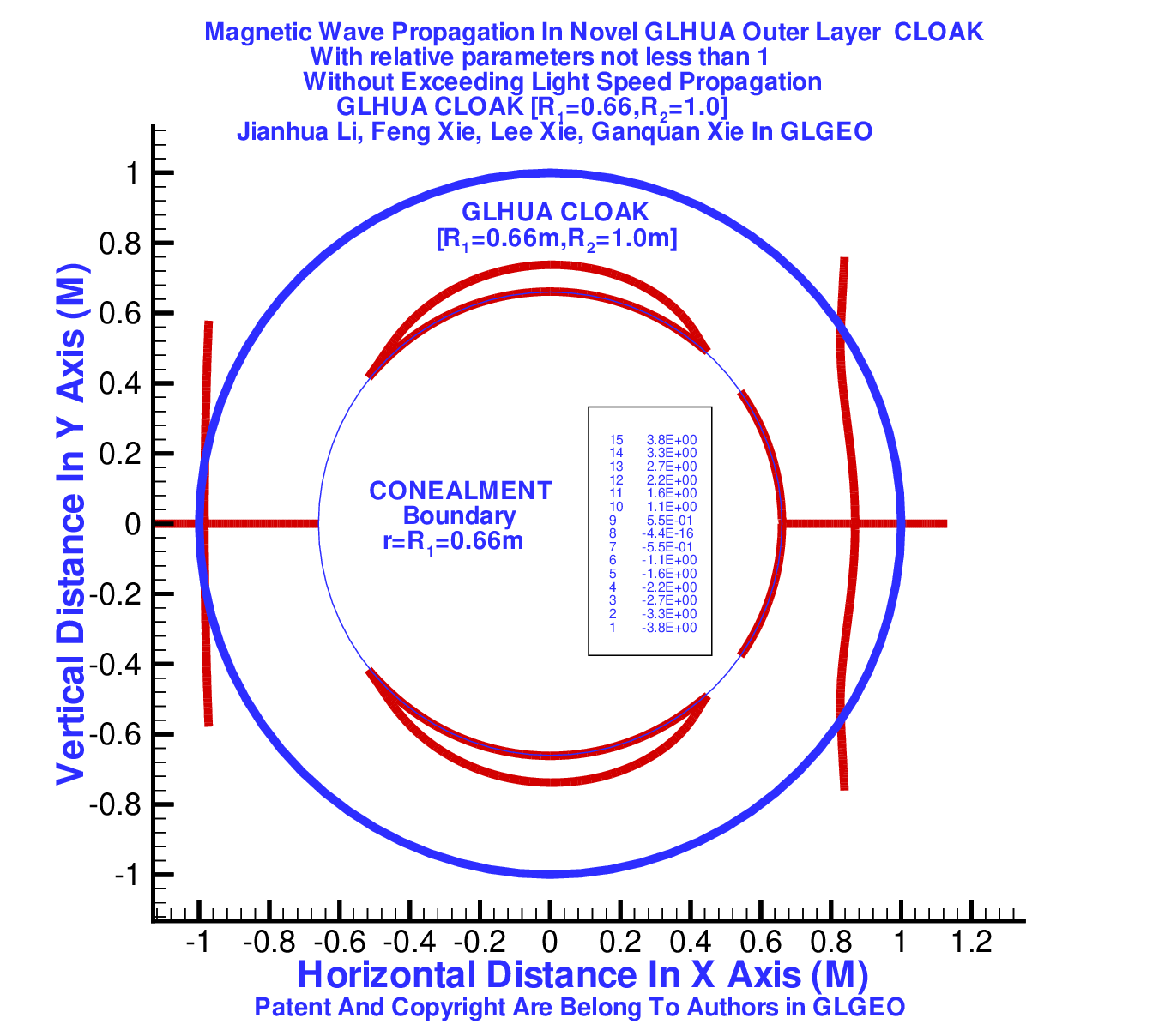}
\caption{ (color online) 
Magnetic wave propagation through GLHUA-1 outer layer cloak, wave front at 195 step.
}\label{fig28}
%\end{minipage}
\end{figure}

The phase image in figure 5 is propagated to left of image in figure 4. In the figure 5, the double crescent GL wave is growing up to a spherical surface GL wave, which is connected with the arc GL wave, the magnetic wave in GLHUA-1 outer layer is propagating to left of it in figure 4, follow and closed to the GL wave along on the inner spherical surface, $r=R_1$. In the figure 6( 180), in the left side of the inner spherical surface, the spherical surface GL wave is growing up to like a wave front on the left side of the inner spherical surface, in the right side of the inner spherical surface, the arc GL wave is growing as half spherical surface GL wave which phase is right part of the inner spherical surface. The right half spherical surface GL wave is connected with the left side spherical surface GL wave. the magnetic wave in GLHUA-1 outer layer is propagating to left of it in figure 5, follow and more closed to the GL wave along on the inner spherical surface, $r=R_1$. The phase image in figure 7( 181) is similar with the phase image in figure 6 and is propagated to left of image in figure 6. The magnetic wave propagation in figure 7 is important step. In the figure 7, the magnetic wave in right side of inner spherical surface meets the spherical surface GL wave in the left side of the inner spherical surface. In figure 8 ( 19), the magnetic wave use the left spherical GL wave as wave front for new propagation, the original wave front of the magnetic wave is disengaged and as new spherical surface GL wave attached in the right side of the inner spherical surface. The magnetic wave change wave front in the figure 7 shows that the magnetic wave front propagation in the GLHA outer layer cloak is discontinuous and the ray is also discontinuous. The phase image in figure 9( 21) is similar with the phase image in figure 8 and is propagated to left of image in figure 8. In the next magnetic wave propagation step in the figure 9, 10, 11,¡­, let P1 be cross point of the wave front and equatorial and P2 be cross point of the outer spherical surface, $r=R_2$ and equatorial. 
 
Let Q1 be cross point of the wave front and outer spherical surface $r=R_2$ and Q2 be the cross point of line $z = z_{q_1 }$ of Q1 and line $x = x_{p_2 } $ of P2. Let d1 be distance between P1 and P2 , d2 be the distance between Q1 and Q2 . In the figure 9, 10, 11,¡­,the magnetic wave is propagating forward to left, on the magnetic wave front, d1 is minimal and phase velocity at point P1 on the wave front is also minimal; . The d2 is maximum and phase velocity at point Q1 on the wave front is also maximum; the d1 < d2, the difference of d2- d1 in monotone decreasing. In the figure 17 , d1=d2=0. the every point on the magnetic wave front are propagating out of the GLHUA-1 cloak simultaneously, the magnetic wave is recovered to the incident wave in free space at same time. In the figure 18, the magnetic wave front is recovered incident wave front in free space that show GLHUA-1 outer layer cloak is invisible. In the figure 9, 10, 11,¡­, along inner spherical surface, $r=R_1$, the upper GL wave is propagating counterclockwise and the lower GL wave is propagating clockwise with some discontinuous.  In figure 9 ( 21), the right GL wave attached on the right inner spherical surface is deformed. In figure 10 (127), the right GL wave attached on the right inner spherical surface is more deformed to double crescent, and left arc GL wave is disengaged from the right GL wave. In figure 11(145), the right GL wave attached on the right inner spherical surface is more deformed to double crescent, and left arc GL wave is disengaged from the right GL wave and is more discontinuous. In figure 12 (151), the right GL wave attached on the right inner spherical surface is discontinuously split to the upper crescent GL wave and lower crescent GL wave. In figure 13 (163). along inner spherical surface, $r=R_1$, the upper crescent GL wave and arc GL wave is propagating counterclockwise and the lower crescent GL wave and arc GL wave is propagating clockwise with some discontinuous. The phase image of the magnetic wave front and GL wave in figure 14 (175) is similar with the phase image in the figure 13, moreover, the left arc GL wave is shrink to very short. The new magnetic wave front is appear in right side in the new period.  The distance between two magnetic wave front is wave length  .

The phase Image of the magnetic wave front and GL wave along the inner spherical surface, $r=R_1$ in figure 15(176) and figure 16 (186) is same as phase image in figure 1 and figure 2, respectively, Summary, the incident magnetic wave in free space never be disturbed by GLHUA-1 outer cloak, GLHUA-1 cloak is invisible; The magnetic wave can not propagating penetrated into inside of the sphere $r < R_1$.The phase velocity of magnetic wave propagation is less than light speed. The exact analytic GLHUA-1 wave propagation simulation shows that the GL wave along the inner spherical surface $r=R_1$ is generated by the GLHUA-1 outer cloak material. The GL wave is a bridge for the EM wave propagation through the GLHUA-1 outer cloak without exceeding light speed. The GL wave is a cloaking making that the EM wave propagation can not penetrate into the concealment.  Summary, the incident magnetic wave in free space never be disturbed by GLHUA-1 outer cloak, GLHUA-1 cloak is invisible; The magnetic wave can not propagating penetrated into inside of the sphere $r < R_1$.The phase velocity of magnetic wave propagation is less than light speed. The exact analytic GLHUA-1 wave propagation simulation shows that the GL wave along the inner spherical surface $r=R_1$ is generated by the GLHUA-1 outer cloak material. The GL wave is a bridge for the EM wave propagation through the GLHUA-1 outer cloak without exceeding light speed. The GL wave is a cloaking making that the EM wave propagation can not penetrate into the concealment.
\section{Discussion and Conclusion ON EXACT ANALYTIC EM WAVE IN GLHUA-2 CLOAK
}Substitute the inverse transformation to incident analytical EM wave, we can obtain the analytical EM wave propagation in the $0$ to $R_1$ spherical radial transformation cloak. It is totally different from transformation optic,  without using transformation, our GLHUA-1 analytical expansion can also be directly used to find analytical EM wave
propagation in the $0$ to $R_1$ spherical radial transformation cloak. For example, we propose 
GLHUA-1 analytical expansion method to find analytical EM wave propagation in Pendry cloak.

\subsection{GLHUA-1 analytical expansion method to find analytical EM wave in Pendry Cloak}

Substitute relative EM parameters in Pendry cloak
\begin{equation}
\varepsilon _r  = \mu _r  = \frac{{R_2 }}{{R_2  - R_1 }}\frac{{(r - R_1 ){}^2}}{{r^2 }},
\end{equation}

\begin{equation}
\varepsilon _\theta   = \mu _\theta   = \varepsilon _\phi   = \mu _\phi   = \frac{{R_2 }}{{R_2  - R_1 }},
\end{equation}        
into the equation ( 98), the ( 98) become to
\begin{equation}
\begin{array}{l}
 \frac{\partial }{{\partial r}}(\frac{{R_2  - R_1 }}{{R_2 }}\frac{{\partial H_l (r)}}{{\partial r}}) \\ 
  - \frac{{R_2  - R_1 }}{{R_2 }}\frac{{l(l + 1)}}{{(r - R_1 )^2 }}H_l (r) \\ 
  + k^2 \frac{{R_2 }}{{R_2  - R_1 }}H_l (r) = 0, \\ 
 \end{array}
\end{equation}
We create GLHUA-1 analytical expansion of radial magnetic wave $l$ component $H_{l,p}$,
$p=1,2$,  

\begin{equation}
\begin{array}{l}
 H_{l,p}  = H_{l,1,p} (r)\cos \left( {k\frac{{R_2 (r - R_1 }}{{R_2  - R_1 }}} \right) \\ 
  + H_{l,2,p} (r)\sin \left( {k\frac{{R_2 (r - R_1 }}{{R_2  - R_1 }}} \right), \\ 
 \end{array}
\end{equation}
Substitute ( 144) into ( 143), we have novel GLHUA-1 dual two expansion equations,$p=1,2$,

\begin{equation}
\begin{array}{l}
 \frac{\partial }{{\partial r}}(\frac{{R_2  - R_1 }}{{R_2 }}\frac{{\partial H_{l,1,p} (r)}}{{\partial r}}) + 2k\frac{{\partial H_{l,2,p} (r)}}{{\partial r}} \\ 
  = \frac{{R_2  - R_1 }}{{R_2 }}l(l + 1)\frac{1}{{(r - R_1 )^2 }}H_{l,1,p} (r), \\ 
 \end{array}
\end{equation}

\begin{equation}
\begin{array}{l}
 \frac{\partial }{{\partial r}}(\frac{{R_2  - R_1 }}{{R_2 }}\frac{{\partial H_{l,2,p} (r)}}{{\partial r}}) - 2k\frac{{\partial H_{l,1,p} (r)}}{{\partial r}} \\ 
  = \frac{{R_2  - R_1 }}{{R_2 }}l(l + 1)\frac{1}{{(r - R_1 )^2 }}H_{l,2,p} (r), \\ 
 \end{array}
\end{equation}
%\subsection {GLHUA-1 analytical expansion method to find analytical EM wave in %Pendry Cloak}

We propose negative power series expansion 
  \begin{equation}
 H_{l,1,p} (r) = \sum\limits_{j = 0}^\infty  {a_{l,j,p} } (r - R_1 )^{ - j} ,
\end{equation}
\begin{equation}
H_{l,2,p} (r) = \sum\limits_{j = 0}^\infty  {b_{l,j,p} } (r - R_1 )^{ - j} ,
\end{equation}
where $p=1,2$, when $p=1$ , $a_{l,0,p}=1$, $b_{l,0,p}=0$, when $p=2$ , $a_{l,0,p}=0$, $b_{l,0,p}=1$.  Substitute ( 147) and ( 148) into GLHUA-1 dual equations ( 145) and ( 146), we obtain

\begin{equation}
\begin{array}{l}
 H_{l,1}  = H_{l,1,1} (r)\cos \left( {k\frac{{R_2 (r - R_1 )}}{{R_2  - R_1 }}} \right) \\ 
  + H_{l,2,1} (r)\sin \left( {k\frac{{R_2 (r - R_1 )}}{{R_2  - R_1 }}} \right) \\ 
  = \frac{{R_2 (r - R_1 )}}{{R_2  - R_1 }}j_l \left( {k\frac{{R_2 (r - R_1 )}}{{R_2  - R_1 }}} \right), \\ 
 \end{array}
\end{equation}
\begin{equation}
\begin{array}{l}
 H_{l,2}  = H_{l,1,2} (r)\cos \left( {k\frac{{R_2 (r - R_1 )}}{{R_2  - R_1 }}} \right) \\ 
  + H_{l,2,2} (r)\sin \left( {k\frac{{R_2 (r - R_1 )}}{{R_2  - R_1 }}} \right) \\ 
  = \frac{{R_2 (r - R_1 )}}{{R_2  - R_1 }}n_l \left( {k\frac{{R_2 (r - R_1 )}}{{R_2  - R_1 }}} \right), \\ 
 \end{array}
\end{equation}

             For arbitrary constants $A$ and $B$,
\begin{equation}
\begin{array}{l}
 H_l (r) = AH_{l,1} (r) + BH_{l,2} (r) \\ 
  = A\frac{{R_2 (r - R_1 )}}{{R_2  - R_1 }}j_l \left( {k\frac{{R_2 (r - R_1 )}}{{R_2  - R_1 }}} \right) \\ 
  + B\frac{{R_2 (r - R_1 )}}{{R_2  - R_1 }}n_l \left( {k\frac{{R_2 (r - R_1 )}}{{R_2  - R_1 }}} \right). \\ 
 \end{array}
\end{equation}
From the continuous boundary condition on $r=R_2$, ( 99) and ( 100) in this paper, we obtained
$B=0$, 
\begin{equation}
\begin{array}{l}
H_l (r) =  \\
- ik\frac{{R_2 (r - R_1 )}}{{R_2  - R_1 }}j_l \left( {k\frac{{R_2 (r - R_1 )}}{{R_2  - R_1 }}} \right){h^(1)}_l (kr_s ),
\end{array}
\end{equation}
\hfill\break \\
${\boldsymbol{Theorem \ 4:}}$ \ In the spherical annular layer $R_1  \le r \le R_2 $, the EM relative parameters is
presented by ( 141) and ( 142), the electric point delta source is denoted by ( 88) in this paper, then analytical
radial GL magnetic wave propagation in Pendry cloak is 
\begin{equation}
\begin{array}{l}
 H(\vec r) =  \\ 
  =  - ik\sum\limits_{l = 1}^\infty  {\frac{{R_2 (r - R_1 )}}{{R_2  - R}}} j_l \left( {k\frac{{R_2 (r - R_1 )}}{{(R_2  - R_1 )}}} \right)h^{(1)} _l (kr_s ) \\ 
 \sum\limits_{m =  - l}^l {} D_h (\theta ,\phi )Y_l^m (\theta ,\phi )Y_l^{m*} (\theta _s ,\phi _s ) \\ 
  = \sum\limits_{l = 1}^\infty  {} H^{(b)} _l \left( {k\frac{{R_2 (r - R_1 )}}{{(R_2  - R_1 )}}} \right) \\ 
 \sum\limits_{m =  - l}^l {} D_h (\theta ,\phi )Y_l^m (\theta ,\phi )Y_l^{m*} (\theta _s ,\phi _s ) \\ 
 \end{array}
\end{equation}
\begin{equation}
\begin{array}{l}
 H_r (\vec r) =  \\ 
  =  - ik\sum\limits_{l = 1}^\infty  {\frac{1}{{(r - R)}}} j_l \left( {k\frac{{R_2 (r - R_1 )}}{{(R_2  - R_1 )}}} \right)h^{(1)} _l (kr_s ) \\ 
 \sum\limits_{m =  - l}^l {} D_h (\theta ,\phi )Y_l^m (\theta ,\phi )Y_l^{m*} (\theta _s ,\phi _s ) \\ 
  = \frac{{R_2  - R_1 }}{{R_2 }}\frac{1}{{(r - R_1 )^2 }}\sum\limits_{l = 1}^\infty  {} H^{(b)} _l \left( {k\frac{{R_2 (r - R_1 )}}{{(R_2  - R_1 )}}} \right) \\ 
 \sum\limits_{m =  - l}^l {} D_h (\theta ,\phi )Y_l^m (\theta ,\phi )Y_l^{m*} (\theta _s ,\phi _s ). \\ 
 \end{array} 
\end{equation}      
\hfill\break \\

\subsection { Comparison GLHUA-1, GLHUA-2,GLHUA-3  cloak and Pendry Cloak}
In GLHUA-1 invisible cloak ( 82) and ( 84),GLHUA-2 invisible cloak(7),(8),GLHUA-3 invisible cloak,(14) and (15),  the refractive index  $n = \sqrt {\varepsilon _\theta  \mu _r } $   is large than 1 and going to infinity when $r$ going to $R_1$. The phase velocity of EM wave propagation in GLHUA-1, GLHUA-2,GLHUA-3   cloak are less than light speed and going to zero when $r$ going to $R_1$. GLHUA-1, GLHUA-2,GLHUA-3   invisible cloak are practicable. In Pendry invisible cloak ( 141) and ( 142) , the refractive index  $n = \sqrt {\varepsilon _\theta  \mu _r } $   is less than 1 and going to zero when $r$ going to $R_1$. The phase velocity of EM wave propagation in Pendry cloak is exceeding light speed and going to infinity when $r$ going to $R_1$. Pendry invisible cloak is very difficulty and can not be practicable.The analytical Magnetic wave in GLHUA-1 cloak ( 135) in theorem 3 and analytical Magnetic  wave in Pendry cloak ( 154) in theorem 4 are 
totally different. Any annular layer Pendry cloaking by Pendry transformation will make the phase velocity exceeding light speed and tends to infinity in $r=R_1$.  Our GLHUA-1, GLHUA-2 and GLHUA-3 invisible cloak without exceeding light speed and  without  infinity, and overcome the fundamental difficuties of Pendry cloak.
Our GLHUA-1 analytical expansion method can be used to find analytical EM
wave  propagation in the GLHUA-1, GLHUA-2 and GLHUA-3 invisible cloak.
Also  analytical EM wave  propagation in the GLHUA-2 and GLHUA-3 invisible cloak
can be found by the GLHUANP-2 and GLHUANP-3 transformation respectively.

\subsection {Analytical EM wave in GLHUA-1,GLHUA-2 and GLHUA-3 cloak are undisputed evidence to prove that GLHUA-1,GLHUA-2 and GLHUA-3 cloak are practicable invisible cloak and super scirnces is being born}
Our analytical EM wave in GLHUA-1,GLHUA-2 and GLHUA-3  cloak are undisputed evidence and rigorous proof to prove that the GLHUA-1,GLHUA-2 and GLHUA-3  double layer cloak are practicable invisible cloak without exceed light speed propagation. Our analytical EM wave in GLHUA-1,GLHUA-2 and GLHUA-3  cloak is undisputed evidence and rigorous proof to prove GLHUA-1,GLHUA-2 and GLHUA-3  double layer invisible cloak and their theoretical proof in  are right. 

GL simulation shows that there are novel GLHUA-1 created wave by cloak materials bridge which is generated by the
GLHUA-1 outer layer invisible cloak materials in ( 82) and ( 84). The GLHUA-1 created wave by cloak materials bridge make the GLHUA-1 analytical EM wave propagation without exceeding light speed. The GLHUA-1 created wave by cloak materials bridge  is obvious for $R_1=0.5R_2$ or $R_1=0.66R_2$,
for $R_> 0.8R_2$, the GLHUA-1 created wave by cloak materials bridge wave become to the arc  GLHUA-1 created wave by cloak materials wave on the inner spherical layer $r=R_1$. 

\begin{figure}[h]
\centering
\includegraphics[width=0.86\linewidth,draft=false]{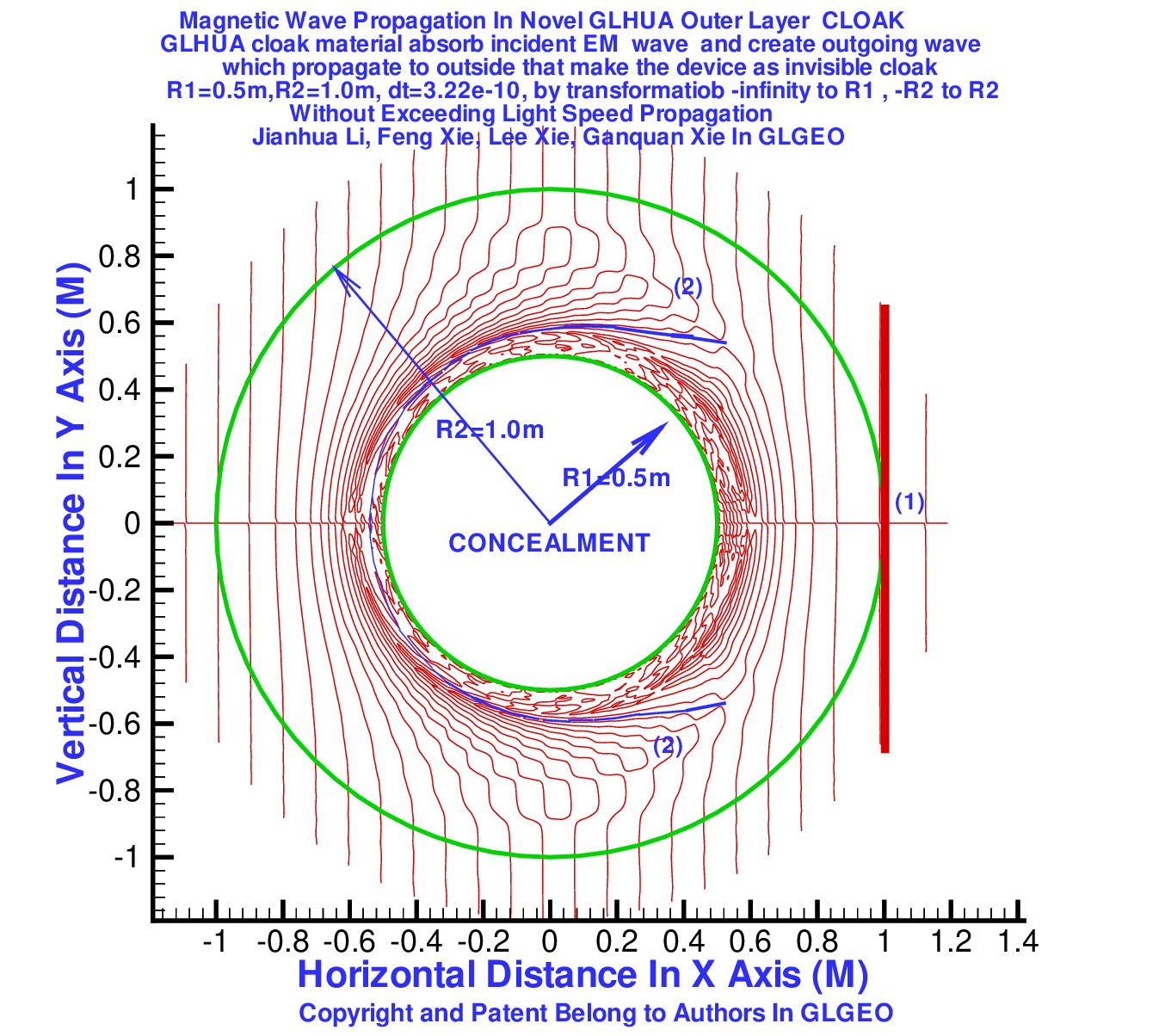}
\caption{ (color online)  At first relative time step, the incident magneic wave front (red line (1)) coming and tangent to the sphere surface  , the novel curve wave front (blue  line (2)) is created by material.}\label{fig29}
%\end{minipage}
\end{figure}
\begin{figure}[h]
\centering
\includegraphics[width=0.86\linewidth,draft=false]{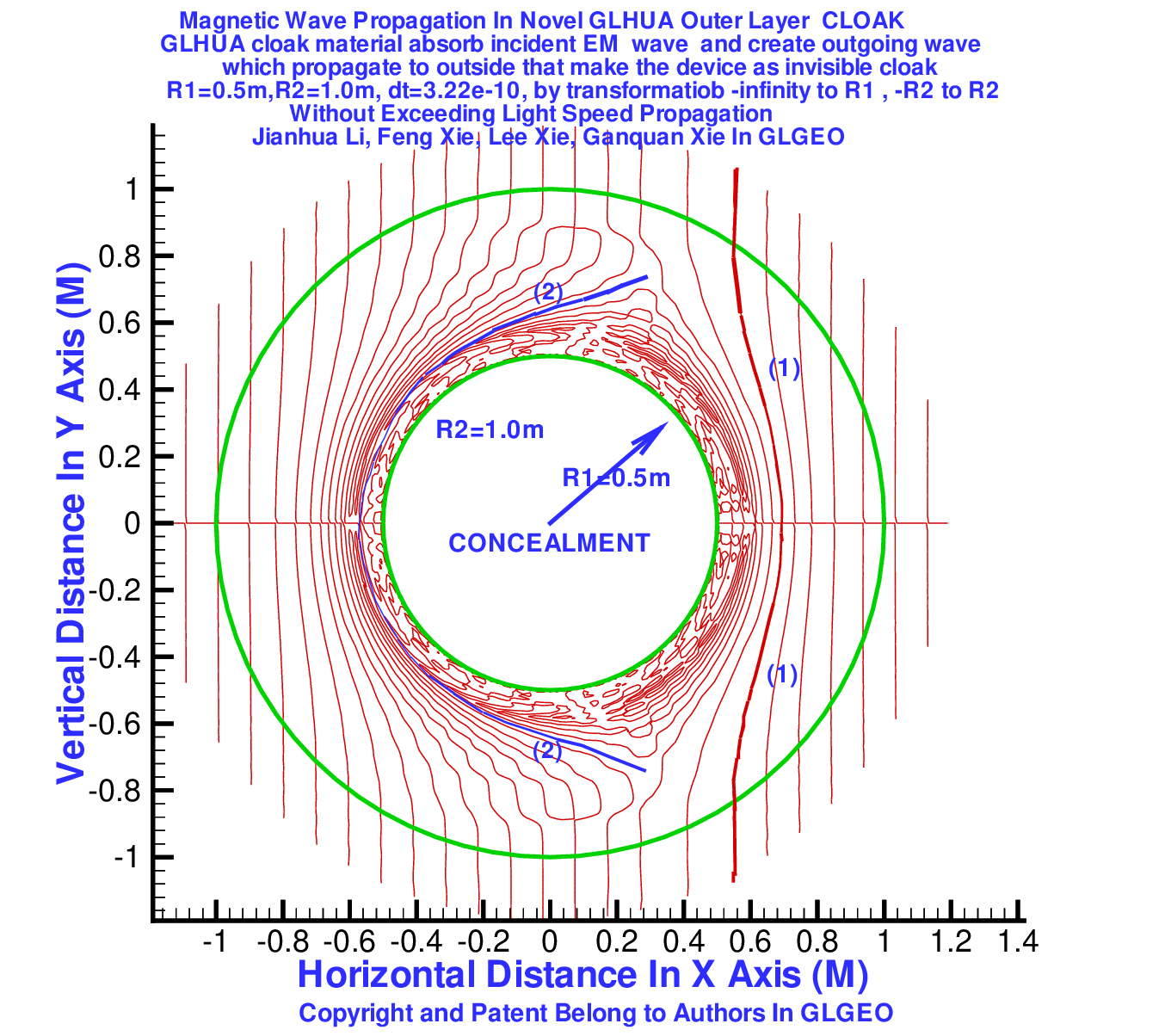}
\caption{ (color online) 
At relative 5th time step, the magneic wave front (red line (1)) incoming to right side of cloak, ,the material created curve wave front (blue  line (2)) expand up and down
}\label{fig30}
%\end{minipage}
\end{figure}  
\begin{figure}[h]
\centering
\includegraphics[width=0.86\linewidth,draft=false]{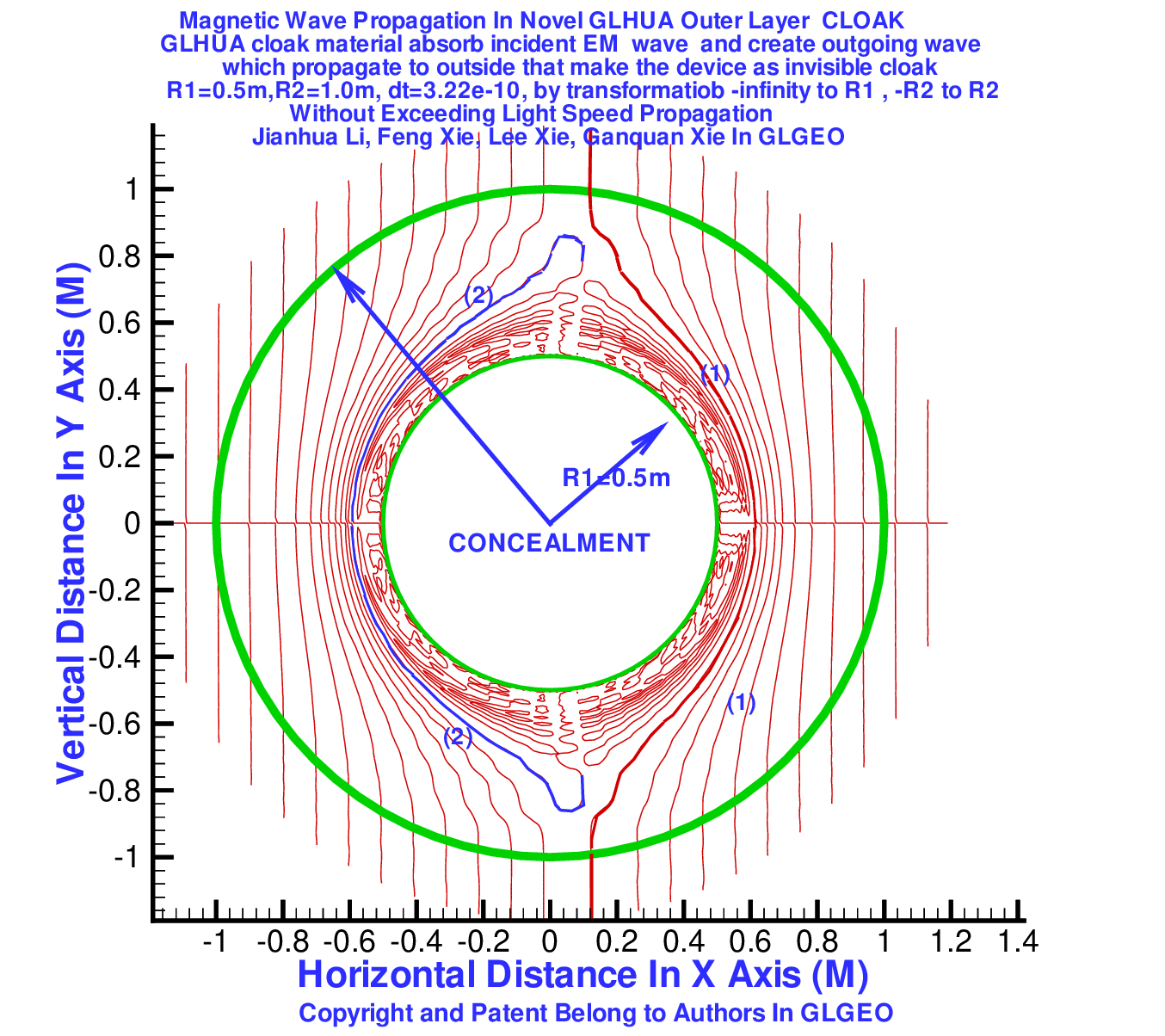}
\caption{ (color online) 
At relative 14 time step, the magnetic wave front (red line (1)) incoming to right side of cloak, the material created curve wave front ( blue line (2)) form wave front on the left side of cloak. Both wave front near each other but no connected.	
}\label{fig31}
%\end{minipage}
\end{figure}
\begin{figure}[h]
\centering
\includegraphics[width=0.86\linewidth,draft=false]{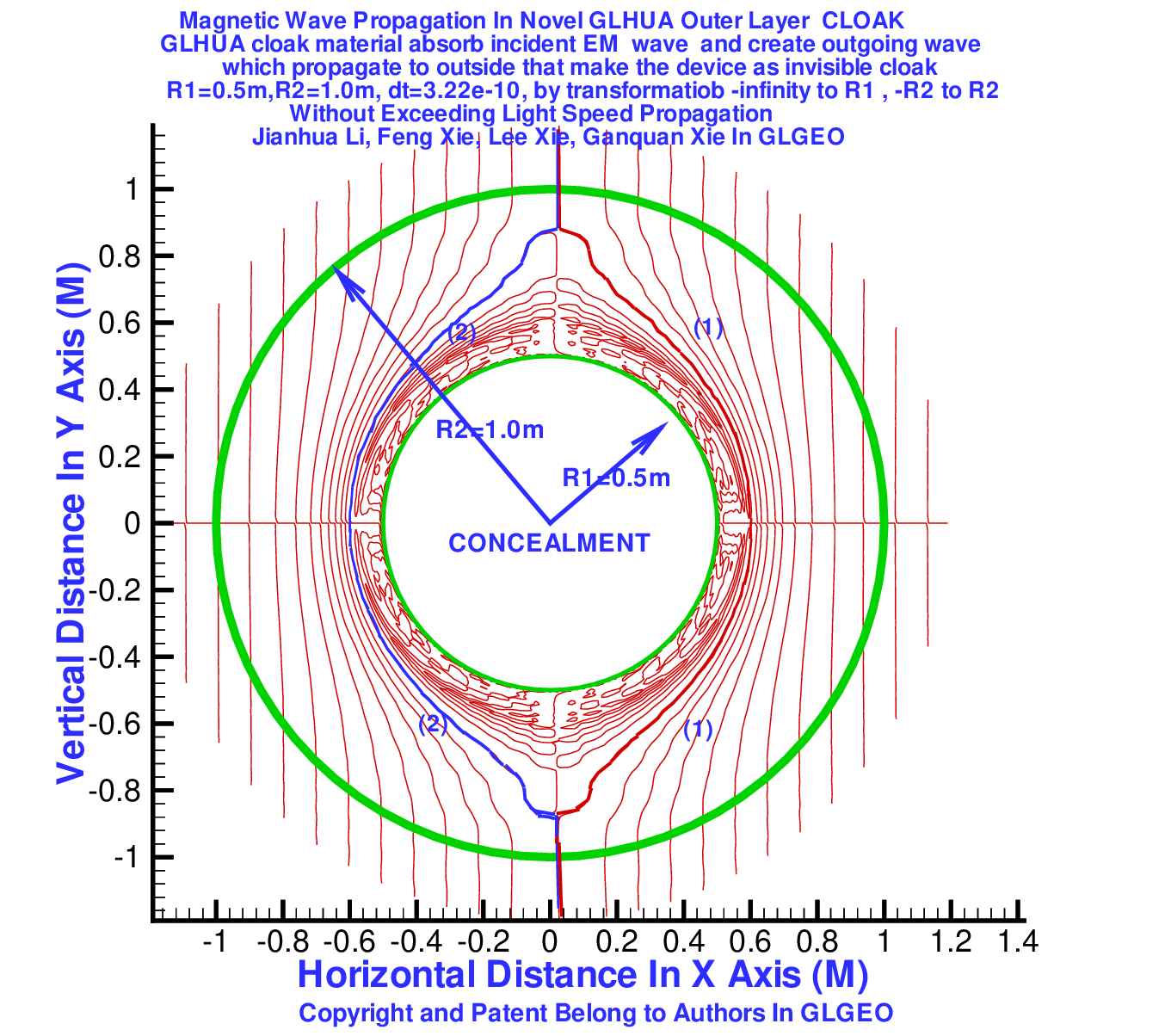}
\caption{ (color online) 
 At relative 15 time step, the magnetic wave front (red line (1)) incoming to right side of cloak, the material created curve wave front (blue line (2)) form wave front in the left side of the cloak, the red incoming wave (1) and blue created wave front (2) are connected.	 
}\label{fig32}
%\end{minipage}
\end{figure}
\begin{figure}[h]
\centering
\includegraphics[width=0.86\linewidth,draft=false]{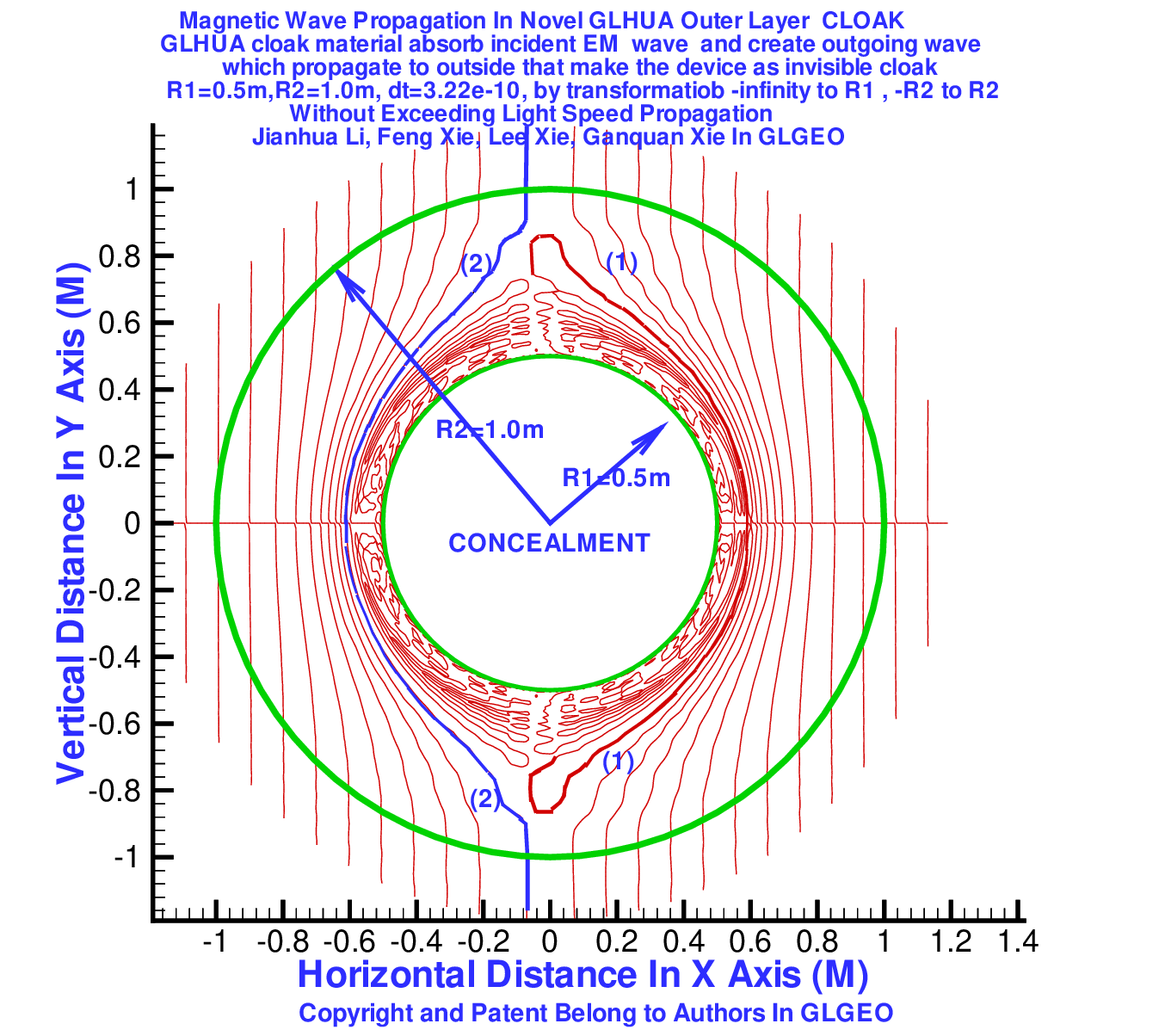}
\caption{ (color online) 
At relative 16 time steps, the curve wave front created by cloak material and incident wave in outside of cloak form wave front in the left side of the cloak  (blue line (2)), the incoming magneic wave front red line (1)) right side of cloak is disconncted with created wave (blue line (2)).
}\label{fig33}
%\end{minipage}
\end{figure}
\begin{figure}[h]
\centering
\includegraphics[width=0.86\linewidth,draft=false]{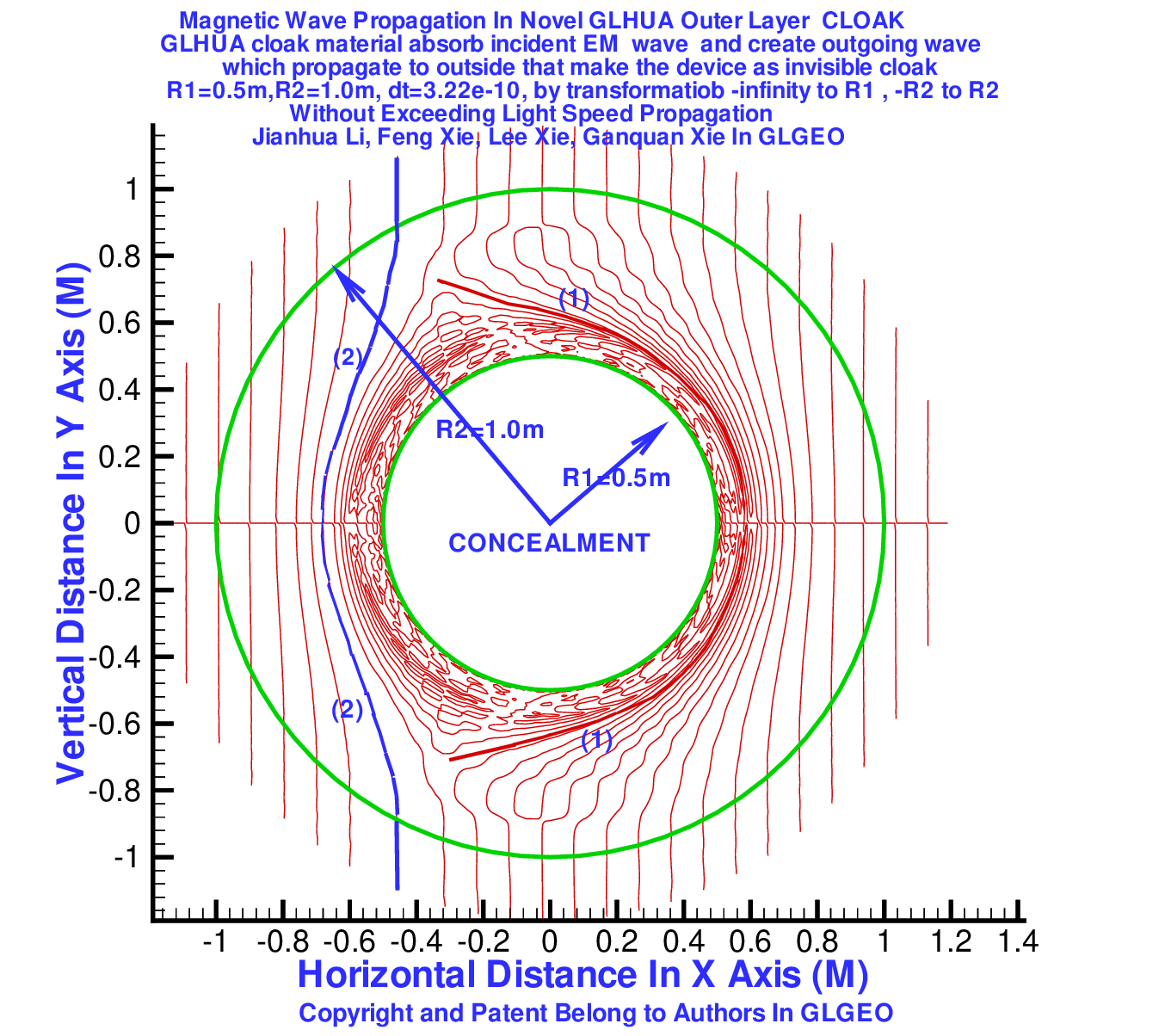}
\caption{ (color online) 
At relative 20 time steps, the curve wave front created by the cloak material is connected to the incident wavefront outside the cloak to form a wavefront and propagate to the left side of the cloak (blue line (2)). The incoming magnetic wave (red line (1)) is separated from the created wave (blue line (2)), and the incoming wavefront (red line (1)) is shrink, and wavefront (red line (1)) is behind from the created wavefront (blue line (2) )¡£	 
}\label{fig34}
%\end{minipage}
\end{figure}
\begin{figure}[h]
\centering
\includegraphics[width=0.86\linewidth,draft=false]{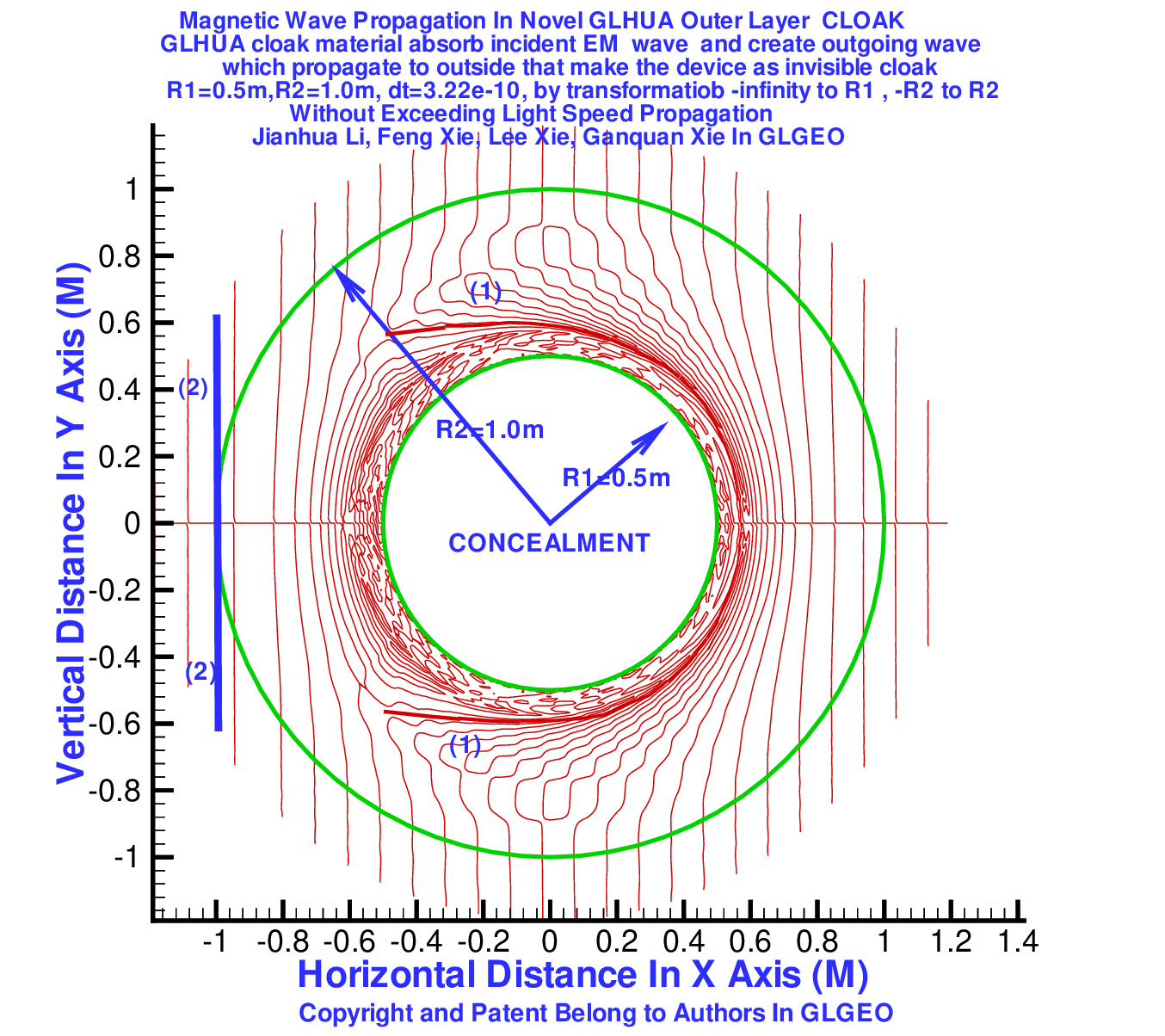}
\caption{ (color online) 
 At relative 25 time steps, the curve wave front created by cloak material and incident wave front in outside of cloak are connected to form wave front that propagate to boundary,    the left side of the cloak ( blue line (2)); The incoming magnetic wave (red line (1)) is separated from the created wave (blue line (2)), and wavefront (red line (1)) is behind from the created wavefront (blue line (2) ), and the incoming wavefront (red line (1)) is absorbed and shrink.	
}\label{fig35}
%\end{minipage}
\end{figure}
\begin{figure}[h]
\centering
\includegraphics[width=0.86\linewidth,draft=false]{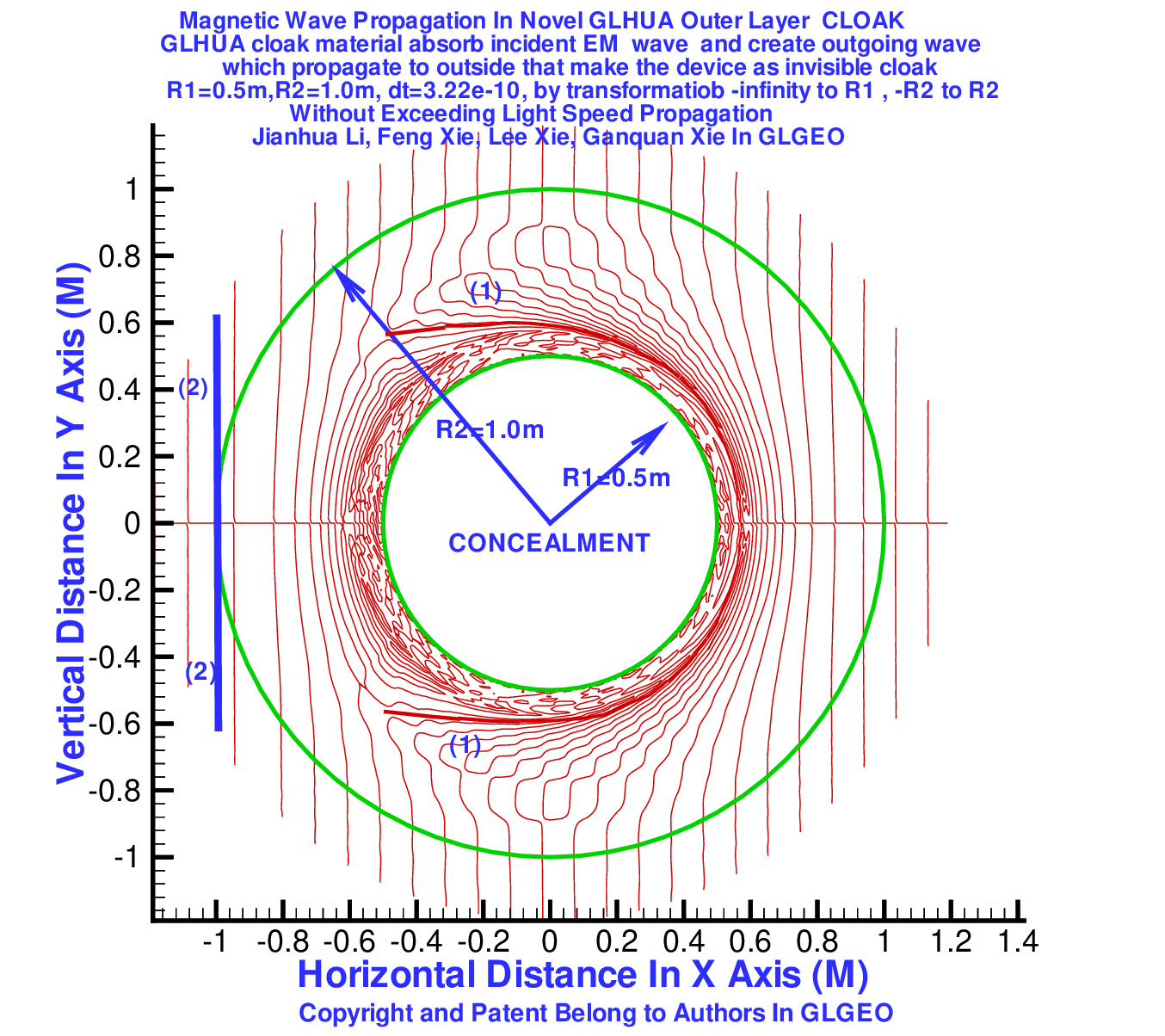}
\caption{ (color online) 
At relative 25 time steps, the material created curve wave front propagation
( blue line (2)) has already been out of cloak and did not disturb incident wave in outside of the cloak and make the cloak is invisible; the incoming maganetic wave front (red line (1)) is shrink to a circle and absorbed and can not be penetrated to the concealment  . The inner sphere   is really cloaked concealment.		
}\label{fig36}
%\end{minipage}
\end{figure}

\begin{figure}[h]
\centering
\includegraphics[width=0.86\linewidth,draft=false]{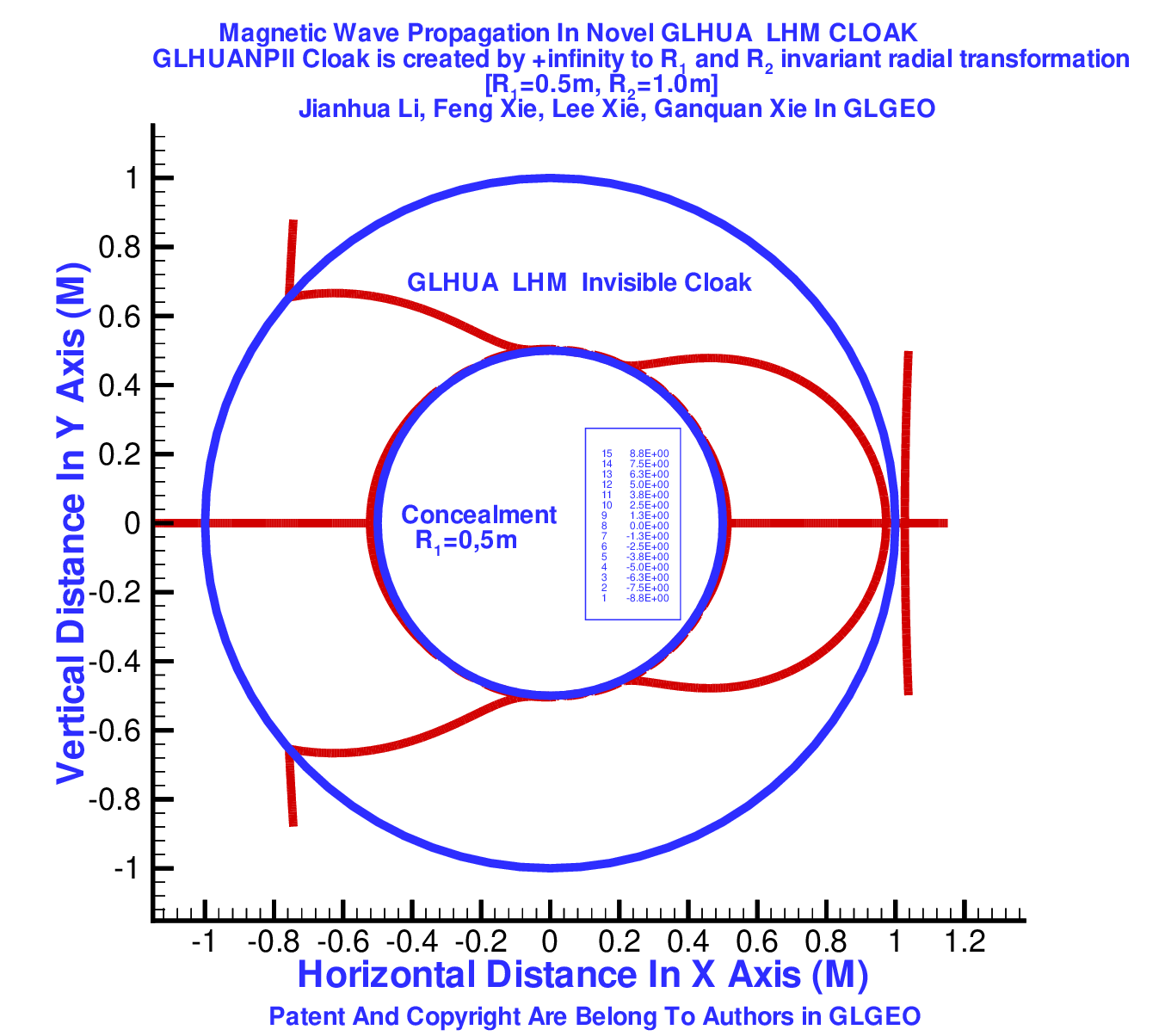}
\caption{ (color online) 
Radial magnetic wave propagation at 1 time step.}\label{fig37}
%\end{minipage}
\end{figure}
\begin{figure}[h]
\centering
\includegraphics[width=0.86\linewidth,draft=false]{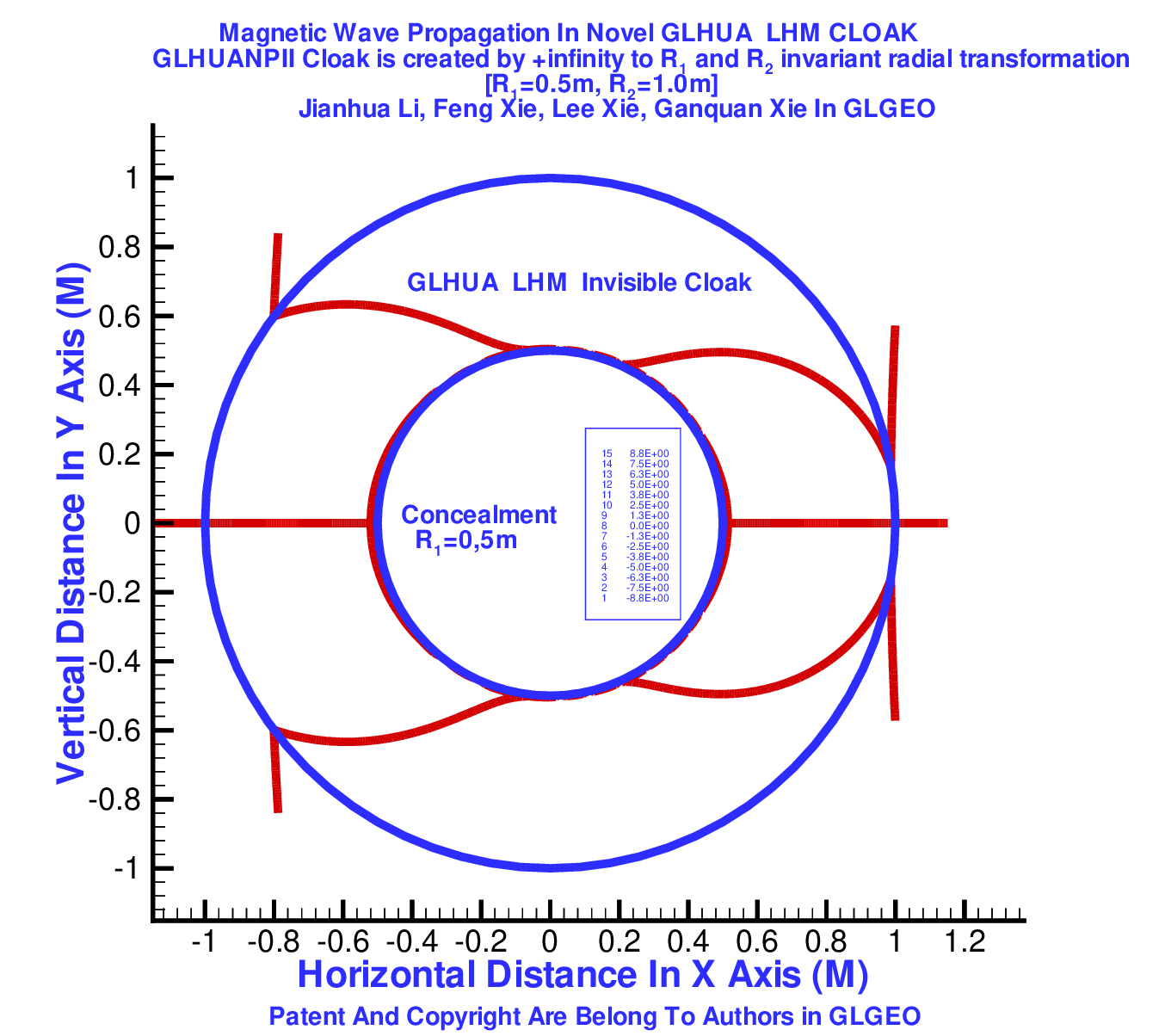}
\caption{ (color online) 
Radial magnetic wave propagation at 5  time step.}\label{fig38}
%\end{minipage}
\end{figure}

\begin{figure}[h]
\centering
\includegraphics[width=0.86\linewidth,draft=false]{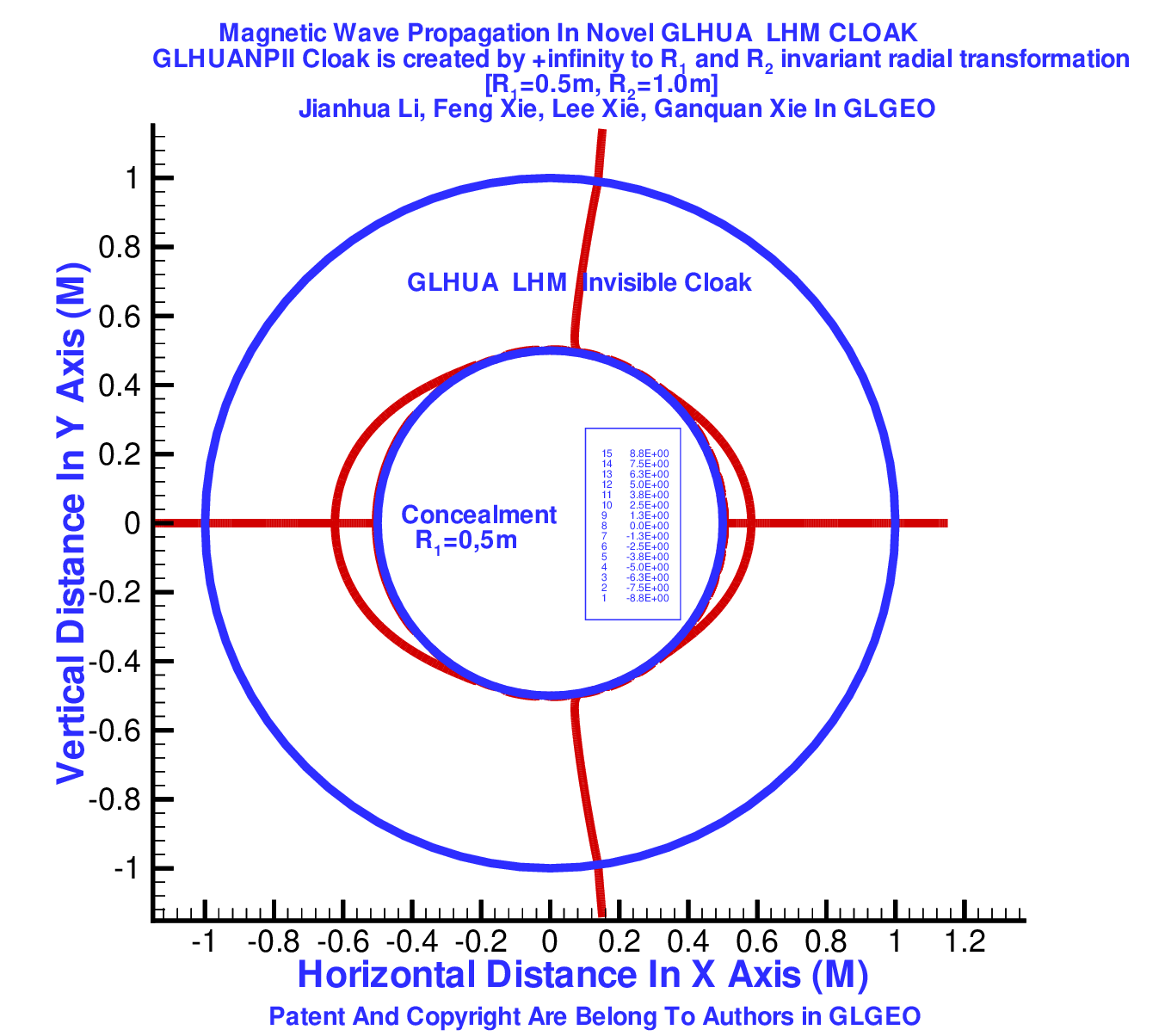}
\caption{ (color online) 
Radial magnetic wave propagation at 90  time step.}\label{fig39}
%\end{minipage}
\end{figure}

\begin{figure}[h]
\centering
\includegraphics[width=0.86\linewidth,draft=false]{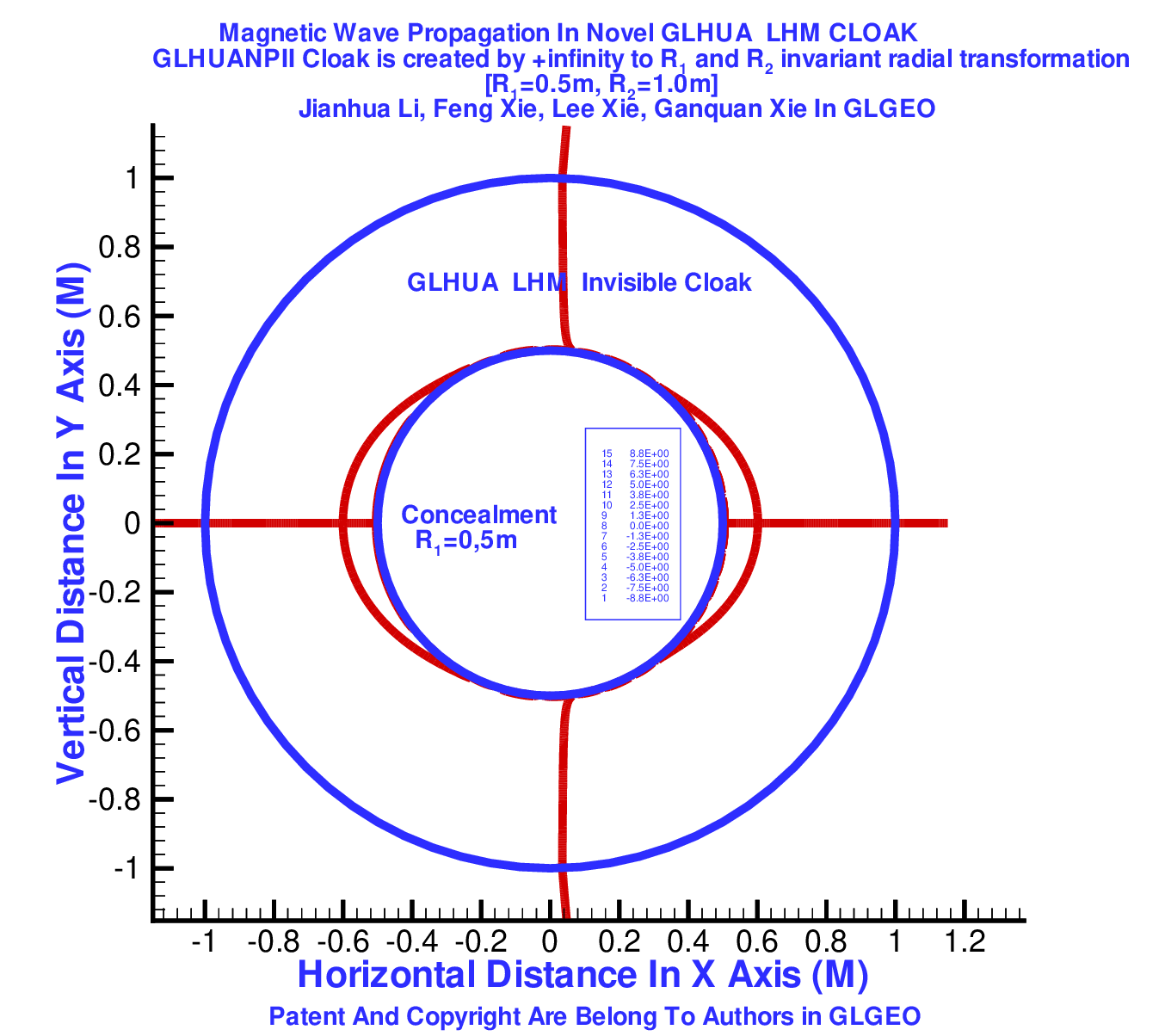}
\caption{ (color online) 
Radial magnetic wave propagation at 100  time step.}\label{fig40}
%\end{minipage}
\end{figure}

\begin{figure}[h]
\centering
\includegraphics[width=0.86\linewidth,draft=false]{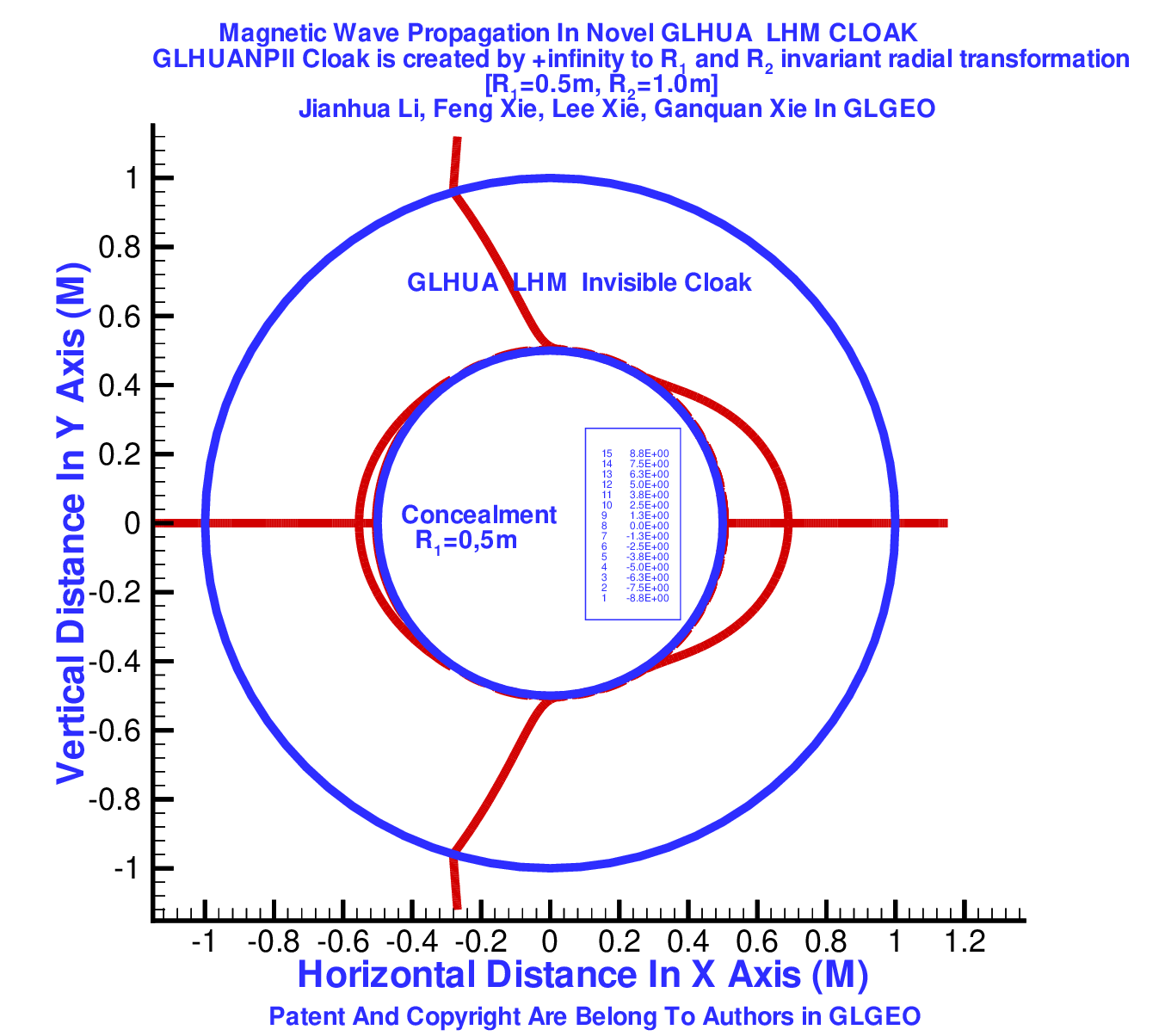}
\caption{ (color online) 
Radial magnetic wave propagation at 130  time step.}\label{fig41}
%\end{minipage}
\end{figure}

\begin{figure}[h]
\centering
\includegraphics[width=0.86\linewidth,draft=false]{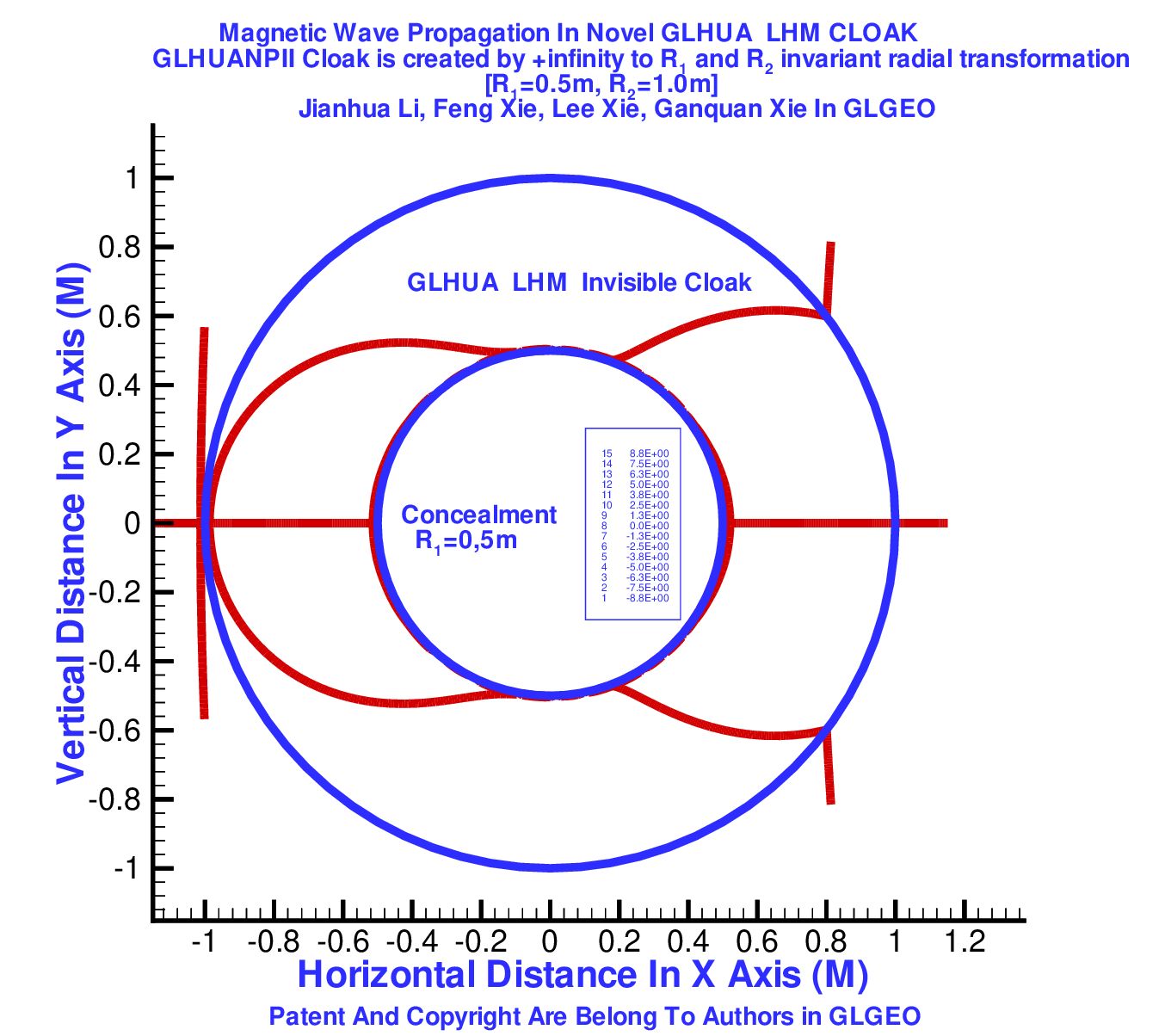}
\caption{ (color online) 
Radial magnetic wave propagation at 196  time step.}\label{fig42}
%\end{minipage}
\end{figure}

\begin{equation}
\begin{array}{l}
 \varepsilon _r  = 1, \\ 
 \varepsilon _\theta   = \varepsilon _\phi   = \frac{1}{2}\left( {\left( {\frac{{r - R_2 }}{{H_1  - R_2 }}} \right)^\alpha   + \left( {\frac{{H_1  - R_2 }}{{r - R_2 }}} \right)^\alpha  } \right), \\ 
 \end{array}
\end{equation}

In our paper [5-6][11-12][15], we proposed GL no scattering modeling and inversion to create practicable GLHUA-1 and GLLH two class of the invisible cloak. Our GLHUA-1 cloak by GL no scattering modeling and inversion is different from GLLH cloak in 2010. In 1971, we used no scattering inversion idea to construct a novel 3D 20 nodes high accurate curve element and developed 3D finite element method first in China and discovered super convergence of 3D iso parameter finite element first in the world [26][27]. GILD and GL scattering and no scattering modeling and inversion idea and method play very important role for research works in invisible cloak. When Pendry cloak in 2006, we had been working in no scattering modeling and inversion for 6 years since 2000. Our idea and method are different from Pendry and other cloak research.

Because there is no experiment to study the complete invisible cloak, therefore the full wave theoretical proof and full wave computational simulation are necessary to verify the invisible clock. GLHUA-1 invisible cloak and GLLH invisible cloak have been verified by the full wave theoretical proof and full wave simulation using GL full wave no scattering modeling and inversion in [1][2]. GLHUA-1 double cloak and created wave by cloak materials in [13] in arXiv:1612.02857 show that GL no scattering modeling and inversion is powerful method to make practicable invisible cloak and to investigate new invisible natural science Inventor Easton LaChappelles mind control robot hands is beginning to investigate visible thinking science and visible social Science. In current general science, visible natural science and invisible thinking science and social science are main object. Now invisible natural science and invisible thinking science and visible social science will be main object for a novel super science. Our GLHUA-1 practicable double cloak and Easton LaChappelles mind control robot hands show that the new novel super science is being born.

\section{ GLHUANP-4 Electromagnetic Invisible Cloak 
 and GLHUANP-4 Spherical Radial Transformation maps the Negative Infinity 
$-\infty $ To $ R_1$ and $ R_2$ to $ R_2 $}

\subsection{Radial magnetic wave equation in the positive and negative sphere} 

When the radial variable in the sphere coordinate space is positive or zero, the radial magnetic wave equation in the sphere coordinate [1][2][3] is
\begin{equation}
\begin{array}{l}
\frac{\partial }{{\partial R}}\frac{\partial }{{\partial R}}R^2 H_r  + \frac{1}{{\sin \theta }}\frac{\partial }{{\partial \theta }}\sin \theta \frac{{\partial H_r }}{{\partial \theta }} + \frac{1}{{\sin ^2 \theta }}\frac{{\partial ^2 H_r }}{{\partial \phi ^2 }} \\
+ k^2 R^2 H_r  = M_s 
\end{array}
\end{equation}

${\boldsymbol{Theorem \ 5:}}$ \ In the normal positive space in which the radial variable in the sphere space is positive, if the wave solution of the  electric wave equation is $H_r (R,\theta ,\phi )$ ,then $H_r (-R,\theta ,\phi )$  is also solution of homogeneous equation ( 82) with $M_s  = 0$ 

Proof, For point source at $(r_s ,\theta _s ,\phi _s )$ , if  $H_r (R,\theta ,\phi )$  is solution of ( 82) with point source. By heave calculus calculation, we prove that the  $H_r (-R,\theta ,\phi )$   satisfies homogeneous equation ( 134) with  $M_s  = 0$ ,
            
${\boldsymbol{Theorem \ 6:}}$ \ The homogeneous radial electric wave equation ( 134) with  $M_s  = 0$,  its solution $H_r (R,\theta ,\phi )$   in positive space can be analytic continuation into negative space with negative radial $R$.

Proof: Let $H_r (R,\theta ,\phi )$    be the solution of the radial electric quation ( 134) with point source, by theorem 5, the $H_r ({-R},\theta ,\phi )$     is also the solution
of the radial elecric wave equation,

\begin{equation}
\begin{array}{l}
\frac{\partial }{{\partial R}}\frac{\partial }{{\partial R}}R^2 H_r ( - R,\theta ,\phi ) + \frac{1}{{\sin \theta }}\frac{\partial }{{\partial \theta }}\sin \theta \frac{{\partial H_r ( - R,\theta ,\phi )}}{{\partial \theta }} \\ 
  + \frac{1}{{\sin ^2 \theta }}\frac{{\partial ^2 H_r ( - R,\theta ,\phi )}}{{\partial \phi ^2 }} + k^2 R^2 H_r ( - R,\theta ,\phi ) = 0, \\ 
 \end{array}
\end{equation}
                                   
For $\xi  < 0$, $ -\xi  > 0$, 
Substitute $ R =  -\xi  $  into the ( 135) , the homogeneou radial electric equation ( 134) with  $M_s  = 0$   and its solution that becomes
         
 \begin{equation}  
\begin{array}{l}
 \frac{\partial }{{\partial ( - \xi )}}\frac{\partial }{{\partial ( - \xi )}}R^2 H_r ( - ( - \xi ),\theta ,\phi ) + \frac{1}{{\sin \theta }}\frac{\partial }{{\partial \theta }}\sin \theta \frac{{\partial H_r ( - ( - \xi ),\theta ,\phi )}}{{\partial \theta }} \\ 
  + \frac{1}{{\sin ^2 \theta }}\frac{{\partial ^2 H_r ( - ( - \xi ),\theta ,\phi )}}{{\partial \phi ^2 }} + k^2 ( - \xi )^2 H_r ( - ( - \xi ),\theta ,\phi ) = 0, \\ 
 \end{array}
\end{equation}     
\begin{equation}             
\begin{array}{l}
 \frac{\partial }{{\partial \xi }}\frac{\partial }{{\partial \xi }}R^2 H_r (\xi ,\theta ,\phi ) + \frac{1}{{\sin \theta }}\frac{\partial }{{\partial \theta }}\sin \theta \frac{{\partial H_r (\xi ,\theta ,\phi )}}{{\partial \theta }} \\ 
  + \frac{1}{{\sin ^2 \theta }}\frac{{\partial ^2 H_r (\xi ,\theta ,\phi )}}{{\partial \phi ^2 }} + k^2 (\xi )^2 H_r (\xi ,\theta ,\phi ) = 0, \\ 
 \end{array}
\end{equation}
By using $R = \xi$, we obtain

\begin{equation}
\begin{array}{l}
 \frac{\partial }{{\partial R}}\frac{\partial }{{\partial R}}R^2 H_r (R,\theta ,\phi ) + \frac{1}{{\sin \theta }}\frac{\partial }{{\partial \theta }}\sin \theta \frac{{\partial H_r (R,\theta ,\phi )}}{{\partial \theta }} \\ 
  + \frac{1}{{\sin ^2 \theta }}\frac{{\partial ^2 H_r (R,\theta ,\phi )}}{{\partial \phi ^2 }} + k^2 (R)^2 H_r (R,\theta ,\phi ) = 0, \\ 
 \end{array}
\end{equation}

Therefore,  homogeneous radial electric wave equation ( 134) with  $M_s  = 0$  and its solution $H_r (R,\theta ,\phi )$   in positive space can be analytic continuation into negative space with negative radial $R$ in equation ( 137).
The theorem 6 is proved.

\subsection{GLHUANP-4 transformation to map the Negative Infinity $ -\infty $ To $ R_1 $ and $ R_2 $ to $ R_2 $}

\subsubsection{GLHUANP-4 transformation}

We create a novel sphere radial transformation GLHUANP-4 to map the negative infinity, $ - \infty  $  in Basic Space (BS) to $ R_1$ in the Physical space (PS) and keep $ R_2 $ invariant that is as follows 
\begin{equation}
r=R_1+(R_2-R_1)e^{(R-R_2)/(R_2-R_1)},     ( 161)
\end{equation}
and its inverse transformation
\begin{equation}
R=(R_2-R_1)log((r-R_1)/(R_2-R_1))+R_2,     ( 162)
\end{equation}
Where $ R $ be radial variable in the Basic Space (BS), $ r $ be radial variable in the Physical Space (PS).

\subsubsection {GLHUANP-4 relative electromagnetic permeability and dielectric are created},
By the GLHUANP-4 transformation, new GLHUANP-4 invisible cloak and GLHUANP-4 relative EM parameters are created as follows:
\begin{equation}
\varepsilon _\theta   = \mu _\theta   = {{(R_2  - R_1 )} \mathord{\left/
 {\vphantom {{(R_2  - R_1 )} {(r - R_1 )}}} \right.
 \kern-\nulldelimiterspace} {(r - R_1 )}}   ( 163) 
\end{equation}
 
\begin{equation}
\varepsilon _r  = \mu _r  = \frac{{\left( {(R_2  - R_1 )\log \left( {\frac{{r - R_1 }}{{R_2  - R_1 }}} \right) + R_2 } \right)^2 }}{{r^2 }}\frac{{r - R_1 }}{{R_2  - R_1 }},  
\end{equation}

the angular electric dielectric and magnetic permeability of GLHUANP-4 are same as the relative  angular EM parameters of GLHUA-1 EM invisible cloak.                           
\subsection{The magnetic wave propagation through the GLHUANP-4 invisible cloak} 
The magnetic wave propagation through the GLHUANP-4 invisible cloak with relative GLHUANP-4 electromagnetic parameters ( 84) in $ R_1  \le r \le R_2 $  and free space in $ 0 \le r \le R_1 $ 
\subsection{GLHUANP-4 Invisible Cloak Borns GLHUA-3 invisible cloak without exceeding light speed propagation and GLHUANP-4 Invisible Cloak With
exceeding light speed propagation }

The GLHUANP-4 radial transformation maps the Negative Infinity $ - \infty $ in BS To $R_1$ in PS and $R_2$in BS to $R_2$ in PS. By the GLHUANP-4 transformation, the relative electric dielectric and magnetic permeability, 
\begin{equation}
\varepsilon _r  = \mu _r  = \frac{{R^2 }}{{r^2 }}\frac{{r - R_1 }}{{R_2  - R_1 }}
\end{equation}
and  
\begin{equation}
\varepsilon _\theta   = \mu _\theta   = \varepsilon _\phi   = \mu _\phi   = \frac{{R_2  - R_1 }}{{r - R_1 }}
\end{equation}
in ( 165)( 166) are installed in the annular layer,$R_1 \le r \le R_2$ ,in addition. let the inner small sphere , $0 \le R \le R_1$, and outside of large sphere ,$r \ge R_2$, be free space, a GLHUANPII invisible cloak is created with
concealment, $0 \le r \le R_1$.  Let 
\begin{equation}
H_1  = R_1  + (R_2  - R_1 )e^{ - \frac{{R_2 }}{{R_2  - R_1 }}}, 
\end{equation}

and 

\begin{equation}
H_0  = R_1  + (R_2  - R_1 )e^{ - \frac{{H_0  + R_2 }}{{R_2  - R_1 }}}. 
\end{equation}
 In the annular layer $H_1 \le r \le R_2$ , the relative EM GLHUANP-4  refractive index  
\begin{equation}
0 \le n = \sqrt {\varepsilon _\theta  \mu _r }  \le 1,
\end{equation}
and 
\begin{equation}
n(H_1 ) = \sqrt {\varepsilon _\theta  \mu _r } (H_1 ) = 0,
\end{equation}
, and 
\begin{equation}
n(R_2 ) = \sqrt {\varepsilon _\theta  \mu _r } (R_2 ) = 1,
\end{equation}
and EM parameters to be 1 in the inner sphere free space $0 \le r \le H_1$ , GLHUANP-4 cloak become an GLHUANP-4 - 3 invisible cloak with exceeding light speed and infinity velocity at $ r = H_1 $,which is similar with, but is different from the Pendry cloak. 
In the annular layer $R_1 \le r \le H_0$, the relative EM GLHUANP-4  refractive index 
\begin{equation}
1 \le n = \sqrt {\varepsilon _\theta  \mu _r }  \le \infty
\end{equation}
 and , 
\begin{equation}
n(H_0 ) = \sqrt {\varepsilon _\theta  \mu _r } (H_0 ) = 1,
\end{equation}
 and EM parameters to be 1 in the inner sphere free space $0 \le r \le R_1$, the GLHUANP-4 cloak becomes GLHUANP-3 invisible cloak which is similar with the GLLHUA-3 invisible cloak with relative EM refractive index parameter not less than one and without exceeding light speed
physical difficulties

\subsection{Analytical exact radial electric and magnetic wave propagation through the GLHUA-1  practicable invisible cloak with EM parameter not less one}
 
\subsubsection{The novel GLHUA-1 practicable invisible cloak with EM parameter not less one}

In our paper arXiv:1706.10147[3], we proposed the novel GLHUA-1  practicable invisible cloak with EM parameter not less one[3]
\begin{equation}
\varepsilon _{\rm r}  = \mu _r  = \frac{{R_2 ^2 }}{{r^2 }},
 \end{equation}                                                    ,                                                    
\begin{equation} 
\varepsilon _\theta   = \mu _\theta   = \varepsilon _\phi   = \mu _\phi   = \frac{{R_2  - R_1 }}{{r - R_1 }},     
\end{equation} 
        
\subsubsection{In our paper arXiv:1706.10147, we create the new GLHUA-1 separate variable method and find  analytical radial electric and radial magnetic wave propagation through GLHUA-1 practicable invisible cloak}

\subsubsection{The analytical radial magnetic wave propagation through GLHUA-1 practicable invisible cloak is shown in the Figure 13 and Figure 27}

\subsection{The electric and magnetic wave propagation through GLHUA-1 practicable invisible cloak with EM parameter not less one}

\subsubsection {GLHUA-1 practicable invisible cloak with EM parameter not less one}

In our papers arXiv:1612.02857[2], arXiv:1701.00534[6], arXiv:1701.002583[7],  we created GLHUA-1  practicable invisible cloak with EM parameter not less one
\begin{equation} 
\varepsilon _r  = \mu _r  = 1,
\end{equation} 
\begin{equation} 
 \begin{array}{l}
\varepsilon _\theta   = \varepsilon _\phi   = \mu _\theta   = \mu _\phi   \\ 
  = \frac{1}{2}\left( {\left( {\frac{{r - R_1 }}{{R_2  - R_1 }}} \right) + \left( {\frac{{R_2  - R_1 }}{{r - R_1 }}} \right)} \right),R_1  < r \le R_2 , \\ 
 \end{array}
\end{equation}                                    ,                     
or 
GLHUA-1 - III practicable invisible cloak with EM parameter not less one
\begin{equation} 
\varepsilon _r  = \mu _r  = 1,
\end{equation}
\begin{equation} 
\varepsilon _\theta   = \mu _\theta   = {{(R_2  - R_1 )} \mathord{\left/
 {\vphantom {{(R_2  - R_1 )} {(r - R_1 )}}} \right.
 \kern-\nulldelimiterspace} {(r - R_1 )}}       
\end{equation}                                                 ,                                            
\subsubsection {By GL super computational sciences no scattering modeling and inversion, we discovered and created GLHUA-1 - I and GLHUA-1 - III practicable invisible cloak with EM parameter not less one, by GL no scattering modeling simulation, we find the numerical electromagnetic wave propagation in the GLHUA-1 - I and GLHUA-1 - III practicable invisible cloak with EM parameter not less one}
            
\subsubsection {By GL no scattering modeling simulation, the radial magnetic wave propagation in
GLHUA-1  practicable invisible cloak }

\subsection {A novel and strange GLHUA-1 cloak wave in $ R_1  \le r \le R_2 $  and around and attached on inner sphere surface  the $ r = R_1$, and the GLHUA-1 wave front is discontinuous}
A novel and strange GLHUA-1 cloak wave in $ R_1  \le r \le R_2 $  and around and attached on inner sphere surface  the $ r = R_1$, and the GLHUA-1 wave front is discontinuous. The novel and strange GLHUA-1 cloak wave is created wave by GLHUA-1 cloak materals in (25)

\subsection {Discussions and Conclusions}

\subsubsection {The two novel wave fronts of GLHUA-1 and GLHUA-2 and GLHUA-3 are discontinuous}

The GLHUA-1  and GLHUA-2 and GLHUA-3  are  practicable invisible cloak with EM parameter not less than one and without exceeding light speed propagation fundamental difficulties.
The two novel wave fronts of GLHUA-1 and GLHUA-2 and GLHUA-3 are discontinuous.
One wave front is absorbing incoming incident EM wave by the part of GLHUA-1,
GLHUA-2 and GLHUA-3 materials. Other wave front is creating outgoing EM wave by the other part of GLHUA-1,GLHUA-2 and GLHUA-3 materials. The novel EM wave front propagation in GLHUA-i,i=1,2,3  are displaying in Figure 2 to Figure 36 that proved that our GLHUA-1 and GLHUA-2 and GLHUA-3 cloak are invisible cloaks
with the refractive index large or equal one.

\subsubsection {The Inner Sphere of Pendry cloak, $0 \le r \le R_1$ is not cloaked concealment, if Pendry EM parameters are put in it}
In the Pendry Cloak, the relative EM parameters are from Pendry transformation in the annular layer $ R_1  \le r \le R_2 $ ,  also. Pendry put free space media with relative EM parameter 1 in the inner sphere $ 0 \le r \le R_1 $ ?that create Pendry invisible cloak in his paper in sciences 2006[4].In my paper[5], we discovered and proved an important problem that If the Pendry relative EM  parameters  
\begin{equation}                             
 \varepsilon _r  = \mu _r  = ({{R_2 } \mathord{\left /
 {\vphantom {{R_2 } {(R_2  - R_1 )}}} \right.
 \kern-\nulldelimiterspace} {(R_2  - R_1 )}}){{(r - R_1 )^2 } \mathord{\left/
 {\vphantom {{(r - R_1 )^2 } {r^2 }}} \right.
 \kern-\nulldelimiterspace} {r^2 }}
\end{equation}
\begin{equation}
\varepsilon _\theta   = \mu _\theta   = \varepsilon _\phi   = \mu _\phi   = {{R_2 } \mathord{\left/
 {\vphantom {{R_2 } {(R_2  - R_1 )}}} \right.
 \kern-\nulldelimiterspace} {(R_2  - R_1 )}}
\end{equation}
are sit in the whole sphere $0 < R_1  < R_2$ which include the annular 
$R_1  \le r \le R_2$ and inner sphere $0 \le r \le R_1$  .  
What happen is for EM wave propagation $ {\rm ?}$  For solving this problem, we discover and propose the novel negative space in which the radial variable of the sphere coordinate is negative. The real 3D space we living with positive or zero radial variable is called positive space. The positive space is visible 3D space we living, the negative space is invisible and From my negative space theory, we prove that 

${\boldsymbol{Theorem \ 7:}}$ \ if the Pendry EM parameters ( 180)-( 181) in [4] are sit in the whole sphere $0 < R_1  < R_2 $. which include the annular $ R_1  \le r \le R_2 $ and inner sphere $ 0 \le r \le R_1 $, then
there are EM wave propagation in the the annular layer $ R_1  \le r \le R_2 $  and also in the inner sphere $ 0 \le r \le R_1 $, the inner sphere is not cloaked concealment. 

In Figure in other paper, we present the Pendry EM parameters (90-91) in [4] are sit in the whole sphere $0 < R_1  < R_2 $. which include the annular $ R_1  \le r \le R_2 $ and inner sphere $ 0 \le r \le R_1 $. The radial magnetic wave excited by point source in $r_s > > R_2$ propagate through whole sphere with Pendry EM parameters ( 180)-( 181) in the paper) in [4] . There is radial magnetic wave propagation in the the annular layer $ R_1  \le r \le R_2 $  and also in the inner sphere $ 0 \le r \le R_1 $, the inner sphere is not cloaked concealment.

\section {ANOTHER NOVEL GLHUAF INVISIBLE CLOAK WITH DOUBLE NEGATIVE EM PARAMETER AND SQUARE OF REFRACTIVE INDEX $n^2 \ge 1$}

\subsection{Another GLHUAF radial transformation}
GLHUANPII radial transformation inspires us to invent another novel radial
Transformation GLHUAF radial transformation.
%\begin {equation}
\[
r = R_1  + (R_2  - R_1 )e^{\frac{{R_2  - R}}{{R_2  - R_1 }}}     ( 19)
\]
%\end {equation}
Which maps positive infinity $ + \infty $ to $R_1$, and $R_2$ is invariant.
$R$ is radial variant in basic space before transformation, $r$ is radial transformation in the physical space after transformation. 

\subsection{GLHUAF Double Negative EM relative parameters}
Use the GLHUAF transformation to transform the radial magnetic equation in free space from positive infinity $ + \infty $ to $R_1$, and $R_2$ is invariant.
%\begin {equation}
\[
\begin{array}{l}
 \frac{\partial }{{\partial r}}\frac{1}{{\mu _\theta  }}\frac{{\partial H}}{{\partial r}} + \frac{1}{{r^2 \mu _r }}\frac{1}{{\sin \theta }}\frac{\partial }{{\partial \theta }}\sin \theta \frac{{\partial H}}{{\partial \theta }} \\ 
  + \frac{1}{{r^2 \mu _r }}\frac{1}{{\sin ^2 \theta }}\frac{{\partial ^2 H}}{{\partial \phi ^2 }} + k^2 \varepsilon _\theta  H = M_s  \\ 
 \end{array},                     ( 20)
%\end {equation}
\]
novel double negative GLHUAF relative EM parameters are created,
%\begin {equation}
\[
\varepsilon _\theta   = \mu _\theta   = \varepsilon _\phi   = \mu _\phi   =  - \frac{{R_2  - R_1 }}{{r - R_1 }},                    ( 21)
\]
%\end {equation}
%\begin {equation}
\[
\varepsilon _r  = \mu _r  =  - \frac{{R^2 }}{{r^2 }}\frac{{r - R_1 }}{{R_2  - R_1 }},\;\;\;\;\;\;\;\;( 22)
\]
%\end {equation}
By the inverse transformation of ( 19), in the ( 21), the
%\begin {equation}
\[
R = R_2  - (R_2  - R_1 )\log \left( {\frac{{r - R_1 }}{{R_2  - R_1 }}} \right),\;\;\;\;\;\;\;\;\;( 23)
\]
%\end {equation}

%\begin {equation}
\[
\begin{array}{l}
 \varepsilon _r  = \mu _r  =  - \frac{1}{{r^2 }}\frac{{r - R_1 }}{{R_2  - R_1 }} \\ 
 \left( {R_2  - (R_2  - R_1 )\log \left( {\frac{{r - R_1 }}{{R_2  - R_1 }}} \right)} \right)^2 , \\ 
 \end{array} ( 24)
\]
%\end {equation}

\subsection{GLHUAF EM invisible cloak}
By installing the GLHUAF relative double negative electric permittivity and magnetic permeability ( 21) and ( 24) in the sphere annular layer $R_1  < r \le R_2 $ and installing the free space in the inner sphere $r \le R_1 $, the GLHUAF invisible cloak is created. 
Novel EM wave propagation through the GLHUAF invisible cloak are presented in Figure 37 to the Figure 42.

\section{GLHUA-3 INVISIBLE CLOAK BY TRANSFORMATION WHICH MAPS NEGATIVE  INFINITY TO $R_1$ AND MAPAS $- R_2$ TO $R_2$}

We discover additional new invisible cloak GLHUA-3 in this paper. GLHUANP-3 radial transformation,
%\begin{equation}
\[
r = R_1  + (R_2  - R_1 )e^{\frac{{R + R_2 }}{{R_2  - R_1 }}} ,     ( 12)
%\end{equation}
\]
which maps negative infinity,   $- \infty $ to $R_1$, and maps $ -R_2 $ to $ R_2 $.

The GLHUANP-3 inverse transformation is
%\begin{equation}
\[
R =  - R_2  + (R_2  - R_1 )\log \left( {\frac{{r - R_1 }}{{R_2  - R_1 }}} \right),       ( 13)
\]
%\end{equation}
From the GLHUANP-3 transformation ( 12), GLHUA-3 relative electric permittivity and magnetic permeability are created as
%\begin{equation}
\[
\varepsilon _\theta   = \mu _\theta   = \varepsilon _\phi   = \mu _\phi   = \frac{{R_2  - R_1 }}{{r - R_1 }},      ( 14)
\]
%\end{equation}

%\begin{equation}
\[
\begin{array}{l}
 \varepsilon _r  = \mu _r  =  \\ 
 \left( { - R_2  + (R_2  - R_1 )\log \left( {\frac{{r - R_1 }}{{R_2  - R_1 }}} \right)} \right)^2  \\ 
 \frac{{r - R_1 }}{{r^2 (R_2  - R_1 )}} \\ 
 \end{array}  ( 15)
\]
%\end{equation}

The GLHUA-3  radial parameters ( 15) is as follows,
By installing GLHUA-3 EM parameters ( 14) and ( 15) in the annular layer 
$R_1  \le r \le R_2 $ and
Installing free space in inner sphere $0 \le r \le R_1 $, the new GLHUA-3 invisible cloak is created.
The properties of GLHUA-3 invisible cloak are as following:

(1) The EM parameters are continuous on the $r=R_2$

%\begin{equation}
\[
\begin{array}{l}
 \varepsilon _\theta  (R_2 ) = \mu _\theta  (R_2 ) = \varepsilon _\phi  (R_2 ) = \mu _\phi  (R_2 ) \\ 
  = \frac{{R_2  - R_1 }}{{r - R_1 }}(R_2 ) = 1, \\  
 \end{array},  ( 16)  
\] 
%\end{equation}
%\begin{equation}
\[
\varepsilon _r (R_2 ) = \mu _r (R_2 ) = 1,  ( 17)
%\end{equation}
\]
(2).	The refractive index 
%\begin{equation}
\[
1 \le n \le  + \infty ,  ( 18)
\]  
%\end{equation}

(3).	The EM wave propagation in the GLHUA-3 cloak without exceeding light speed difficulties.

The radial electric wave propagation through the GLHUA-3 cloak is display in the figure 43-50.

\begin{figure}[h]
\centering
\includegraphics[width=0.86\linewidth,draft=false]{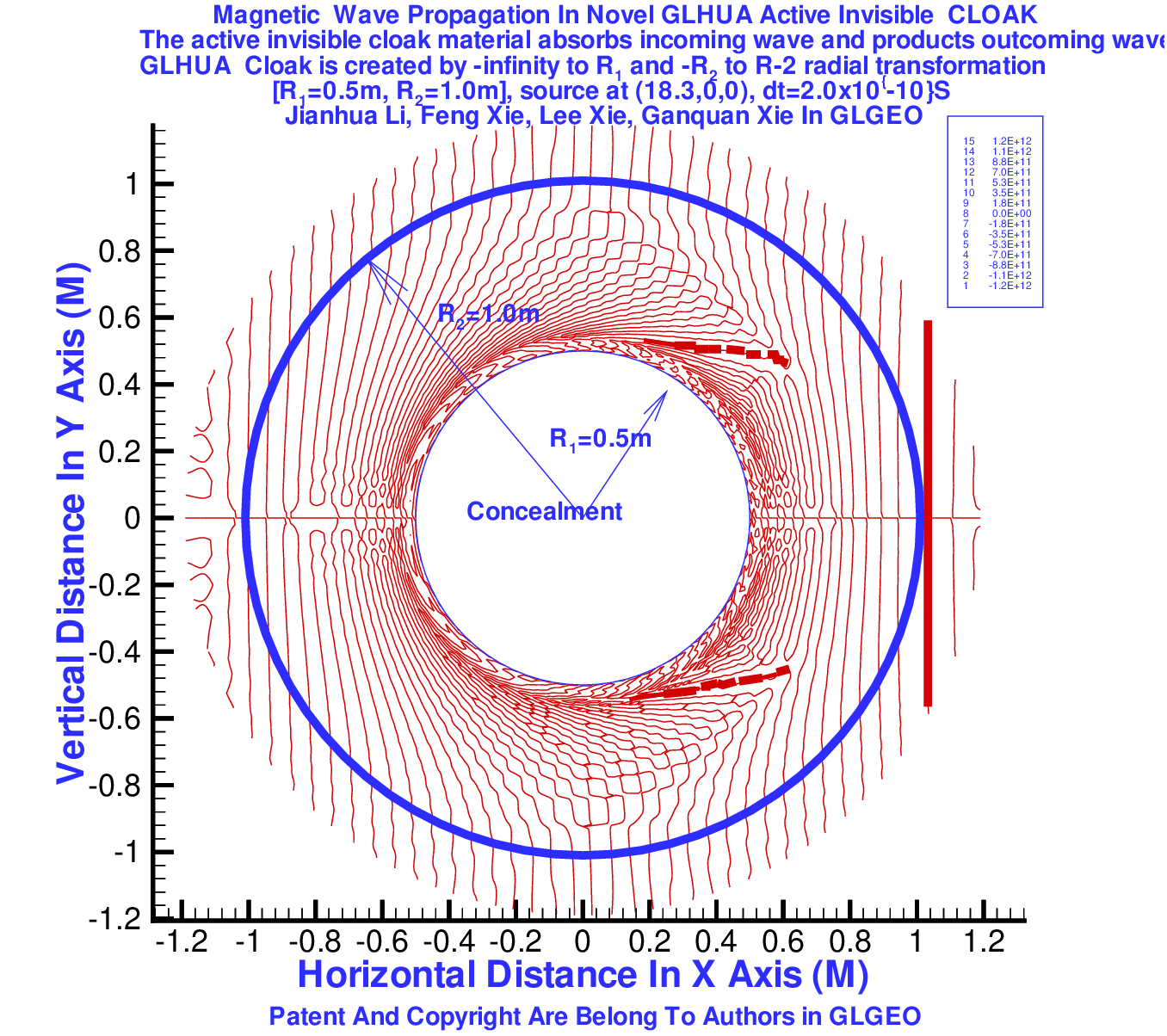}
\caption{ (color online) 
At first relative time step, the incident magneic wave front coming and tangent to the sphere surface  , the novel curve wave front is created by material GLHUA-3 invisible cloak
}\label{fig43}
%\end{minipage}
\end{figure}
\begin{figure}[h]
\centering
\includegraphics[width=0.86\linewidth,draft=false]{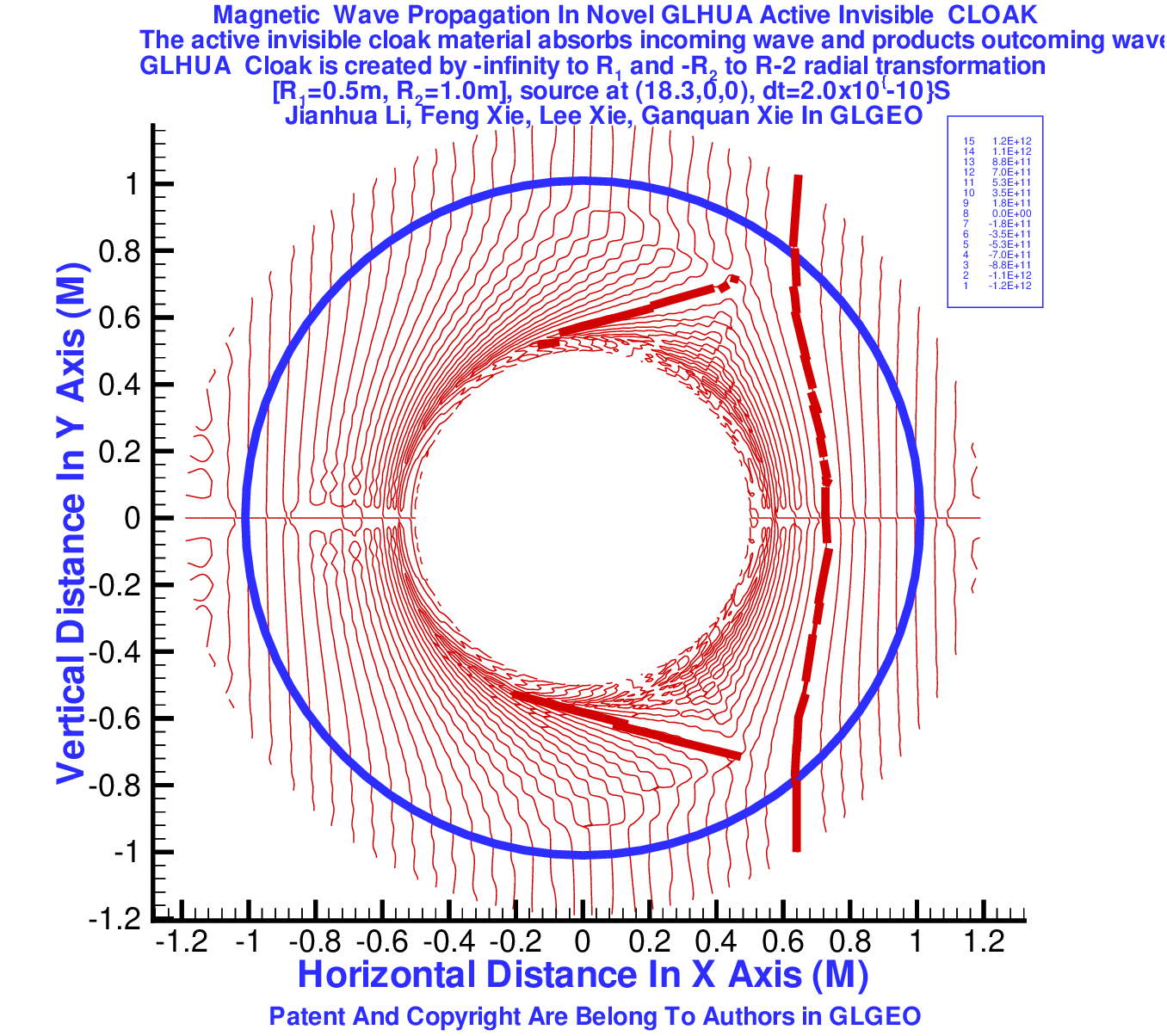}
\caption{ (color online) 
 At relative 5th time step, the magneic wave front incoming to right side of cloak, ,the material created curve wave front expand up and down
}\label{fig44}
%\end{minipage}
\end{figure}  

\begin{figure}[h]
\centering
\includegraphics[width=0.86\linewidth,draft=false]{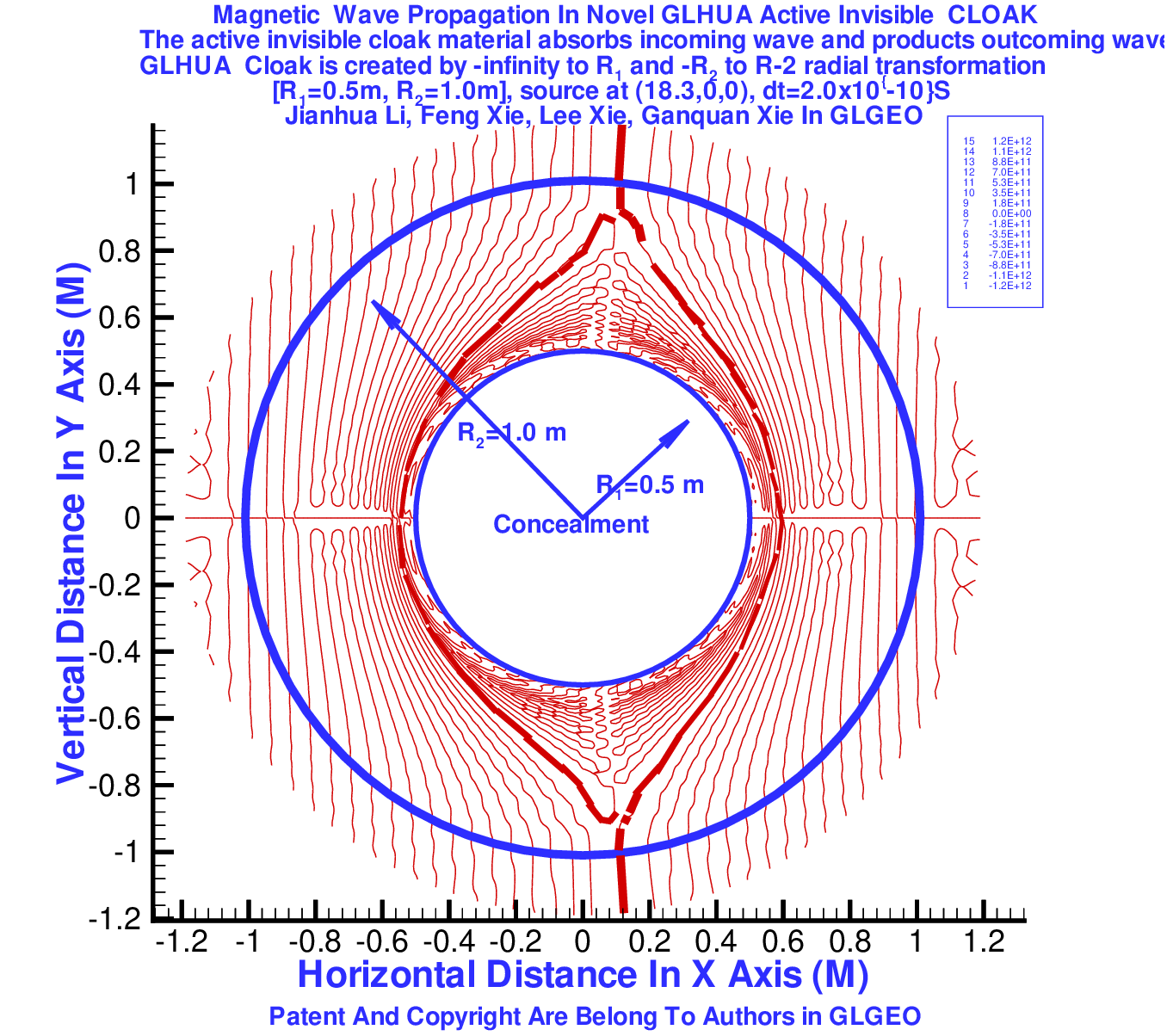}
\caption{ (color online) 
At relative 14 time step, the magnetic wave  incoming to right side of cloak, the material created curve wave front  form wave front on the left side of cloak. Both wave front near each other but no connected.	
}\label{fig45}
%\end{minipage}
\end{figure}
\begin{figure}[h]
\centering
\includegraphics[width=0.86\linewidth,draft=false]{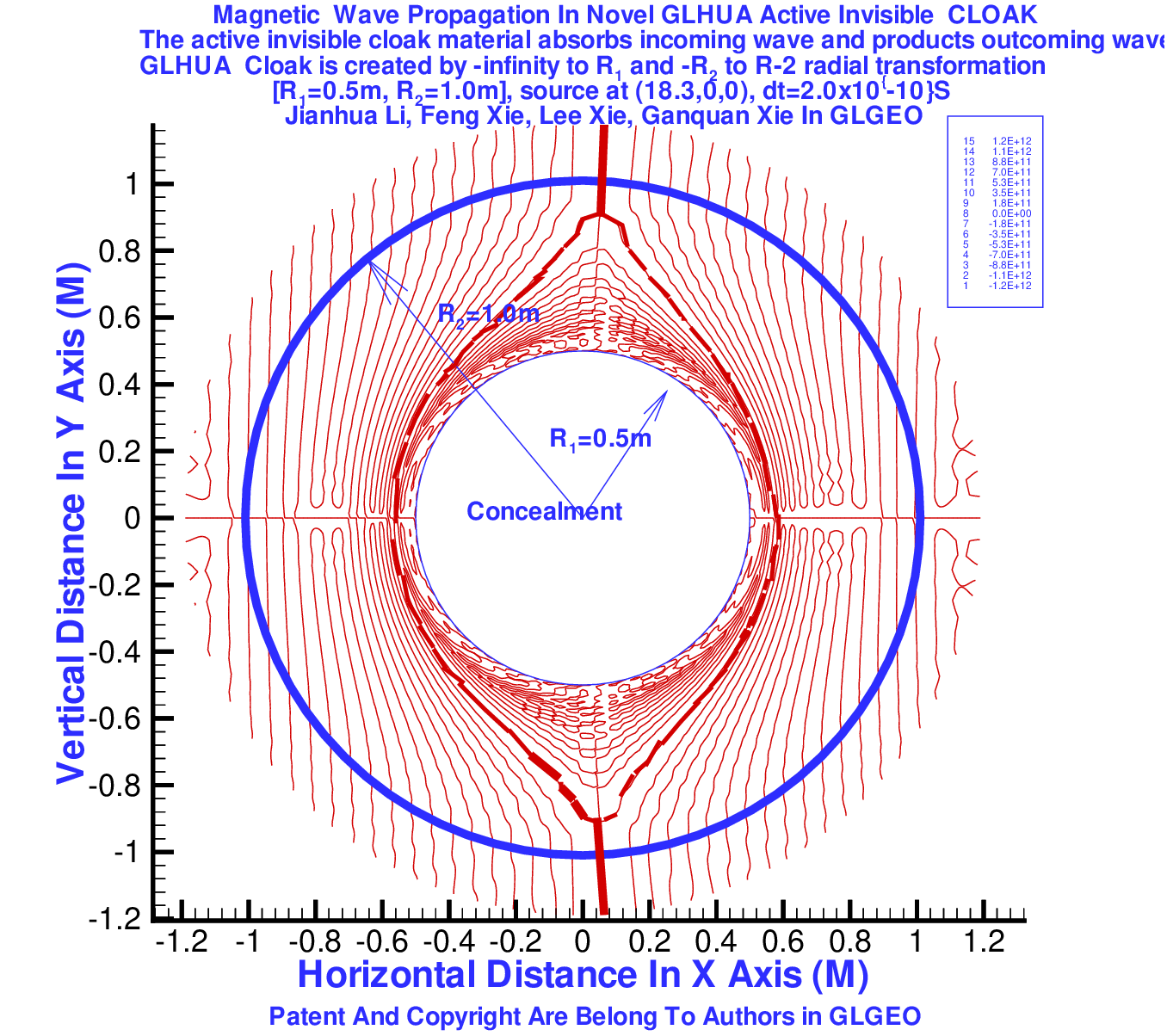}
\caption{ (color online) 
At relative 15 time step, the magnetic wave front incoming to right side of cloak, the material created curve wave frontform wave front in the left side of the cloak, the red incoming wave (1) and blue created wave front (2) are connected.	 
}\label{fig46}
%\end{minipage}
\end{figure}
\begin{figure}[h]
\centering
\includegraphics[width=0.86\linewidth,draft=false]{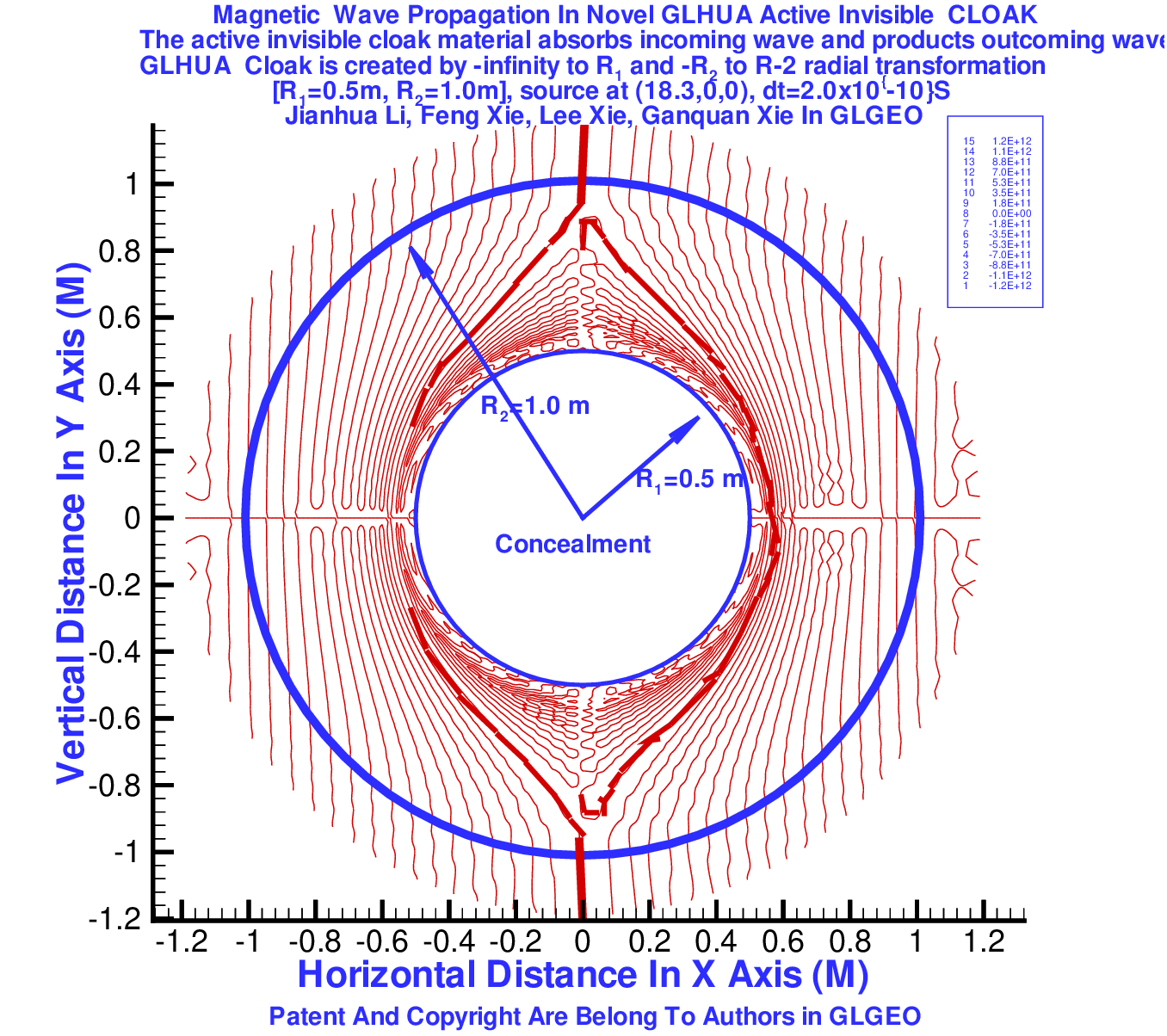}
\caption{ (color online) 
At relative 16 time steps, the curve wave front created by cloak material and incident wave in outside of cloak form wave front in the left side of the cloak, the incoming magneic wave front right side of cloak is disconncted with created wave 
}\label{fig47}
%\end{minipage}
\end{figure}
\begin{figure}[h]
\centering
\includegraphics[width=0.86\linewidth,draft=false]{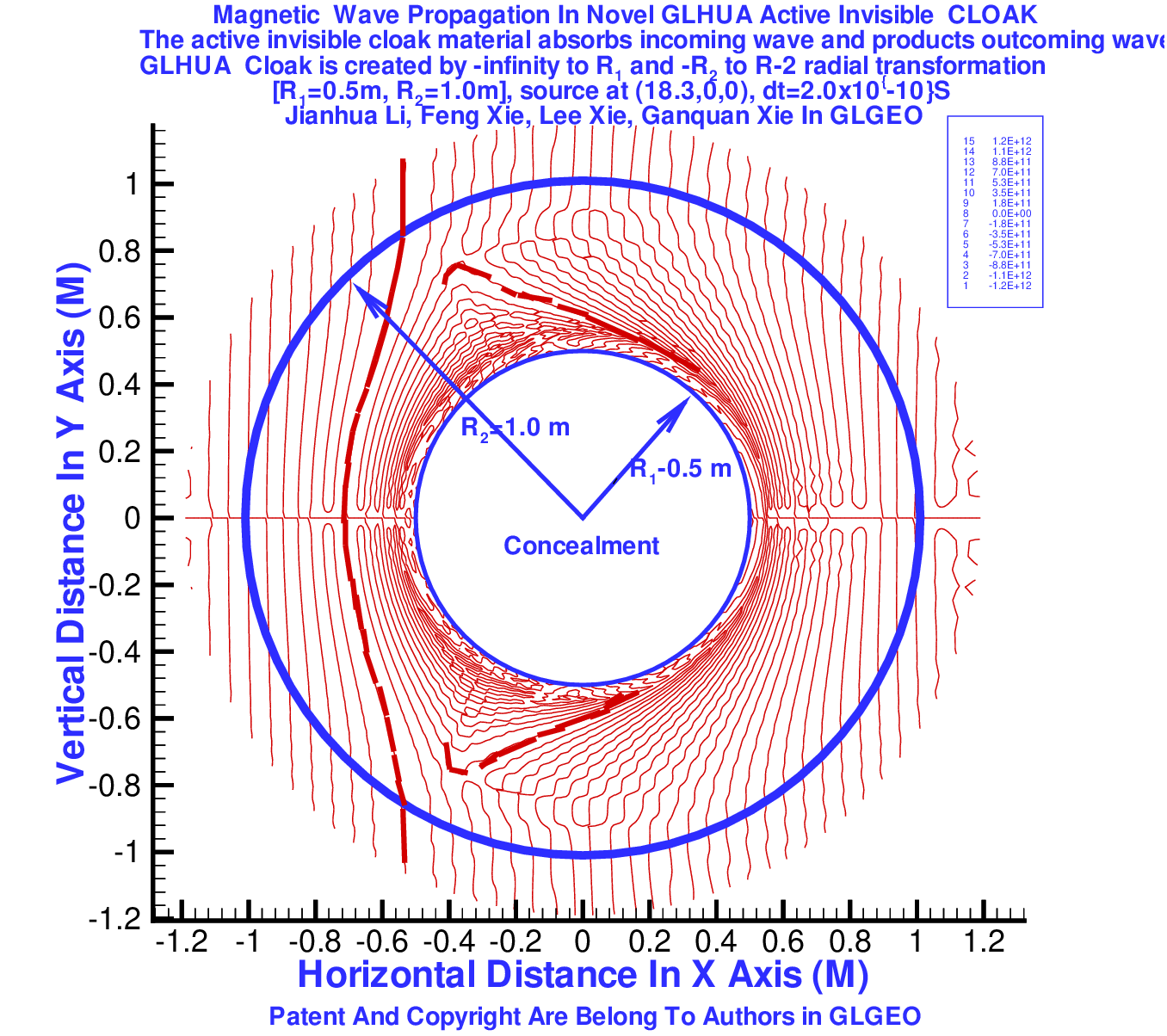}
\caption{ (color online) 
At relative 20 time steps, the curve wave front created by the cloak material is connected to the incident wavefront outside the cloak to form a wavefront and propagate to the left side of the cloak. The incoming magnetic wave is separated from the created wave (blue line (2)), and the incoming wavefront (red line (1)) is shrink, and wavefrontis behind from the created wavefront¡£
}\label{fig48}
%\end{minipage}
\end{figure}
\begin{figure}[h]
\centering
\includegraphics[width=0.86\linewidth,draft=false]{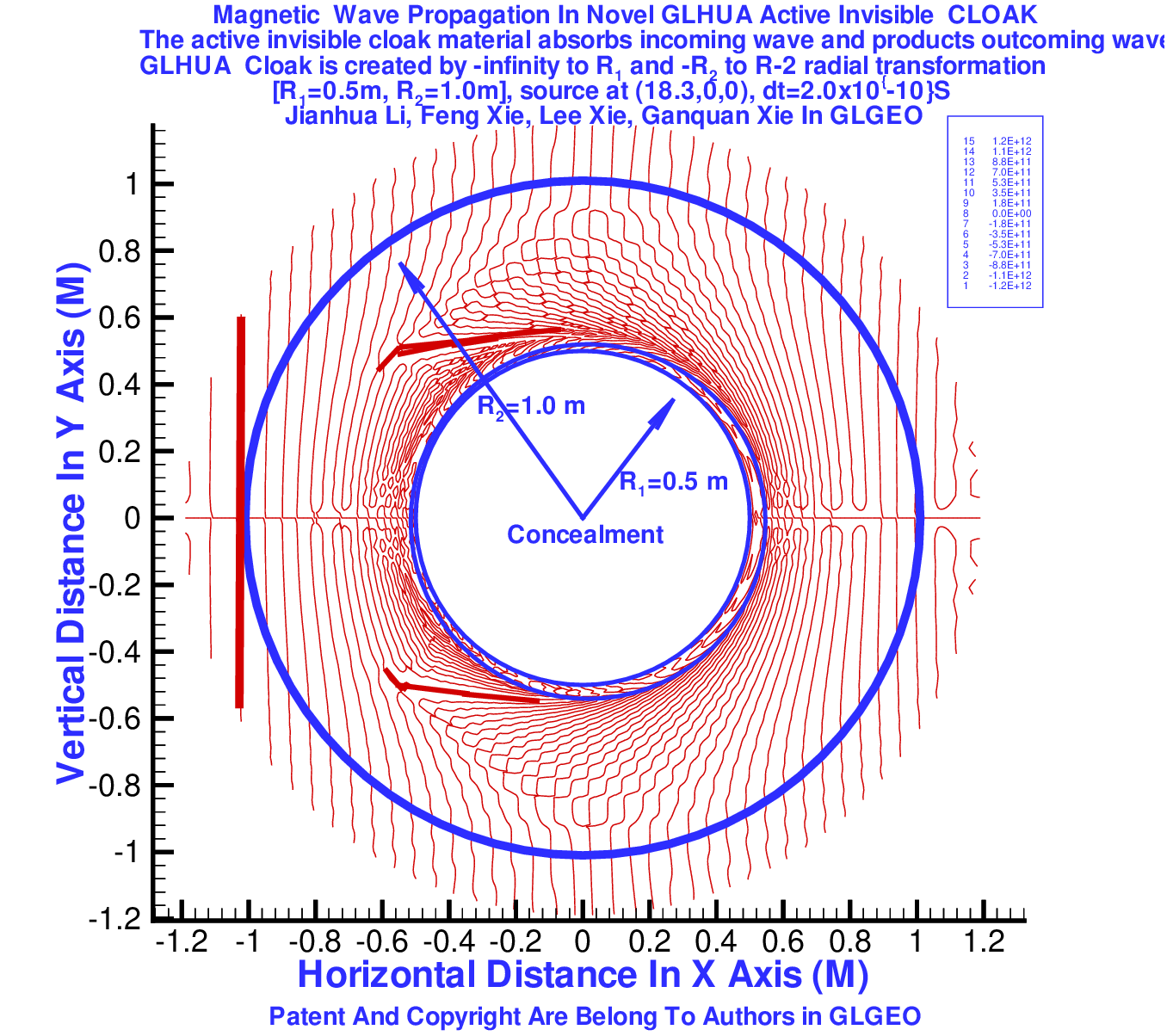}
\caption{ (color online) 
At relative 25 time steps, the curve wave front created by cloak material and incident wave front in outside of cloak are connected to form wave front that propagate to boundary, the left side of the cloak.The incoming magnetic wave is separated from the created wave (blue line (2)), and wavefront is behind from the created wavefront, and the incoming wavefront (red line (1)) is absorbed and shrink.	
}\label{fig49}
%\end{minipage}
\end{figure}
\begin{figure}[h]
\centering
\includegraphics[width=0.86\linewidth,draft=false]{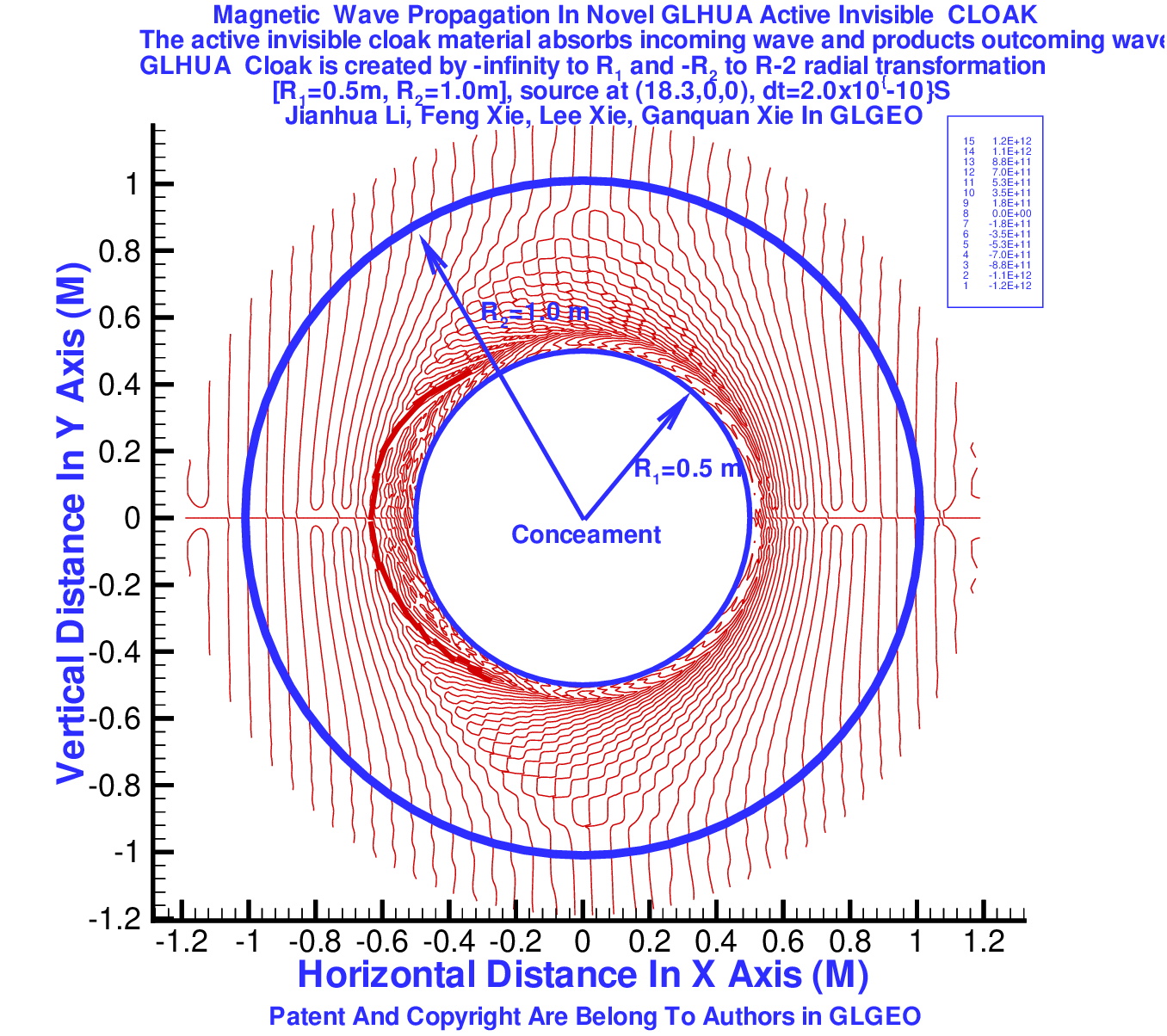}
\caption{ (color online) 
At relative 30 time steps, the material created curve wave front propagation
( blue line (2)) has already been out of cloak and did not disturb incident wave in outside of the cloak and make the cloak is invisible; the incoming maganetic wave front is shrink to a circle and absorbed and can not be penetrated to the concealment  . The inner sphere   is really cloaked concealment.	 
}\label{fig50}
%\end{minipage}
\end{figure}
\begin{figure}[h]
\centering
\includegraphics[width=0.86\linewidth,draft=false]{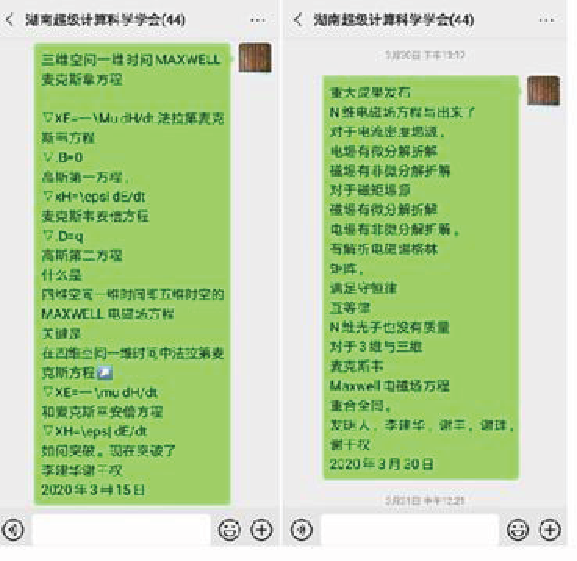}
\caption{ (color online) 
On March 15,2020, Feng Xie, Jianhua Li, Lee Xie, Ganquan Xie created
New N Dimensional Maxwell Equation and made this announcement in WeXIN group of
Hunan Super Computational Sciences Society, China on March 15,2020
}\label{fig51}
%\end{minipage}
\end{figure}

\begin{acknowledgments}
We wish to acknowledge the support of the GL Geophysical Laboratory and thank the GLGEO Laboratory to approve the paper
publication. Authors thank to Professor P. D. Lax for his concern and encouragements  Authors thank to Dr. Michael Oristaglio and Professor You Zhong Guo for their encouragments
\end{acknowledgments}

%=============================================

\end{document}